\documentclass[%
 reprint,
 amsmath,amssymb,
 aps,
]{revtex4-2}

\usepackage{graphicx}
\usepackage{dcolumn}
\usepackage{bm}
\usepackage{caption}
\usepackage{subcaption}
\usepackage{xcolor}

\begin{document}

\preprint{APS/123-QED}

\title{Automated identification and subtraction of gravitational-wave glitches using boundary refinement}

\author{Mohammad Abu Thaher Chowdhury}
 \email{chowdm4@rpi.edu}
 \affiliation{Department of Physics, Applied Physics and Astronomy, Rensselaer Polytechnic Institute.}
 
\altaffiliation[Also at ]{%
 Department of Physics and Astronomy, The University of Texas Rio Grande Valley, One West University Blvd., Brownsville, TX 78520, United States of America
}%





\begin{abstract}
Transient noise artifacts, or glitches, in gravitational wave strain data elevate the false alarm rate of astrophysical searches and degrade parameter estimation when overlapping a signal. No method previously identified a glitch's time boundary: existing detection and classification tools flag and label glitches without resolving their extent, forcing subtraction to run over padded windows that cost time and erase signal beyond the glitch itself. We present three boundary identification methods, AMPS (Amplitude-based Multi-glitch Pulse Segmenter), FLARE (Fitness-based Localization And Refinement Extractor), and CRISP (connected-region identification via spectrogram power), paired with three subtraction techniques, adaptive spline fitting, wavelet shrinkage, and their combination, across five glitches from the GravitySpy database spanning Advanced LIGO's first three observing runs. AMPS sets a boundary from a robust amplitude threshold, FLARE from the best fitness of a segmented spline fit, and CRISP from spectrogram power. Injecting a chirp signal on four broadband glitches, the combined technique recovers 95 to 97 percent of the injected signal-to-noise ratio, against roughly 64 percent for wavelet shrinkage alone. CRISP gives the most uniform boundary width across glitches; AMPS and CRISP both identify a boundary in a fraction of a second, two to three orders of magnitude faster than FLARE. For the glitch overlapping GW170817, a low-frequency residual persists under the default boundary, a deliberate tradeoff against removing signal power. Scattered light glitches remain untested and are the primary direction for future work.

\end{abstract}

\maketitle



\section{Introduction}
\label{sec:introduction}

The false alarm rate of searches for compact binary coalescences and generic short duration gravitational wave bursts is dominated by transient non-astrophysical artifacts, commonly called glitches. Sine-Gaussian glitches can produce significant matched-filter
triggers whose peak times are displaced from the glitch itself,
complicating the association of search triggers with transient
noise~\cite{Canton_2013}. A glitch that overlaps a genuine signal can cause a search pipeline to reject the signal outright, as occurred for GW170817~\cite{GW170817_2017}, or can bias the estimated parameters of the signal even when the overlap is partial.  The effect of glitch modeling on parameter estimation has also been examined for GW191109, where a coincident scattering glitch affected the inferred binary spin~\cite{Udall_2025}. Glitch mitigation for events observed during the first three observing runs is discussed in Ref.~\cite{Davis_2022}. In the third observing run of the Advanced LIGO and Advanced Virgo detectors, a substantial fraction of detected signals overlapped with glitches, owing in part to the elevated glitch rate in Virgo.

A number of methods exist for mitigating glitches, ranging from vetoes computed alongside the primary detection statistic~\cite{Abbott_2018, Girgaonkar_2024}, to gating of the affected time interval~\cite{Usman_2016, Pankow_2018, Stelner_2022, Aubin_2021}, to estimation and subtraction of the glitch waveform itself using either auxiliary sensor data~\cite{Tiwari_2015, Drigger_2019, Was_2021} or the strain time series alone~\cite{Cornish_2015, Merritt_2021, Mohanty_2023}. Adaptive spline fitting, implemented in the algorithm Swarm Heuristics based Adaptive and Penalized Estimation of Splines (SHAPES)~\cite{mohanty_2020, Mohanty_2023}, has been demonstrated to estimate and subtract broadband, short duration glitches with negligible impact on an overlapping astrophysical signal~\cite{Mohanty_2023}, and wavelet shrinkage~\cite{David_1995} provides a related nonparametric approach explored further in this work. Glitch-robust parameter inference has also been studied using normalizing flows~\cite{Sun_2024, sun_2026_robustparameter}.


Reliable glitch subtraction depends first on locating and recognizing a glitch in the data, a task distinct from the subtraction itself. Gravity Spy combines citizen-science visual inspection of Q-transform spectrograms with a convolutional neural network first pass that flags candidate glitches before sorting them into morphological classes such as Blip, Tomte, and Koi Fish \citep{Zevin_2017,BAHAADINI_2018}. Omicron takes a fully automated route, applying a bank of sine-Gaussian templates across the data stream to flag excess-power transients as glitch triggers without relying on a pre-labeled training set, and similarity-learning approaches push this further by embedding glitches in a learned feature space, which allows a previously unseen morphology to be flagged as anomalous before it has a name \citep{ROBINET_2021,Coughlin_2019}. This open-ended identification proved useful in practice: detector characterization efforts combining citizen science with machine learning uncovered new glitch classes during the third observing run, including fast scattering and low-frequency blip morphologies that had not been catalogued before \citep{Soni_2021}. Other methods identify transients through statistical or geometric departures from the detector background.


A t-SNE based approach embeds Q-transform features into a low-dimensional space and tracks outlier transients across successive days of observing time \citep{Ferreira_2025}, an autoencoder trained on auxiliary-channel data encoded through fractal dimension flags departures from the learned background distribution, uncovering glitches and glitch overlaps in about 6.6 percent of the data examined \citep{Laguarta_2024}, and Fisher information velocity models the detector noise floor as a point on a Riemannian manifold, flagging a glitch when the noise floor drifts or warps sharply across frequency without requiring a spectrogram or waveform template at all \citep{Kennington_2026}. The distinction between detecting a transient, assigning its morphological class, and selecting an interval for subtraction is important here. Our focus is on the last task. The subtraction methods applied to the selected intervals are described in Section~\ref{sec:glitch_subtract}.


Earlier work on individual glitch instances instead confirms a boundary by starting from the glitch's peak and scanning the surrounding time series visually until the strain tapers to the noise floor \cite{Mohanty_2023}. A recurring limitation across these identification methods is that the actual duration of a glitch is rarely known with precision. Most existing subtraction pipelines apply their algorithm over a generously padded window around a rough glitch location, wasting computation and risking removal of power from ordinary background. If the true glitch boundary could be located more tightly, the subtraction algorithm would only need to run on the glitch segment itself rather than across the surrounding data, which would cut down runtime substantially and make large-scale application across an observing run far more practical.

Recent work has also addressed the subtraction of scattered-light glitches. An adaptive time-varying filter approach decomposes the strain to isolate and subtract scattering arches without requiring a template for the arch shape, and reports recovery within a fraction of a percent of the injected glitch amplitude on synthetic data built to resemble O3-era scattering \cite{Longo_2026}. This is the identification and subtraction gap our own comparison leaves open, and we return to it in Section \ref{sec:conclusion}.


In this paper we present three boundary identification methods: AMPS, the Amplitude-based Multi-glitch Pulse Segmenter; FLARE, the Fitness-based Localization And Refinement Extractor; and CRISP, connected-region identification via spectrogram power, each described in Section~\ref{sec:glitch_id_methods}. We compare the boundaries these methods produce, and the subsequent subtraction results obtained using SHAPES and wavelet shrinkage, across Blip, Tomte, Koi Fish, the low frequency blip, and the glitch overlapping GW170817, drawn from the first, second, and third observing runs of Advanced LIGO.

\section{Glitch Identification Methods}
\label{sec:glitch_id_methods}

We describe three methods for selecting the time interval passed
to a glitch subtraction algorithm. AMPS uses amplitude thresholding,
sample clustering, and padding. FLARE uses spline fitness to select
candidate segments and a spectrogram to refine their boundaries.
CRISP uses spectrogram-based localization followed by amplitude
confirmation. Each method returns a start and end index for the
selected interval.


\subsection{AMPS: an amplitude-threshold identification method}
\label{sec:amps_method}

\begin{figure}[htbp!]
     \centering
     \captionsetup{justification=centering}
     \begin{minipage}{0.48\columnwidth}
        \includegraphics[width=\linewidth, height=1in]{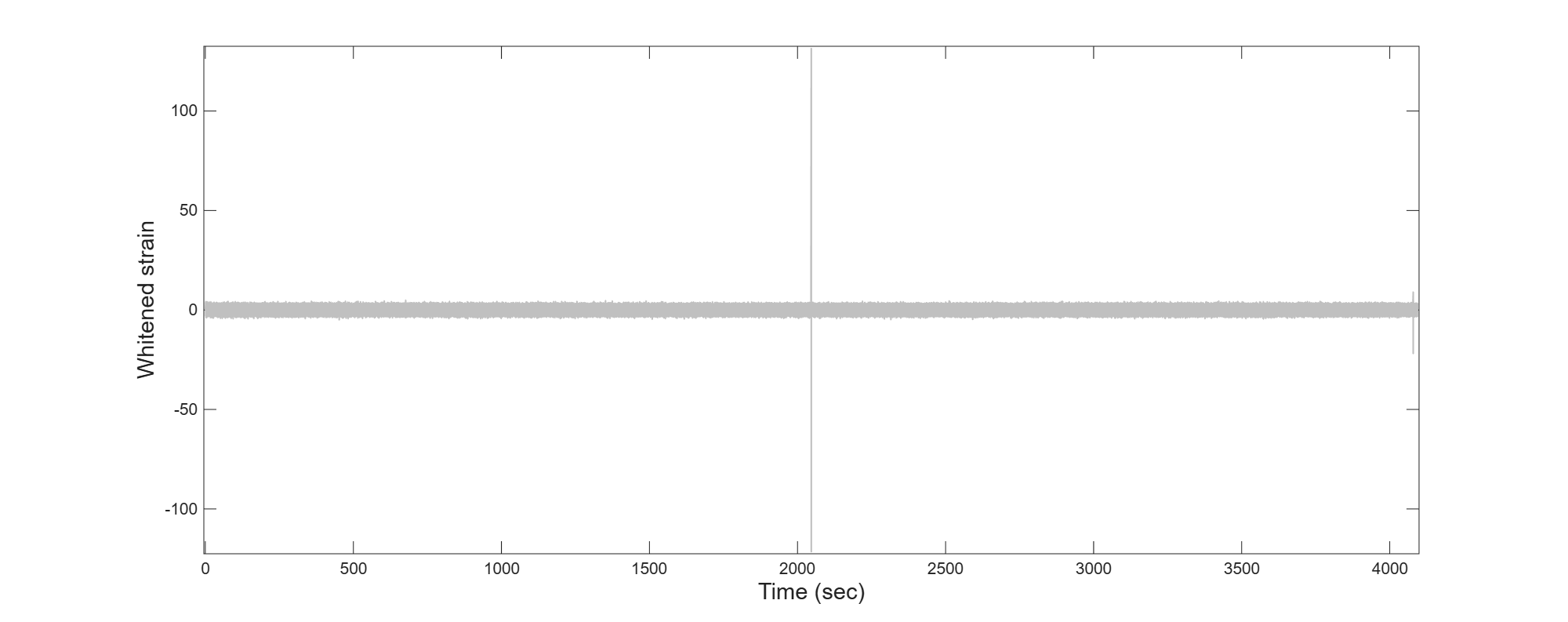}
    \end{minipage}
    \hfill
    \begin{minipage}{0.48\columnwidth}
        \includegraphics[width=\linewidth, height=1in]{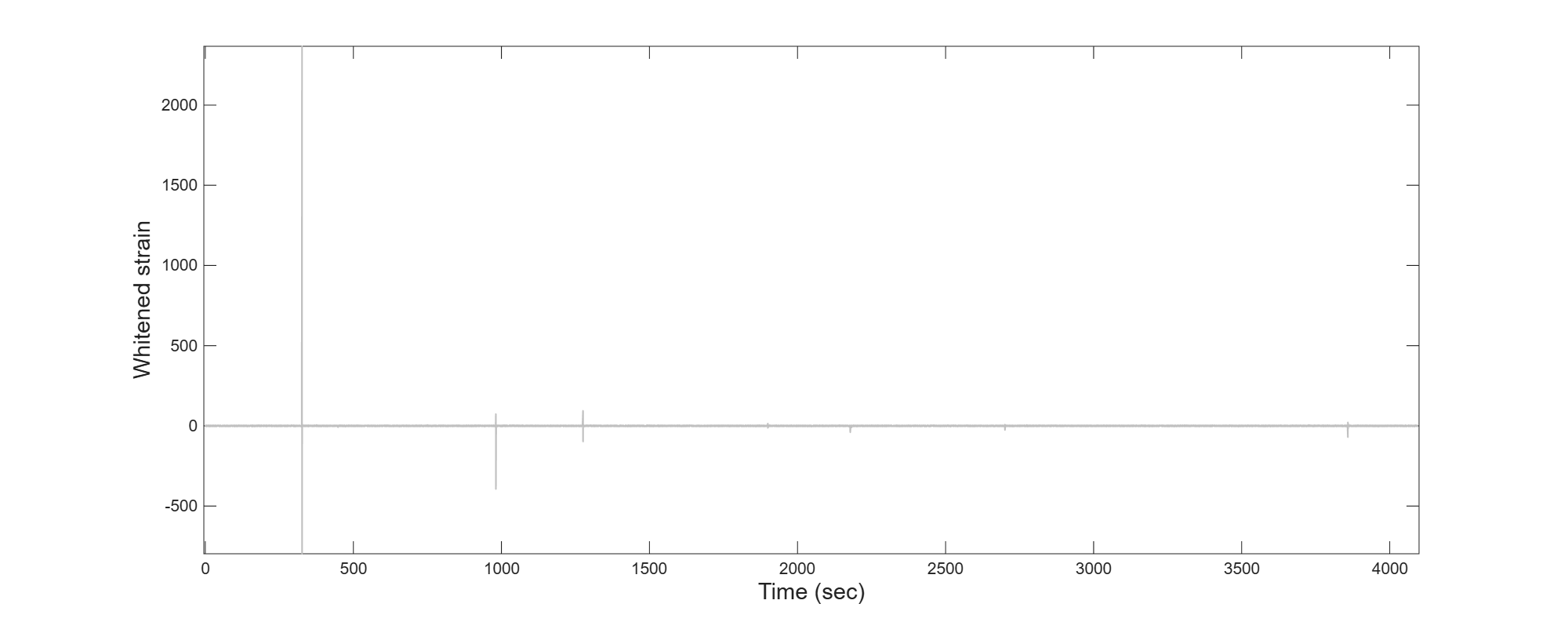}
    \end{minipage}

    \vspace{0.5em}

    \begin{minipage}{0.48\columnwidth}
        \includegraphics[width=\linewidth, height=1in]{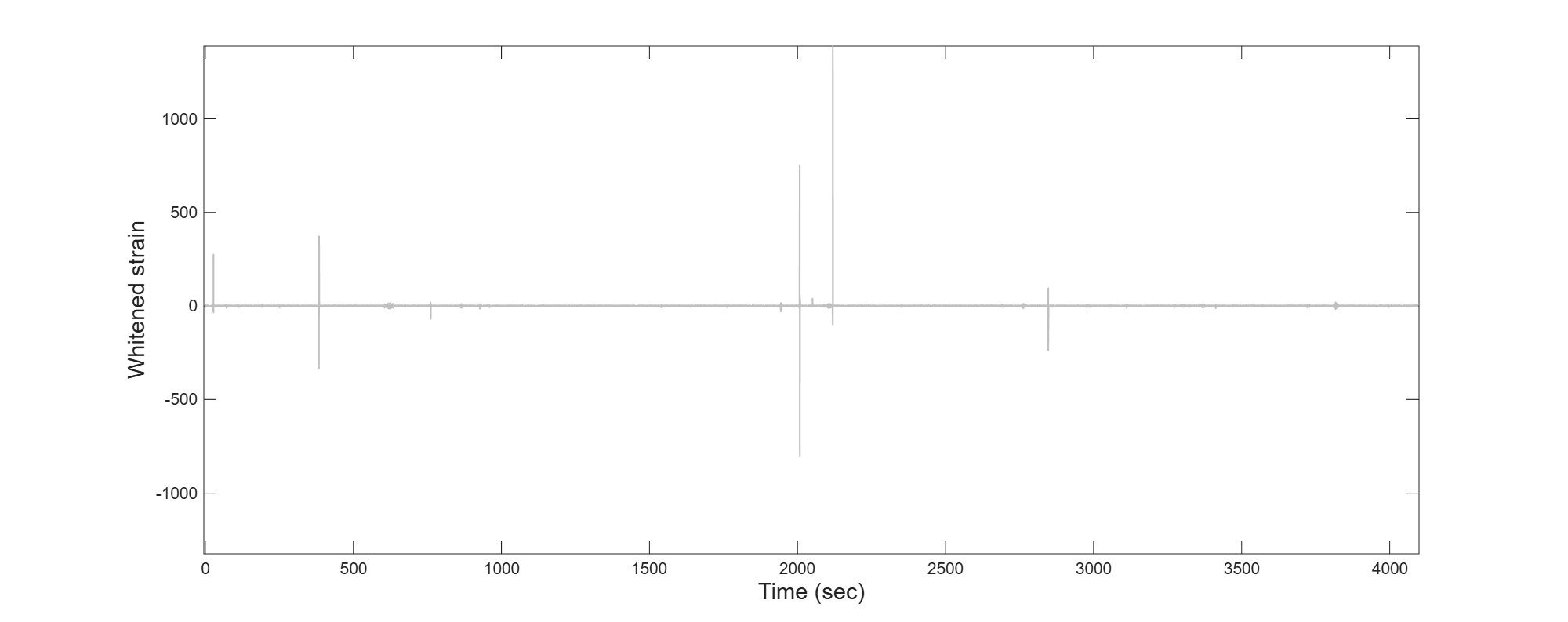}
    \end{minipage}
    \hfill
    \begin{minipage}{0.48\columnwidth}
        \includegraphics[width=\linewidth, height=1in]{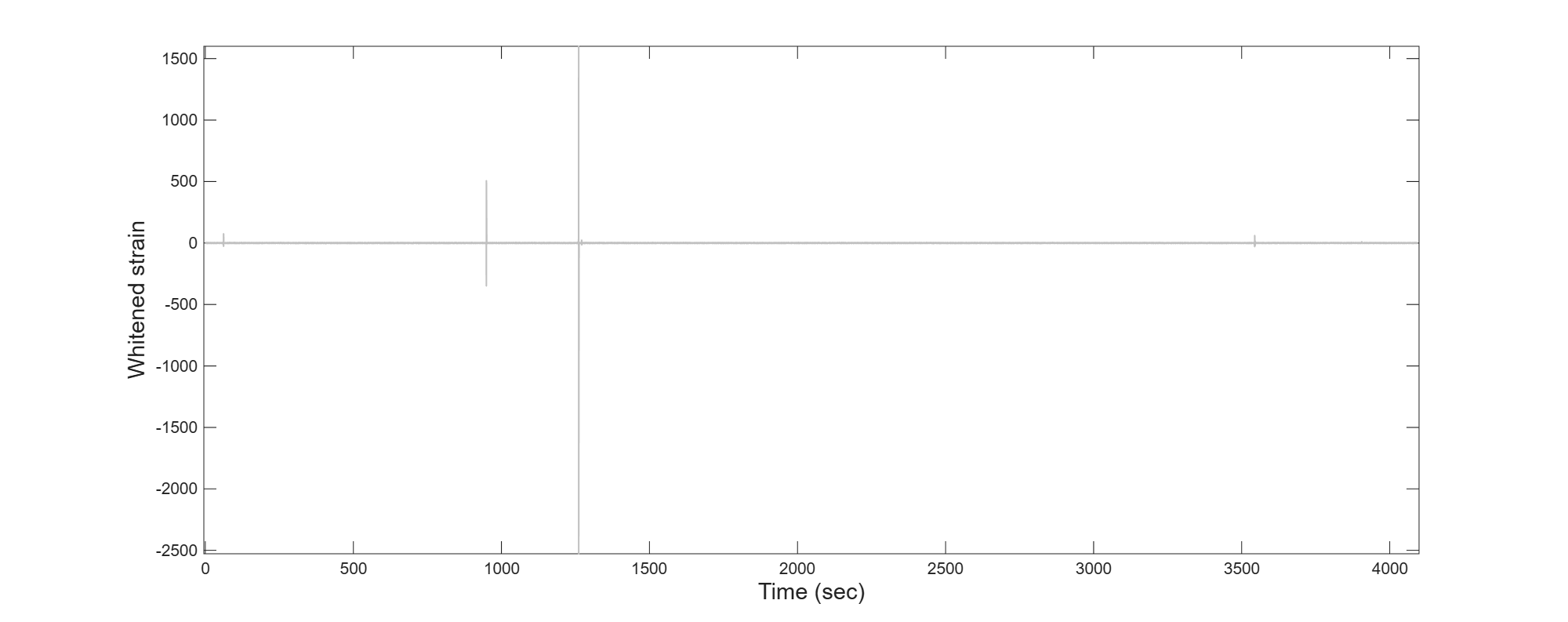}
    \end{minipage}
     \caption{Whitened strain amplitude (vertical axis, dimensionless) against time in seconds (horizontal axis) for four segments, each containing a different glitch class: GW170817 (top left), Blip (top right), Tomte (bottom left), and Low Frequency Blip (bottom right). Each segment also contains several other glitches. The vertical axis range differs from panel to panel, scaled to the amplitude of the glitch shown, so peak heights should not be compared across panels. The sharp amplitude excursion above the surrounding background is visible for every glitch shown, motivating the amplitude threshold method described below.}
     \label{fig:ampVar}
\end{figure}

The amplitude threshold method rests on a simple empirical observation: glitches of the classes considered in this work (Blip, Tomte, Koi Fish, low frequency blip, and GW170817) all show noticeably higher sample amplitude than the surrounding Gaussian background. This is visible in Figure~\ref{fig:ampVar}, where the whitened strain amplitude rises sharply above the flat background level at the location of each glitch, whether it is the glitch near GW170817, the Tomte glitch, or the blip glitch.

We refer to the identification procedure built on this observation as AMPS, the Amplitude-based Multi-glitch Pulse Segmenter. AMPS operates on the absolute value of the whitened strain and proceeds in three stages: setting a robust amplitude threshold, flagging and clustering the samples that exceed it, and padding the resulting clusters into working data segments.Amplitude thresholding has also been used to select non-Gaussian transients for gating in continuous-wave searches ~\cite{Stelner_2022}. In contrast, AMPS returns padded sample-index intervals for subsequent subtraction.

\begin{figure}[htbp!]
\centering
\captionsetup{justification=centering}
\includegraphics[width=\columnwidth]{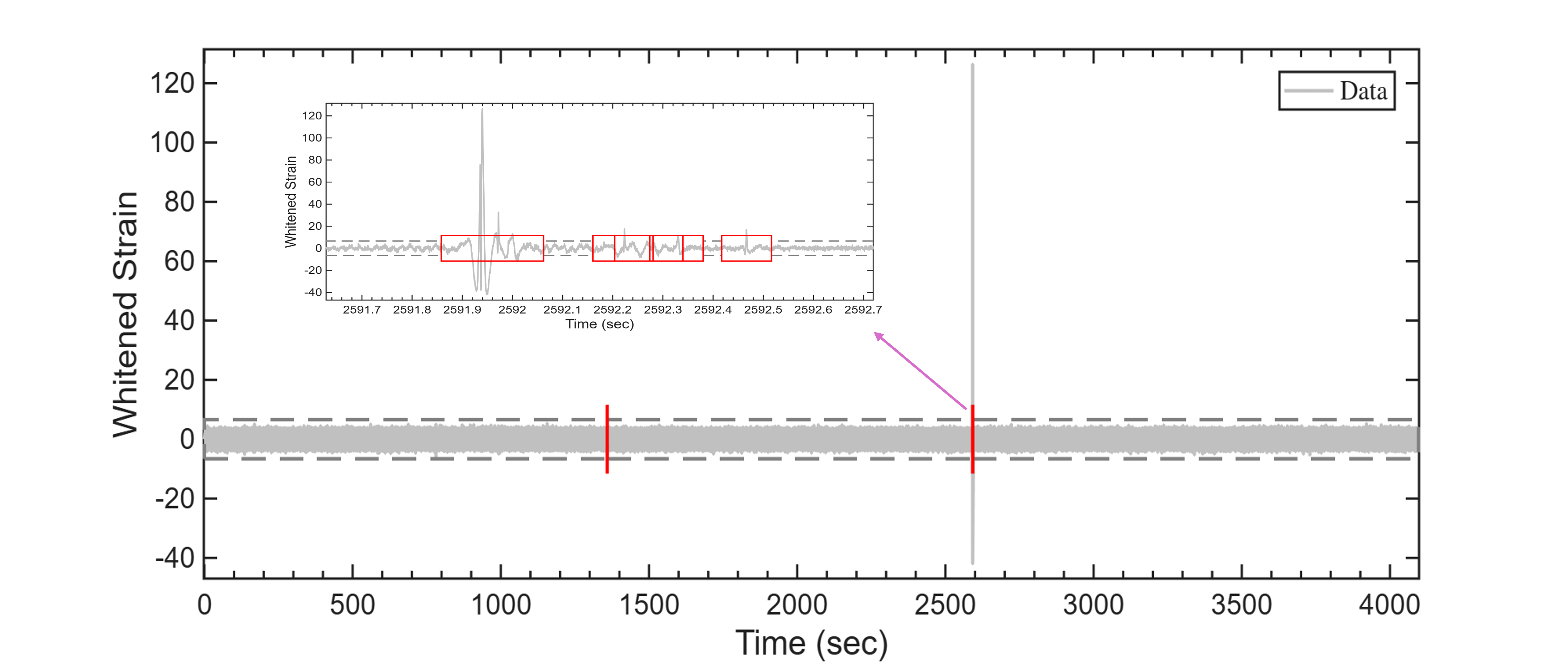}
\caption{Identification of multiple glitches by AMPS in a representative segment of O3a data. In the lower panel, the horizontal axis is time in seconds and the vertical axis is whitened strain amplitude, dimensionless. The dashed gray lines mark the amplitude threshold $y_{thr}$ of Eq.~\ref{eq:amps_threshold}, and red boxes enclose the segments where this threshold is exceeded. The pink arrow points to the region reproduced in the inset, spanning roughly 2591.7 to 2592.7~s on the same two axes. Within this short window, five separate glitches are each enclosed in their own red box, showing that AMPS separates closely spaced glitches into distinct detections rather than merging them into one, provided the amplitude drops back below threshold between them.}
\label{fig:gltch_idntfctn_ampth}
\end{figure}

The threshold is set from the median and the median absolute
deviation (MAD) of the absolute strain rather than from the sample
mean and standard deviation. This reduces the influence of loud
glitches on the estimated background level~\cite{Leys_2013}. Denoting the whitened strain samples by $y_i$, the threshold is
\begin{equation}
\begin{split}
    \sigma_{\mathrm{r}} = 1.4826 \cdot \mathrm{MAD}(|y_i|), \\
    y_{\mathrm{thr}} = \mathrm{median}(|y_i|) + k \sigma_{\mathrm{r}},
\end{split}
\label{eq:amps_threshold}
\end{equation}
where the scale factor 1.4826 converts the MAD into an estimate of the standard deviation under a Gaussian assumption, and $k$ is a user-set multiplier that controls the sensitivity of the detector. A default of $k=10$ was used for the raw amplitude check across all five glitch classes in this work, and this value cleared the background cleanly without flagging spurious noise fluctuations for any of them. Under the Gaussian assumption used to whiten the strain and underlying the MAD-based threshold construction~\cite{Leys_2013}, this threshold corresponds to a two-sided false-alarm probability of order $10^{-23}$
per sample. Combined across the full exposure of the O1, O2, and O3 observing runs used in this work, the expected number of false-flagged samples remains below $10^{-11}$ , leaving substantial margin against non-Gaussian tails in real detector noise while remaining sensitive to genuine glitches. Every sample with $\left|y_i\right|\geq y_{thr}$ is flagged and its index recorded. Figure~\ref{fig:gltch_idntfctn_ampth} marks this threshold on a segment of O3a data as a pair of dashed gray lines, one at $y_{thr}$ and one at $-y_{thr}$, against which the flagged excursions can be compared.

The flagged indices are then partitioned into clusters using the gap between consecutive flagged sample indices, not the gap between their amplitude values. A gap of more than 100 samples between two consecutive flagged points was used to start a new cluster in this work, and this setting reliably separated distinct glitches without splitting a single glitch whose amplitude dipped briefly below threshold partway through. This index-based partitioning is what lets AMPS separate two glitches that occur close together in time from a single, larger glitch behaving the same way. The inset of Figure~\ref{fig:gltch_idntfctn_ampth} shows this directly: five glitches occurring within about one second of each other are each enclosed in their own red box, and the narrow gap between the third and fourth boxes marks a brief return of the strain below $y_{thr}$, which is what keeps AMPS from merging these two into a single detection.

Each cluster is finally extended by a fixed total pad, split evenly across its two ends so that an equal number of samples is prepended before the cluster start and appended after the cluster end. A pad of 200 samples on each side was used throughout this work to build the wing regions flanking each glitch, and this padding consistently captured enough quiescent background on both sides for the subtraction stage without extending far enough to risk overlapping a neighbouring glitch. Because the same number of samples is added on each side, the working segment carries a matched stretch of quiescent background both before and after the flagged excursion. Boundary cases are handled independently at each end of the data: the padded start of the first cluster is clipped to the first sample if it would otherwise fall before the start of the time series, and the padded end of the last cluster is clipped to the last sample under the equivalent condition at the other end. The output of AMPS is a list of start and end index pairs, one per identified glitch, that can be passed to a subtraction or characterization pipeline.

The strength of AMPS is its simplicity and speed: it requires no training, no auxiliary channel, and no iterative fitting. Its main weakness is that for weaker glitches the raw amplitude may not clear the threshold, which requires a low-pass filtering step before the check can succeed, and the resulting boundary from padding is comparatively coarse next to the best-fit spectrogram method described next.

\subsection{FLARE: a spline-fitness-guided boundary refinement method}
\label{sec:bestfit_threshold}

The second method leans on the \textsc{SHAPES}~(explained in Section \ref{subsec:shapes}) adaptive spline-fitting algorithm itself, rather than on the raw whitened amplitude used by AMPS. We refer to this identification procedure as FLARE, the Fitness-based Localization And Refinement Extractor. FLARE fits a spline~\cite{deBoor_2001, mohanty_2020} model with a fixed number of knots to each of a sequence of consecutive, overlapping segments spanning the full time series, using particle swarm optimization~\cite{mohanty_2012} to search over knot placements, and repeating the search for several independent optimizer runs per segment before keeping the run with the best fitness as the representative model for that stretch of data. Prior to fitting, the strain may optionally be passed through a short moving-average filter and resampled to a different sampling rate; both steps are optional, and the data is used unmodified when they are skipped. The trailing segment at the end of the series is merged inward with its predecessor whenever the data left over would otherwise fall short of a usable minimum length.

\begin{figure*}[htbp]
\centering

\begin{subfigure}{0.31\textwidth}
\centering
\includegraphics[width=\textwidth]{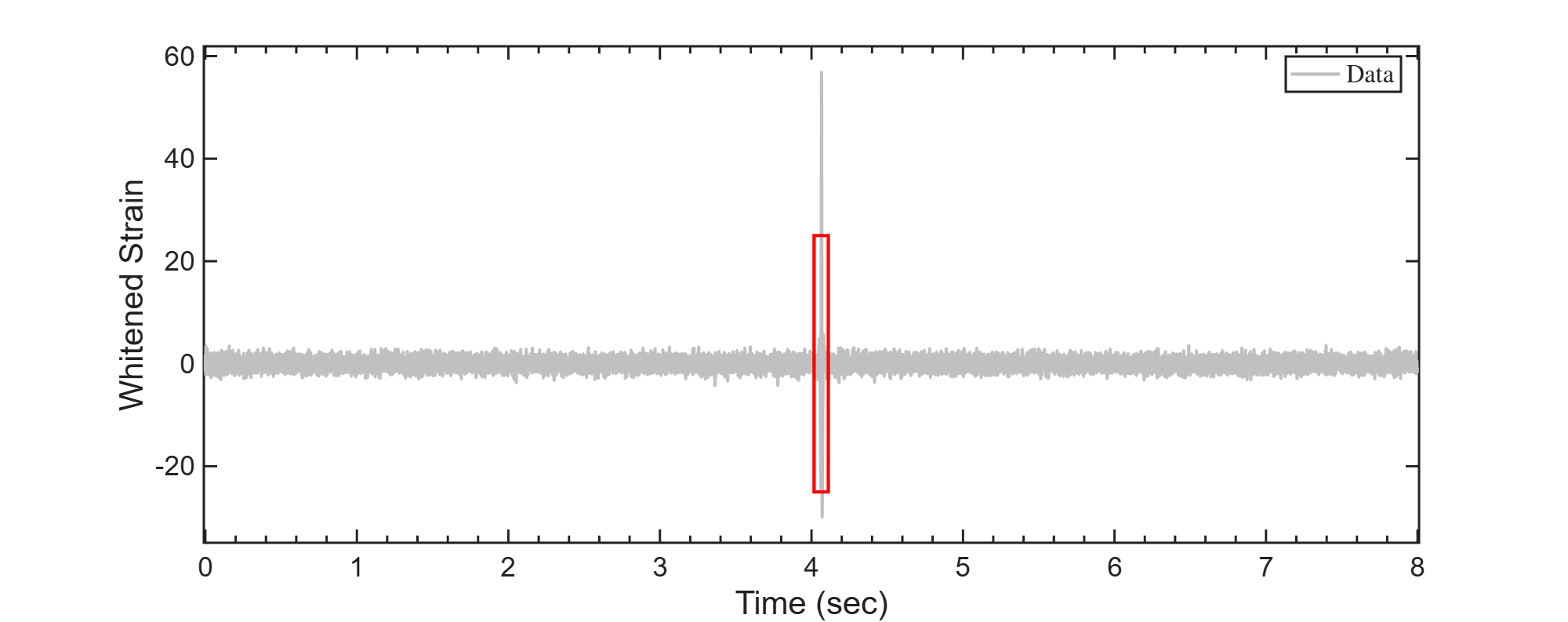}
\caption{}
\label{fig:kf_ident_a}
\end{subfigure}
\hfill
\begin{subfigure}{0.31\textwidth}
\centering
\includegraphics[width=\textwidth]{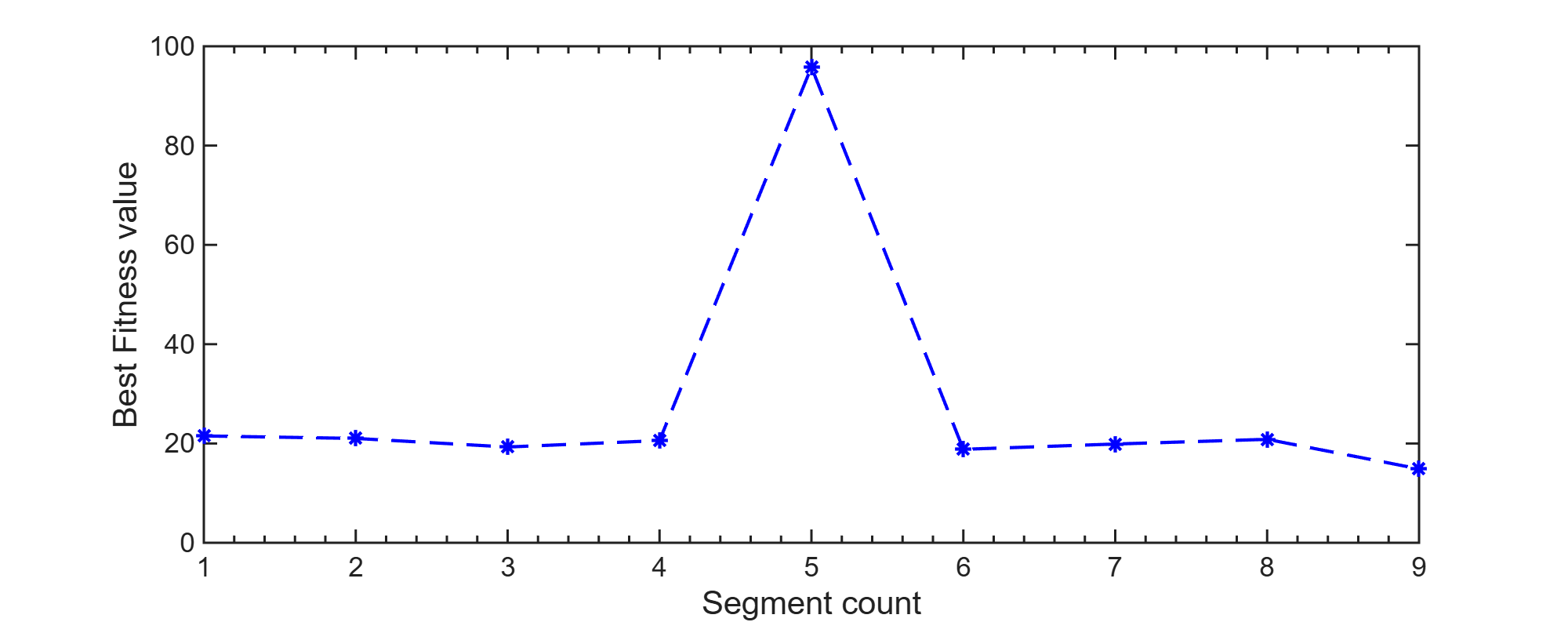}
\caption{}
\label{fig:kf_ident_b}
\end{subfigure}
\hfill
\begin{subfigure}{0.31\textwidth}
\centering
\includegraphics[width=\textwidth]{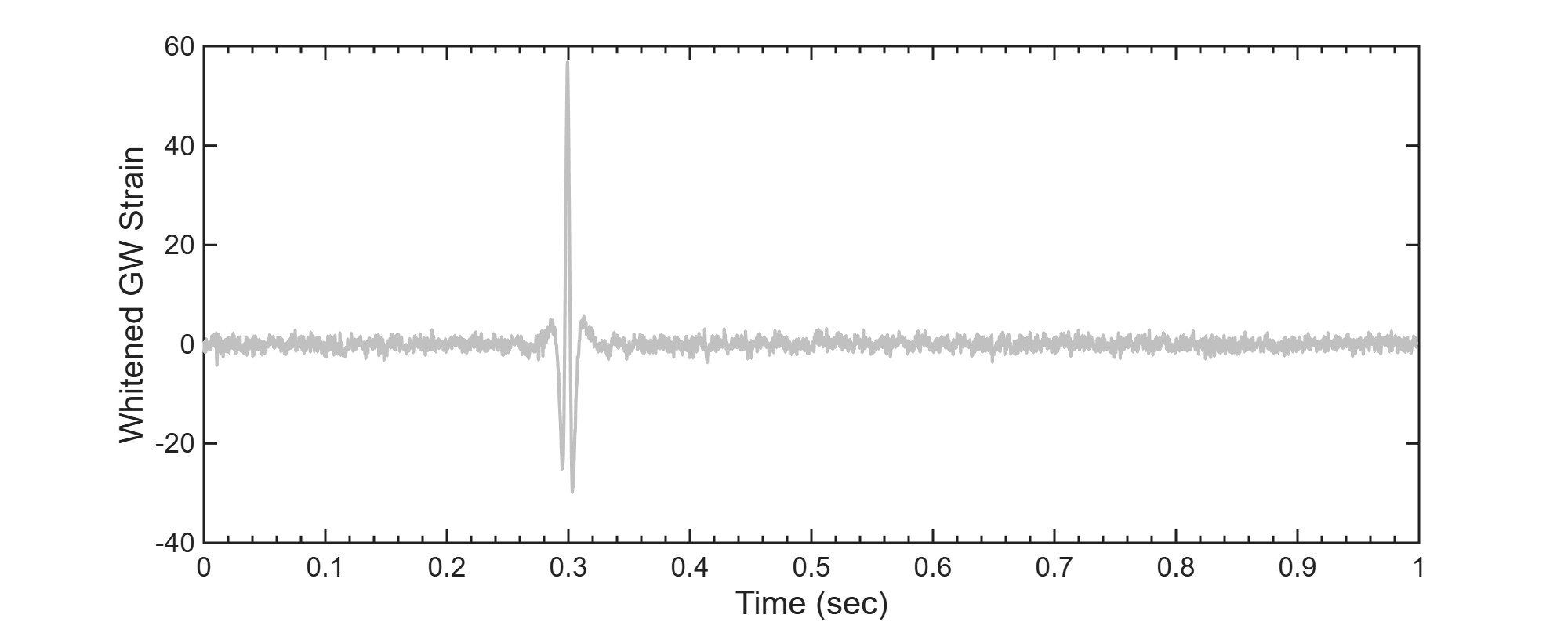}
\caption{}
\label{fig:kf_ident_c}
\end{subfigure}

\vspace{0.5cm}

\begin{subfigure}{0.31\textwidth}
\centering
\includegraphics[width=\textwidth]{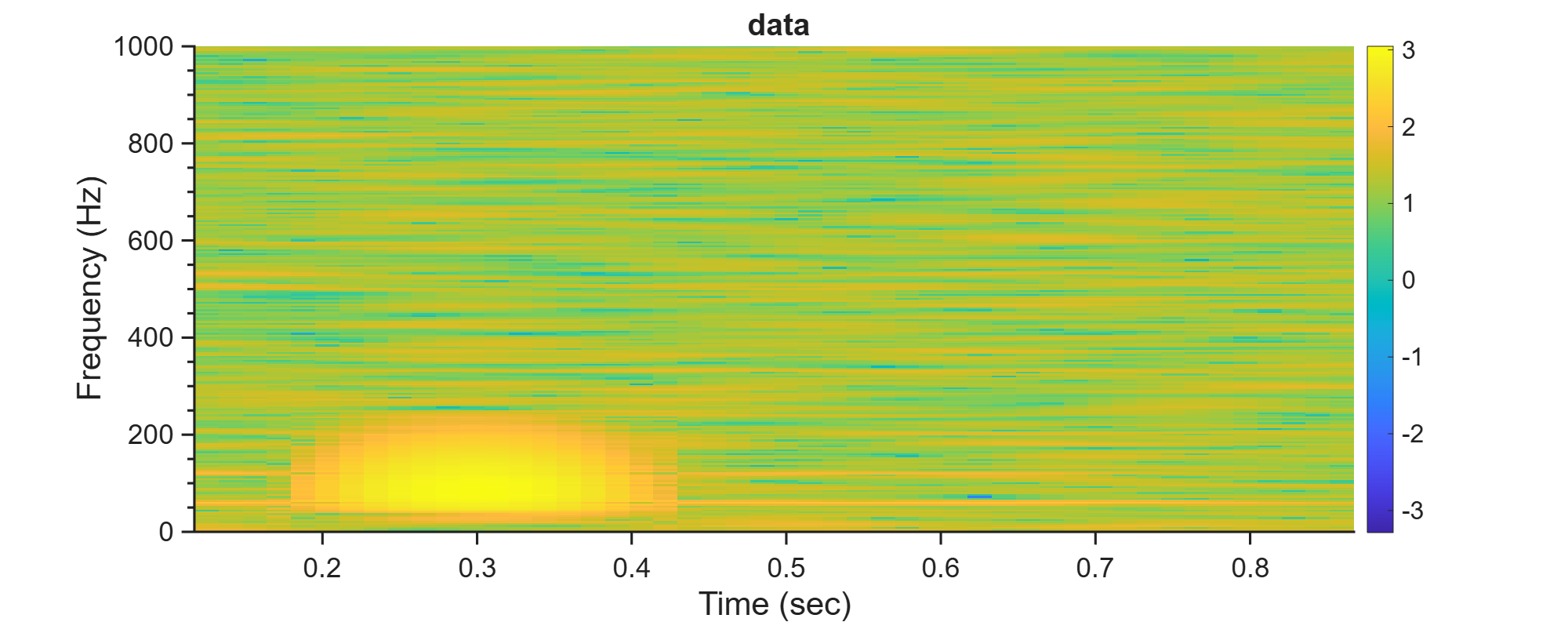}
\caption{}
\label{fig:kf_ident_d}
\end{subfigure}
\hfill
\begin{subfigure}{0.31\textwidth}
\centering
\includegraphics[width=\textwidth]{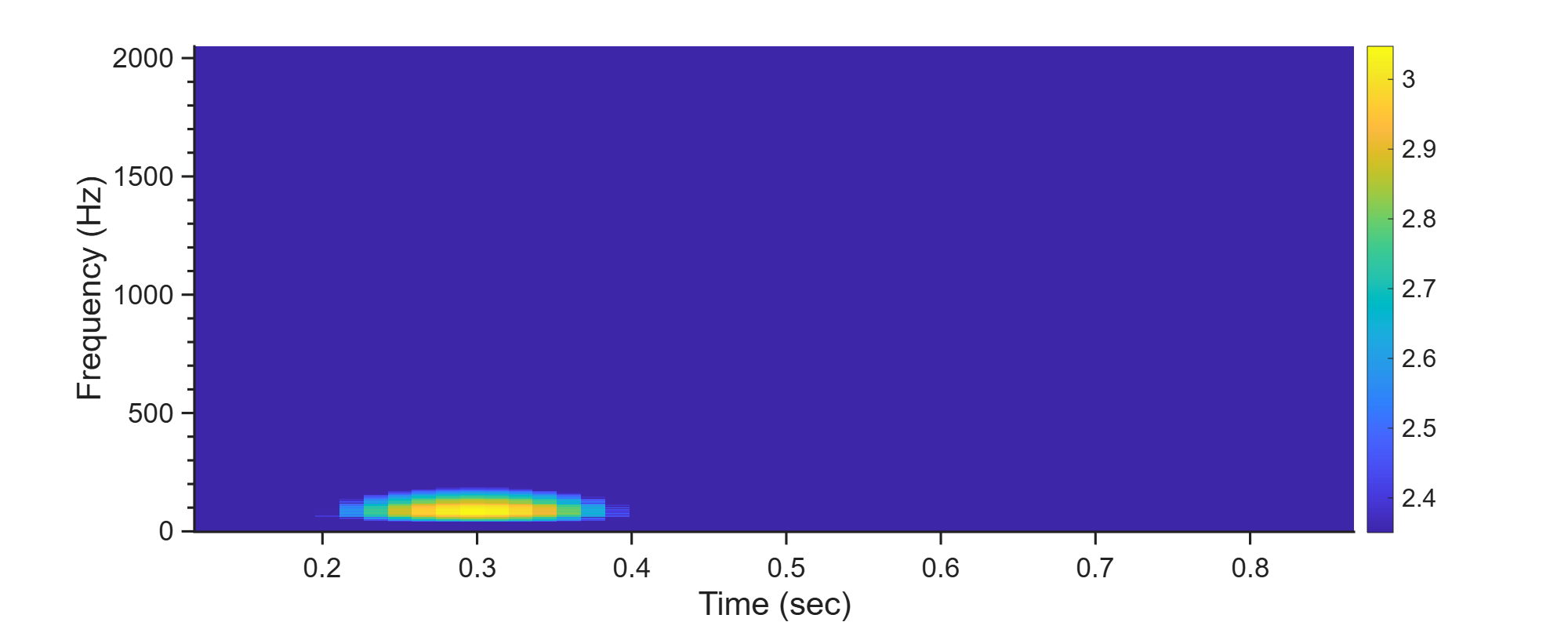}
\caption{}
\label{fig:kf_ident_e}
\end{subfigure}
\hfill
\begin{subfigure}{0.31\textwidth}
\centering
\includegraphics[width=\textwidth]{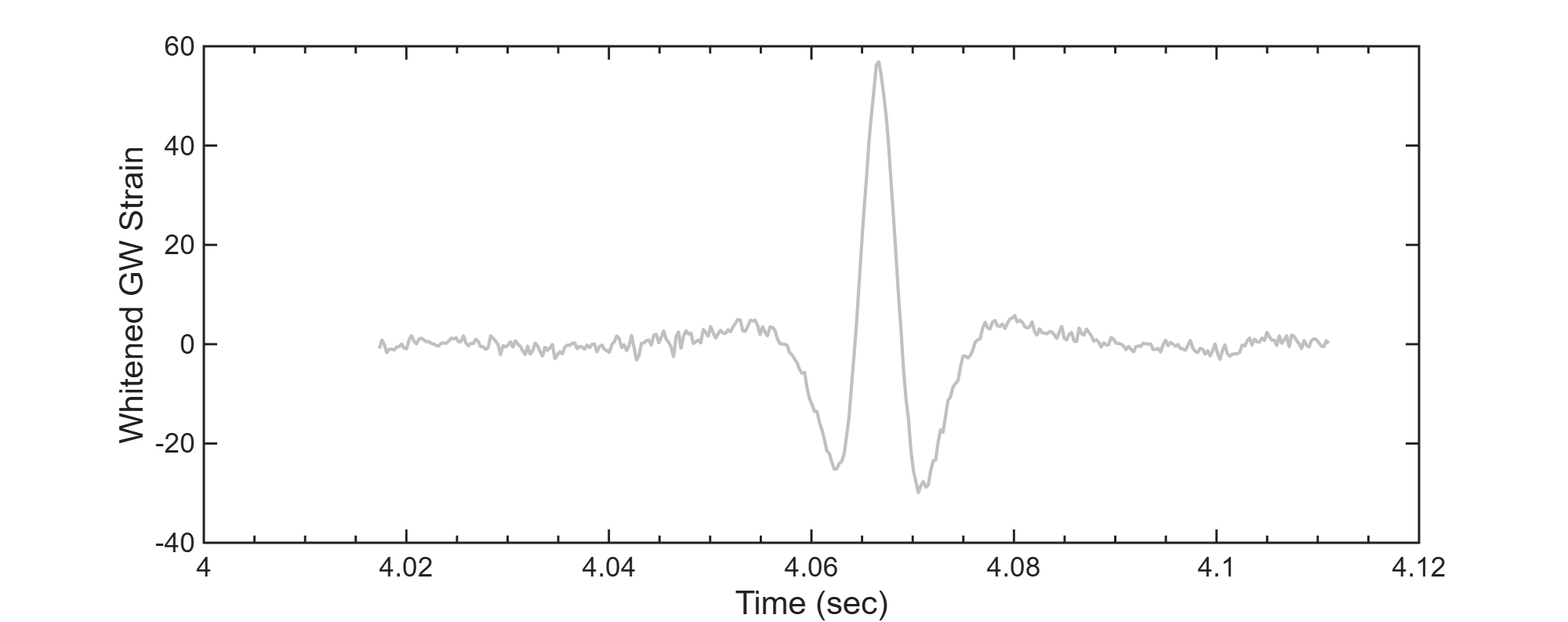}
\caption{}
\label{fig:kf_ident_f}
\end{subfigure}
\caption{Identification of a Koi Fish glitch using the bestfitness-spectrogram threshold method. (a) Whitened strain (vertical axis) against time in seconds (horizontal axis) for an 8-second stretch of O1 data, with the region FLARE identifies as the glitch marked by the red box. (b) Bestfitness value against segment count for the same stretch; the horizontal axis indexes the nine overlapping segments produced by splitting the 8-second stretch as described in the text, and the vertical axis is the bestfitness value of Eq.~\ref{eq:flare_threshold}, dimensionless. (c) Whitened strain (vertical axis) against time in seconds (horizontal axis) for the one-second segment selected by the bestfitness value in (b), with the glitch spike visible near \(t=0.3\)~s. (d) Spectrogram of the segment in (c), with time in seconds on the horizontal axis, frequency in Hz on the vertical axis, and normalized energy on a logarithmic color scale. (e) The same spectrogram after column-wise power thresholding, leaving only the pixels associated with the glitch, on the same three axes as (d). (f) Whitened strain (vertical axis) against time in seconds (horizontal axis) for the identified glitch boundary, obtained by mapping the surviving columns of (e) back onto the time series. Panels (a)-(c) illustrate the coarse detection step, and (d)-(f) illustrate the boundary refinement step.}

\label{fig:kfIdent}
\end{figure*}

Figure~\ref{fig:kf_ident_a} shows an 8-second stretch of O1 data containing a Koi Fish glitch, with the region FLARE eventually identifies marked by the red box. Applying the method to this stretch splits it into nine overlapping segments rather than eight, since each segment's overlap with its predecessor shifts the start of the next segment back by less than a full segment length; the segment overlap used throughout this work is 30 samples out of a nominal segment length set by the sample count in use. Figure~\ref{fig:kf_ident_b} plots the bestfitness value against this segment count, one point per segment, for the same stretch of data.

Because Gaussian background is comparatively easy to fit, its bestfitness value stays close to a stable baseline across the whole time series, while a segment containing a glitch produces a markedly higher bestfitness value since the spline has to work harder to track the sharp features. This is visible in Figure~\ref{fig:kf_ident_b}: the bestfitness value stays between roughly 15 and 25 for eight of the nine segments, then jumps to nearly 95 at segment 5, the segment that contains the glitch marked in panel (a). The working threshold is set from the fitness distribution itself rather than from a fixed cutoff. Writing \(f_j\) for the best-fit fitness of segment \(j\), the threshold is
\begin{equation}
f_{\mathrm{thr}} = \mathrm{mode}(f_j) + \Big(\max_{j} f_j - \min_{j \neq N} f_j\Big),
\label{eq:flare_threshold}
\end{equation}
where the minimum is taken over all segments except the final one, since a merged or undersized trailing segment can otherwise distort the spread estimate. Any segment whose fitness exceeds \(f_{\mathrm{thr}}\) is flagged as a glitch candidate, and its boundaries are mapped from the (possibly resampled) fitting grid back onto the original sampling grid before the refinement stage. Segment 5 clears this threshold and is passed forward as the sole candidate; Figure~\ref{fig:kf_ident_c} shows the corresponding one-second data window on its own, with the glitch spike visible near \(t=0.3\)~s against an otherwise flat background.

Once a candidate segment is flagged, a spectrogram is built for that segment alone, and the maximum power within each time column, taken across frequency, forms an envelope describing how the signal content varies over the segment. Figure~\ref{fig:kf_ident_d} shows this spectrogram for the Koi Fish segment, computed with the MATLAB \texttt{spectrogram} function~\cite{matlab_spectrogram}, returns the square root of the power of the short-time Fourier transform (STFT) of the signal~\cite{matlab_spectrogram_computation, boashash_2016}, plotted here on a base-10 logarithmic scale. A broadband patch of elevated energy, extending from near 0~Hz to above 200~Hz, sits between roughly 0.2 and 0.4~s, well above the flat, low-level background that fills the rest of the panel. Two separate fractions of this envelope are used at this stage, and it is worth keeping them distinct: one fraction governs the display thresholding shown in Figure~\ref{fig:kf_ident_e}, in which time-frequency pixels below the fraction are zeroed out, leaving only the patch just described; a second fraction, applied to the envelope rather than to the full time-frequency plane, determines which time columns are taken to belong to the glitch, and it is this second criterion that sets the tight glitch boundary read off from the surviving columns and mapped back onto the time series in Figure~\ref{fig:kf_ident_f}. Both fractions are reported in Section~\ref{sec:results} together with the other method parameters.

Panels (a) through (c) of Figure~\ref{fig:kfIdent} therefore correspond to the coarse detection step, while panels (d) through (f) correspond to the boundary refinement step, so this single figure closes the loop between the two mechanisms described above. Panel (f) shows the glitch confined to a window of about 0.01~s width, with the strain returning to background level cleanly on both sides.

FLARE additionally applies a downsampling factor of 4 to the fitting grid before the particle swarm search, a step not used by AMPS or CRISP. Fitting a spline through particle swarm optimization at the native sampling rate of the strain is computationally demanding, and repeating that search once for every segment across a series that is mostly Gaussian background spends far more effort than the bulk of the data warrants. On an 8-second stretch of data originally sampled at 32768 points, running FLARE at the native rate takes close to 30 seconds to reach the best-fitness values for every segment, and pushing the sampling rate higher still, to twice the native length through upsampling, raises this past 240 seconds. Reducing the sample count instead brings the timing down sharply: 22 seconds at 8192 samples and roughly 16 seconds at 4098 samples, a four- to eightfold count reduction from the native rate.

This speed carries two costs. More samples mean more work for the particle swarm optimizer on every segment, so fitting time grows with data length, as the timings above show. Less obviously, the strain at native sampling, and more so under upsampling, carries additional high-frequency noise that the spline model must absorb along with the glitch itself, which degrades how cleanly FLARE isolates the glitch segment: the best-fitness curve grows noisier and less sharply peaked at the true glitch location. Downsampling suppresses this noise as a side effect of reducing the sample count, which is part of why boundary identification holds up well despite the coarser data. The reduction has a floor, since the segment-boundary bookkeeping that maps flagged segments back onto the original time grid is tied to the sampling rate in use; lowering it further would require adjusting that bookkeeping rather than rerunning the same pipeline. A sample count in the low thousands, rather than the native tens of thousands, gives the best balance of fitting time and boundary noise for the glitch classes considered here, and is the operating point used throughout Section~\ref{sec:results}.

Compared to the amplitude threshold method, this approach identifies a noticeably tighter boundary because it draws on the frequency content of the glitch rather than on amplitude alone, at the cost of an additional spectrogram-construction step and a considerably heavier per-segment fitting cost.

\subsection{CRISP: a spectrogram-localized, amplitude-confirmed identification method}
\label{sec:crisp_method}

The third identification method combines spectrogram-based localization with an amplitude-based confirmation step. We refer to this method as CRISP. CRISP is built on the strengths of the two methods above: it borrows AMPS's amplitude confirmation for speed, and FLARE's use of a spectrogram for a tight, frequency-informed boundary, without carrying FLARE's per-segment spline-fitting cost. CRISP scans a full data stretch rather than requiring a pre-selected glitch segment. It determines the start and stop times from the magnitude of the STFT, computed as a spectrogram; AMPS is applied only to confirm a candidate inside the resulting boundary and never redefines that boundary.CRISP operates on the automatically whitened strain data. Whitening removes much of the unwanted noise and improves the visibility of the glitch classes considered here, including the low frequency blip.

\begin{figure}[htbp!]
    \centering
    \captionsetup{justification=centering}
    \begin{subfigure}[b]{0.48\columnwidth}
        \includegraphics[width=\linewidth]{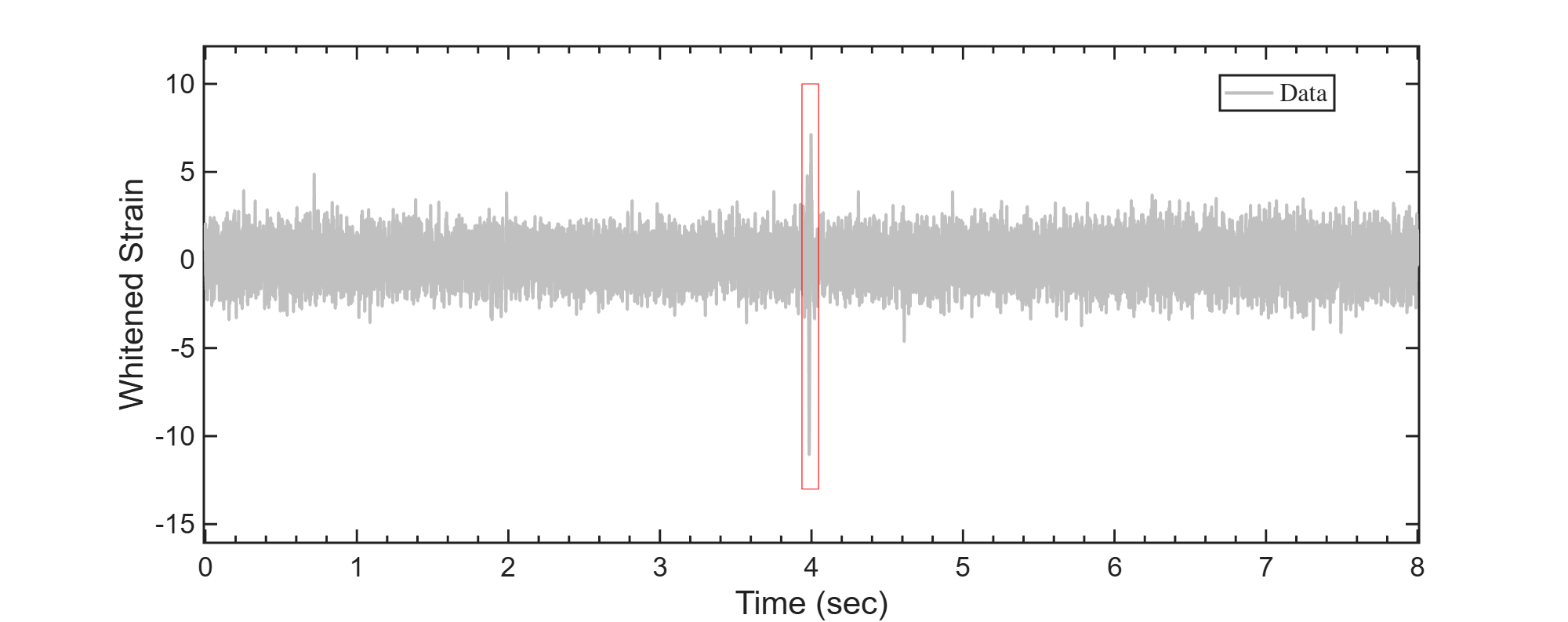}
        \caption{}
        \label{fig:crisp_a}
    \end{subfigure}
    \hfill
    \begin{subfigure}[b]{0.48\columnwidth}
        \includegraphics[width=\linewidth]{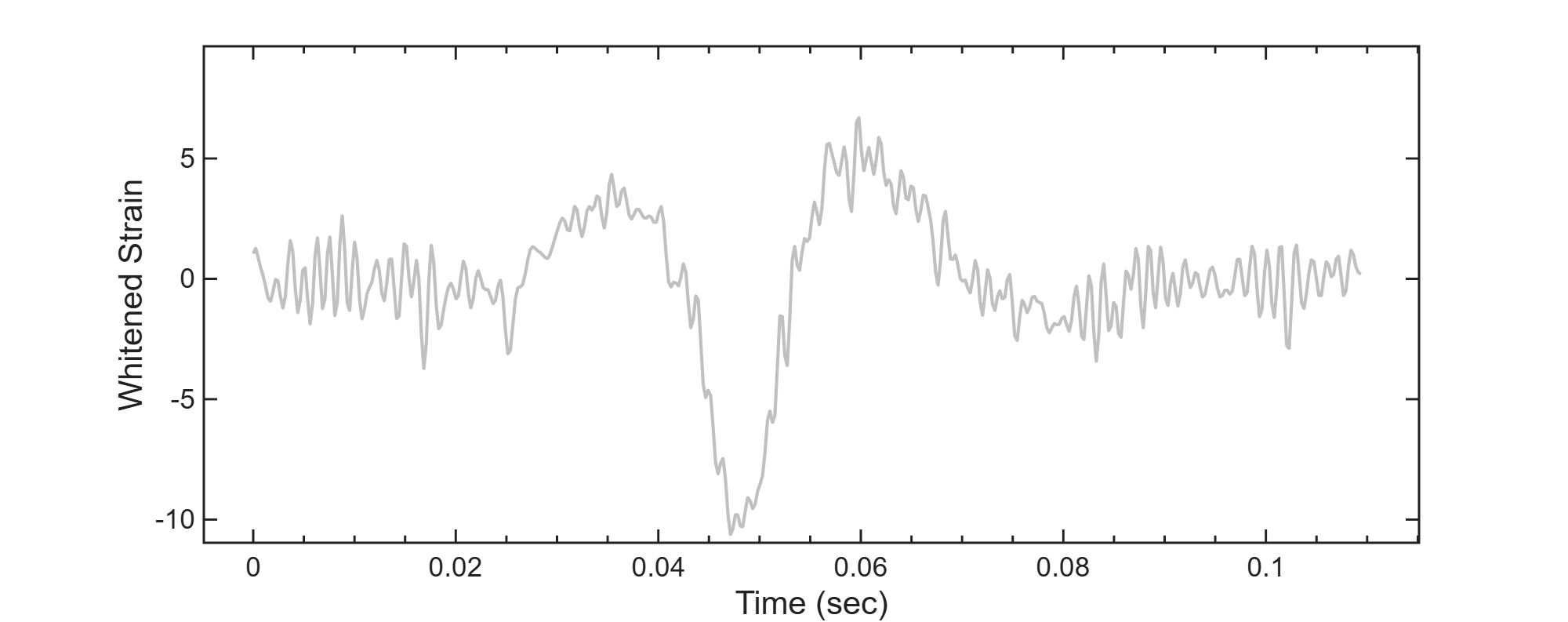}
        \caption{}
        \label{fig:crisp_b}
    \end{subfigure}
    \\[1ex]
    \begin{subfigure}[b]{0.48\columnwidth}
        \includegraphics[width=\linewidth]{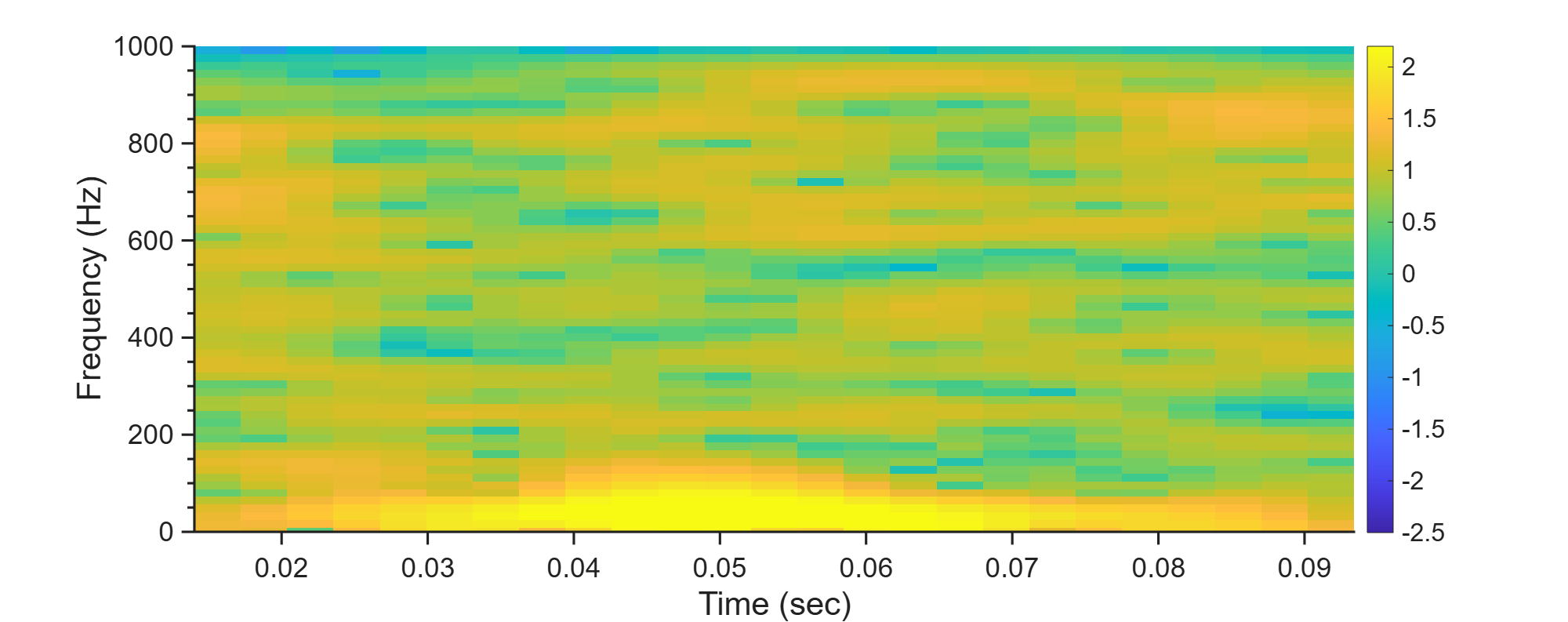}
        \caption{}
        \label{fig:crisp_c}
    \end{subfigure}
    \hfill
    \begin{subfigure}[b]{0.48\columnwidth}
        \includegraphics[width=\linewidth]{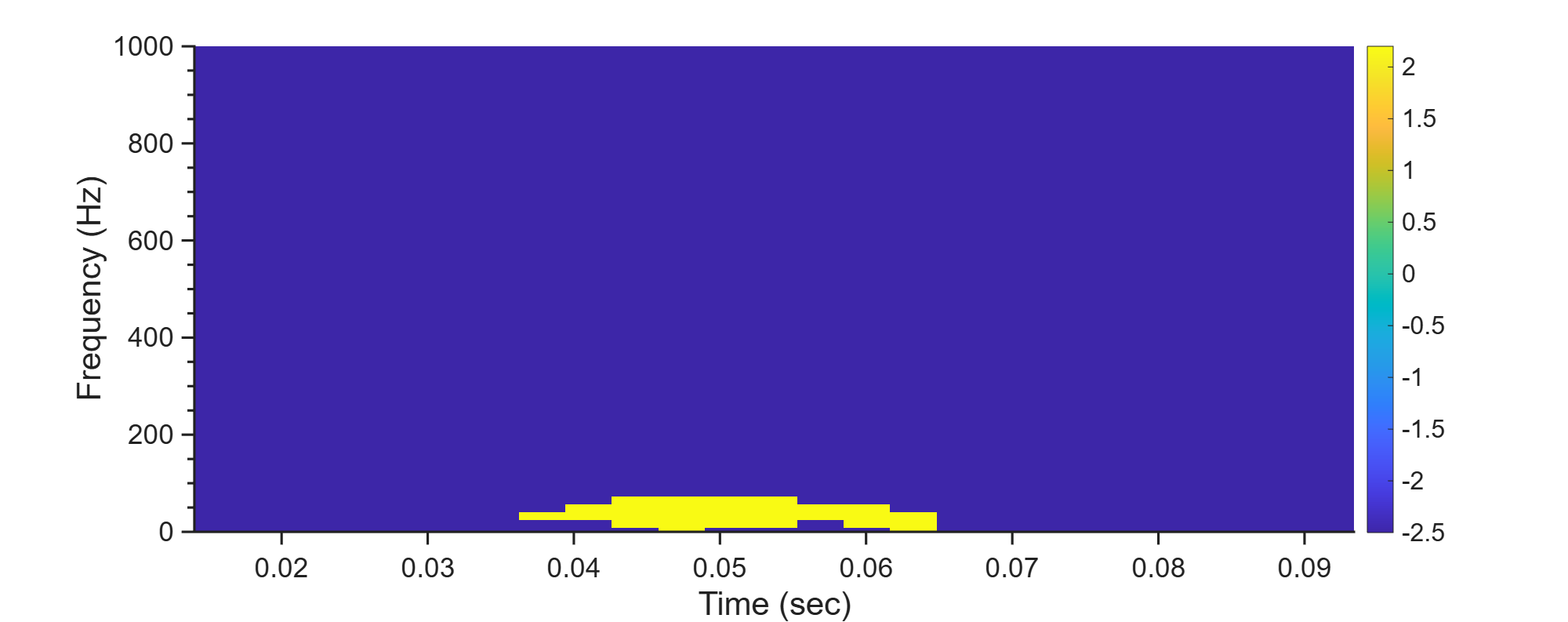}
        \caption{}
        \label{fig:crisp_d}
    \end{subfigure}
    \caption{CRISP applied to a Tomte glitch drawn from O3b data. (a) Whitened strain (vertical axis) against time in seconds (horizontal axis) for an 8-second stretch, with the interval CRISP identifies marked by the red box. (b) Whitened strain (vertical axis) against time in seconds (horizontal axis) for the identified boundary, zoomed in, showing the glitch spike against the surrounding background. (c) Spectrogram of the candidate segment before masking; the horizontal axis is time in seconds, the vertical axis is frequency in Hz, and the color scale is the base-10 logarithm of the short-time Fourier transform magnitude. (d) The same spectrogram after the second, stricter power mask, on the same three axes as (c); the start and stop columns of the surviving region set the boundary shown in (a) and (b).}
    \label{fig:crisp_pipeline}
\end{figure}

CRISP begins with a coarse scan over the complete time series. The data are split into overlapping blocks of 64~s duration with a 2~s overlap. Each block is turned into a spectrogram with a 512-sample Hann window, 480-sample overlap, and a 4096-point Fourier transform. The maximum STFT magnitude in each time column is computed, and a column is flagged as a candidate when this magnitude exceeds a robust threshold built from the median and median absolute deviation of the column maxima, with a default z-score of 3. Flagged columns within 2~s of each other are merged, giving the rough candidate intervals of Stage~A. Figure~\ref{fig:crisp_a} shows this stage applied to an 8-second stretch of O3b data containing a Tomte glitch: the red box marks the interval CRISP eventually identifies, against the whitened strain over the full stretch.

Each Stage~A candidate interval is extended by 0.5~s at its start and 0.5~s at its end, and the extended segment is re-examined with a finer spectrogram. Figure~\ref{fig:crisp_c} shows this finer spectrogram for the same glitch: the horizontal axis is time in seconds, the vertical axis is frequency in Hz, and the color scale is the base-10 logarithm of the short-time Fourier transform magnitude. The strong STFT magnitude near the bottom of the panel is already visible in this unmasked spectrogram. A loose STFT magnitude mask, set at 20\% of the largest column maximum, is passed to an eight-connected component search on this spectrogram to isolate this region formally. A component survives only if it has sufficient area, frequency span, and time span, and is rejected if it is line-like or spans too large a fraction of the candidate window; these conditions keep narrowband spectral lines and other persistent structures from being mistaken for transient glitches. As with the AMPS threshold, the z-score and mask fractions used here were tuned against the five instances in this study rather than derived from a general false-alarm argument; Section~\ref{sec:conclusion} outlines a machine-learning-based search intended to fix these values against a larger sample.

The connected-component step confirms a candidate blob is real without giving a tight boundary; CRISP tightens it with a second, stricter STFT magnitude mask, set at three times the loose mask's fraction, applied only within pixels the loose mask already validated. Figure~\ref{fig:crisp_d} shows the result of this stricter mask: the surviving region narrows relative to the unmasked spectrogram in Figure~\ref{fig:crisp_c}, and the start and stop columns of this tightened region define the final CRISP boundary. This boundary comes from the resolved time extent of a spectrogram-power component, not from an approximate sample count. Figure~\ref{fig:crisp_b} shows the resulting boundary zoomed in on the whitened time series, confirming that the selected region lines up with the visible glitch spike.

This excess-power premise is the same one behind Omicron's tile-based triggering \cite{Robinet_2020}, but CRISP resolves a full start and stop boundary through its two-tier masking step rather than stopping at the triggering tile, and it does not use the cross-detector coherence test that Coherent WaveBurst relies on to separate signal from glitch \cite{Klimenko_2016}, since CRISP's task is to localize a glitch already known to be present rather than to classify it.

The bounded segment is then tested with the AMPS procedure of Section~\ref{sec:amps_method}, using the whole-dataset median and median absolute deviation of the whitened strain computed once up front, rather than recomputed on each short segment, so that a short segment with unusually flat statistics cannot bias its own threshold.   Running AMPS on the bounded segment confirms the glitch through an independent amplitude-based check. A full best-fit boundary refinement of the kind FLARE performs would still be costly to run on every Stage~A candidate, so confirming with a fast amplitude check instead keeps CRISP close to FLARE's boundary accuracy without its per-candidate cost.

%
\section{Glitch subtraction}
\label{sec:glitch_subtract}

A number of methods have been developed for glitch subtraction, each resting on a different assumption about what separates a glitch from the surrounding background. \textsc{BayesWave} models a glitch as a superposition of wavelets and separates it from an astrophysical signal using coherence across the detector network \citep{Cornish_2015}. \textsc{gwsubtract} estimates a linear transfer function from an auxiliary witness channel and subtracts the resulting noise estimate from the strain \citep{Davis_2019}. \textsc{glitchschen} applies probabilistic principal component analysis to a library of known glitches to build a parametric model \citep{Merritt_2021}, while \textsc{DeepClean} and related convolutional architectures learn a nonlinear mapping from auxiliary channels to the noise component of the strain \citep{Ormiston_2020,Mogushi_2021}. More recently, \textsc{DeepExtractor} has reframed subtraction as a background-noise reconstruction problem, using a U-Net to predict and remove the noise component directly without a glitch template \citep{Dooney_2025}, and a null-stream-based method has been proposed for triangular third-generation detectors such as the Einstein Telescope, where any transient in the null stream can be attributed to a glitch by construction \citep{Narola_2025}. A fuller treatment of these methods and their relative strengths is given in \citep{Chowdhury_2024}. In this work, glitch subtraction is carried out using adaptive spline fitting, followed by wavelet shrinkage, both described below.

\subsection{Adaptive spline fitting}
\label{subsec:shapes}

Adaptive spline fitting, implemented in the algorithm SHAPES (Swarm Heuristics based Adaptive and Penalized Estimation of Splines), estimates a signal $s(t)$ buried in white Gaussian noise by modeling $s(t)$ as a cubic spline with a free number and placement of knots~\cite{mohanty_2020}. For sampled data
\begin{equation}
y_i = s(t_i) + \epsilon_i, \quad i = 0, 1, \ldots, N-1,
\label{eq:noise-model}
\end{equation}

with $\epsilon_i$ drawn from $\mathcal{N}(0,1)$, the spline is written as a linear combination of cubic B-spline basis functions,
\begin{equation}
s(t) = \sum_{j=0}^{P-5} \alpha_j B_{j,4}(t),
\label{eq:spline-model}
\end{equation}
where $\boldsymbol{\tau} = (\tau_0, \tau_1, \ldots, \tau_{P-1})$ is a sequence of $P$ knots and $\boldsymbol{\alpha}$ the corresponding spline coefficients. The B-spline basis functions $B_{j,4}(t)$ are computed efficiently using the Cox--de Boor recursion relations~\cite{DEBOOR_1972,deBoor_2001}. Knots may repeat up to three times, allowing the spline to represent point discontinuities in value or in derivatives up to second order, a feature required to fit both the smooth and rapidly varying parts of a glitch waveform within a single model.

The best fit parameters $\boldsymbol{\alpha}$ and $\boldsymbol{\tau}$ minimize a penalized least squares objective,
\begin{equation}
L(\boldsymbol{\alpha}, \boldsymbol{\tau}) = \sum_{i=0}^{N-1}
\left(y_i - s(t_i)\right)^2 + \gamma R(\boldsymbol{\alpha}),
\label{eq:penalized-ls}
\end{equation}
where
\begin{equation}
R(\boldsymbol{\alpha}) = \sum_{j=0}^{P-5} \alpha_j^2
\label{eq:penalty-term}
\end{equation}
suppresses spurious clustering of knots around noise outliers, and the penalty gain $\gamma$ controls the smoothness of the fit. Since $\boldsymbol{\alpha}$ enters the model linearly, the minimization over $\boldsymbol{\alpha}$ for a fixed $\boldsymbol{\tau}$ reduces to a closed form ridge regression solution. The harder problem, optimizing the placement of the knots $\boldsymbol{\tau}$~\cite{wold_1974,burchard_1974,jupp_1978,
luo_1997}, is solved in SHAPES using particle swarm optimization~\cite{Kennedy_1995}, a metaheuristic that explores the knot placement space with a population of particles updated according to their own and their neighbors' best located solutions. Multiple independent runs of the swarm are carried out in parallel, and the run with the lowest objective value is retained. The number of knots $P$ is chosen by minimizing the Akaike information criterion~\cite{akaike_1998} over a fixed, user-specified range of candidate values, while the penalty gain $\gamma$ remains a free parameter set by the user, though SHAPES has been found to be fairly insensitive to its exact value across a range of glitch waveforms~\cite{Mohanty_2023} (see~\cite{mohanty_2020,Mohanty_2023} for a complete derivation, including model selection and the particle swarm scheme).

Despite this penalization, SHAPES can still overfit a segment of data on occasion, placing a knot near a stretch of nothing but noise and fitting that stretch more closely than is warranted, which shows up as a spurious local feature in the estimated glitch waveform. This behavior motivates the wavelet shrinkage step applied on top of the SHAPES estimate, described next, which is designed to suppress exactly this kind of spurious feature while leaving a well-behaved fit largely unchanged.

The spline model selection is carried out over five candidate knot-count groups, \(\{5,6\}\), \(\{7,8,9\}\), \(\{10,12,14,16,18\}\), and \(\{20,25,30,35,40\}\), with the regulator gain fixed at \(\gamma = 0.1\). Four independent particle swarm optimization runs are carried out per segment, each with a swarm size of 40 particles, a maximum of 100 iterations, cognitive and social acceleration coefficients \(c_1 = c_2 = 2\), a maximum particle velocity of 0.5, and inertia weight annealed linearly from 0.9 to 0.4 over the run. The data is upsampled by a factor of two before fitting, segments overlap by 30 samples, and the exponential decay rate used to average estimates in the overlap region is set to 16.

\subsection{Wavelet shrinkage}
\label{subsec:waveshrink}

Wavelet shrinkage~\cite{David_1995} provides a nonparametric, multiresolution alternative to spline fitting for estimating a signal from noisy data. Unlike the Fourier transform, whose basis functions are localized in frequency alone, the wavelet transform decomposes the data on an orthogonal basis of functions localized in both time and frequency, which makes it well suited to analyzing the nonstationary structure of a glitch. The transform produces a set of wavelet coefficients $\mathbf{w} = \mathbf{W} \mathbf{y}^{T}$, where $\mathbf{W}$ contains the wavelet basis functions sampled at each data point and arranged on a dyadic scale indexed by resolution level $j$ and translation $k$. A soft threshold function,
$T_\lambda(x) = \mathrm{sgn}(x)(|x| - \lambda)$ for $|x| > \lambda$ and $0$ otherwise, is then applied to each coefficient, shrinking coefficients near the noise floor toward zero while leaving coefficients associated with genuine signal structure largely intact. The estimate of the signal is recovered by inverting the thresholded coefficients. In this work, wavelet shrinkage is applied using the WaveShrink implementation in the WaveLab package~\cite{Huo_2000}, with the threshold set by the Hybrid
method and the coarsest resolution level fixed at $L = 1$, a
choice found to have little effect on the result across the
values tested. We refer to this technique as WS for the remainder of this paper.

Applied to the SHAPES estimate, this thresholding step targets exactly the overfitting described above: the wavelet coefficients produced by a spurious knot sit near the noise floor of the decomposition, making them the coefficients most affected by soft thresholding. When SHAPES has not overfit a given segment, the coefficients of its estimate already sit well above this floor, and wavelet shrinkage leaves the fit largely unchanged. This makes the combination of the two methods self-correcting: wavelet shrinkage only intervenes where SHAPES has actually overfit, and otherwise defers to the spline estimate. Wavelet shrinkage uses the Hybrid threshold rule with the coarsest resolution level fixed at \(L=1\) and the default Symmlet 8 quadrature mirror filter.

\section{Data Description}
\label{sec:demo_data}

\begin{table}[htbp!]
\centering
\caption{Glitch instances considered in this work, with the detector and observing run in which each was recorded and the signal-to-noise ratio of the glitch itself, prior to any subtraction. Event times, detectors, and glitch SNR values are drawn from the Gravity Spy glitch catalog~\cite{Glanzer_2021}.}
\label{tab:glitch_data}
\resizebox{\columnwidth}{!}{%
\begin{tabular}{|l |c |c |c |c|}
\hline
Glitch type & Event time & Glitch SNR & Detector & Observing run \\
\hline
GW170817            & 1187008882 & 646.86 & L1 & O2  \\
Blip                & 1257429123 & 35.972 & H1 & O3b \\
Koi Fish            & 1136368106 & 68.184 & H1 & O1  \\
Tomte               & 1257472072 & 16.51  & L1 & O3b \\
Low frequency blip  & 1267778808 & 101.49 & H1 & O3b \\
\hline
\end{tabular}%
}
\end{table}

The strain segments used to demonstrate the three identification methods are drawn from the public data releases of the LIGO detectors, spanning the O1, O2, and O3b observing runs~\cite{O1O2_2021, O3_2023,GW170817_2017}. Each segment is a 4kHz frame downloaded in its native HDF5 format, containing the standard GWOSC(Gravitational Wave Open Science Center)~\cite{Vallisneri_2015} metadata alongside the strain channel itself. The glitch type, event time, and detector for each segment, along with an estimate of the glitch's own signal-to-noise ratio prior to any subtraction, are taken from the Gravity Spy glitch catalog release~\cite{Glanzer_2021}, which provides both a morphological label and an amplitude-based SNR estimate for each cataloged glitch.

Every method presented in this work assumes stationary, white background noise, so each raw segment is conditioned before identification or subtraction. Conditioning proceeds through a bandpass stage that removes seismic and other low-frequency contamination outside the band of interest, a robust estimate of the noise floor that resists distortion from narrow spectral lines, a whitening filter built from that estimate and applied in the time domain, and a final notch-filtering pass that suppresses the strongest remaining line features. This sequence follows the conditioning logic already standard across LIGO search pipelines, so it is not repeated here beyond this summary.

Table~\ref{tab:glitch_data} lists the five glitch instances considered in this work. Koi Fish, Tomte, Blip, and the low frequency blip differ in time-frequency morphology from one another. The glitch overlapping GW170817 is included separately: Gravity Spy confirms it as a glitch and reports its SNR, but does not assign it a morphological class, so it is treated here as confirmed but unclassified rather than grouped with the other four. Its value lies in coinciding with a real astrophysical signal, and here the comparison is limited to visual inspection of how each identification and subtraction combination affects the signal, rather than the quantitative recovery test used for the injected chirp of Section~\ref{subsec:sigVec}. A supplementary matched-filter check for this instance is given in Appendix~\ref{app:gw170817_mf}. Table \ref{tab:snr_comparison} accordingly reports its boundary only, leaving the SHAPES, WS, and Combined recovery columns empty for this row. The glitch SNR across the five instances spans roughly a factor of forty, from Tomte at the low end to the GW170817 glitch at the high end, so the sample probes differing noise conditions, signal strength, and label certainty rather than establishing general applicability from so small a set.

\subsection{Injected signal}
\label{subsec:sigVec}

To assess subtraction quality on a signal of known parameters, a chirp of controlled amplitude is injected into each conditioned segment rather than a full compact-binary waveform. The injected signal takes the form
\begin{equation}
    \begin{split}
        s(t) = A \sin\!\big(2\pi\phi(t)\big), \\
        \phi(t) = f_0 (t - t_a) + \tfrac{1}{2}k(t-t_a)^2 + \frac{\theta_0}{2\pi},
    \end{split}
    \label{eq:signal}
\end{equation}
where \(t_a\) marks the onset of the injection, \(f_0\) and \(f_1\) set the starting and ending instantaneous frequency over the injection window, \(k = (f_1 - f_0)/(t_b - t_a)\) is the resulting sweep rate, and \(\theta_0\) is an initial phase offset. The amplitude \(A\) is fixed by normalizing the raw waveform to unit norm and rescaling to a chosen target value, so that in whitened, unit-variance noise this target value equals the optimal matched-filter SNR of the injected signal, following the convention used for the injected binary neutron star signal in the source method this work builds on~\citep{Mohanty_2023}. A linear chirp keeps the injection simple to reproduce and interpret, at the cost of not capturing the amplitude and frequency evolution of a genuine compact-binary waveform; whether this simplification changes the recovered SNR comparison is left for the results rather than assumed here. This is also why the GW170817 instance is treated separately in Section~\ref{sec:demo_data} rather than folded into the same recovery test: no synthetic chirp was injected there, so the comparison for that glitch relies on the real inspiral signal already present in the data, sidestepping the amplitude and frequency mismatch a linear chirp would introduce against an actual compact-binary waveform. The starting frequency \(f_0\) is fixed at 30~Hz for every glitch class. The ending frequency \(f_1\) is 300~Hz for Koi Fish, Tomte, and the low frequency blip, and 600~Hz for Blip and GW170817, both broadband glitches. The injection window spans 1.5~s, beginning approximately 0.5~s before the glitch is flagged so the two overlap. The injected SNR for each glitch class is reported in Table~\ref{tab:snr_comparison}.

Recovery of the injected signal is quantified with a single matched-filter statistic rather than a sliding search, since the injection time is already known from the signal generation step. Writing \(r(t)\) for the residual after subtraction and \(\hat{s}(t) = s(t)/\|s(t)\|\) for the injected template normalized to unit norm, the reported SNR is
\begin{equation}
    \rho = \sum_i r(t_i)\,\hat{s}(t_i),
    \label{eq:matched_filter}
\end{equation}
computed once over the full residual with the template held at its injected alignment. This is equivalent, for whitened unit-variance noise, to the peak of a sliding matched filter evaluated exactly at zero lag. A separate FFT-based sliding correlation with power-spectral-density weighting was checked against this value and found to agree, so only the simpler form is retained here.

\subsection{Time-frequency representation}
\label{subsec:q_transform}

Time-frequency panels throughout this work are produced with a continuous wavelet transform implementation in the MATLAB Wavelet Toolbox~\cite{matlab_wavelet}, using a generalized Morse wavelet~\cite{Lilly_2012, Othelde_2002} evaluated at 48 voices per octave, rather than the sine-Gaussian tile bank of the Q-transform proper~\cite{Chatterji_2004}. For each frequency row, the transform is evaluated at five values of the quality factor spanning the requested range, and the value producing the largest total energy in the segment is kept, giving a multiresolution time-frequency map in the spirit of a Q-transform without its tile decomposition. Energy in each frequency row is normalized by its median value across time before display and shown on a logarithmic color scale, so a fixed color range remains meaningful across panels of differing absolute energy. A narrow band near 60~Hz, a residual of the powerline notch filter applied during whitening, is interpolated from neighbouring frequency rows where noted in a figure caption.

\section{Results and Discussion}
\label{sec:results}

The recovered-SNR percentages reported below are computed from one representative instance per glitch class, selected from a small number of instances checked per class during development to span a range of glitch SNR, consistent with the SNR spread already noted in Section~\ref{sec:demo_data}, so they should be read as characterizing this specific sample rather than as class-level averages. Section~\ref{sec:conclusion} discusses this limitation and a planned remedy. The recovered-SNR averages reported in Sections~\ref{sec:results_amps} through \ref{sec:results_crisp} carry an intrinsic spread of roughly one SNR unit, quantified directly in Section~\ref{sec:comparison_subtraction}; readers should treat the reported figures as accurate to a few percentage points rather than to the tenth of a percent implied by their stated precision. Upsampling was applied identically to all five instances and does not distinguish the two groups. What differs between them is the rule used to choose a segment length. Segment length is drawn from a fixed set of candidate lengths, each a power of two from 256 to 32768 samples. The baseline rule selects the smallest candidate in this set that is large enough to contain the full data stretch. For Koi Fish, Tomte, and the low frequency blip, this baseline rule was used directly. For GW170817 and Blip, both of which carry substantial power above 500 Hz, the position of the baseline candidate within the set was halved before selection, so the segment length used is a shorter entry drawn from earlier in the set rather than the one the baseline rule would otherwise pick. This gives a shorter segment, and so a higher density of knots per unit of data, for the two glitches whose structure extends to higher frequency. Once a segment length is fixed this way, the data is divided into overlapping segments of that length, and any leftover segment at the end shorter than a minimum length is dropped rather than kept as an undersized final segment.

\begin{table*}[htbp!]
\centering
\caption{Glitch boundary and residual signal-to-noise ratio of the injected signal after subtraction, compared across the three identification methods (AMPS, FLARE, CRISP) and three subtraction techniques (SHAPES, WS, Combined) for each glitch instance considered in this work. Koi Fish recovers the largest fraction of injected SNR among the four glitches under every method; Appendix~\ref{app:koi_fish_reposition} shows that recovered SNR at this injection position can exceed the injected value at other positions within the overlap window, a chance effect of the noise realization rather than of subtraction~\cite{wang2014ligosurf, Gerosa_2024}.}
\label{tab:snr_comparison}
\resizebox{\textwidth}{!}{%
\begin{tabular}{|l|c|c|c|c|c|c|c|c|c|c|c|c|c|}
\hline
& & \multicolumn{4}{c|}{AMPS} & \multicolumn{4}{c|}{FLARE} & \multicolumn{4}{c|}{CRISP} \\
\cline{3-6} \cline{7-10} \cline{11-14}
Glitch type & Signal SNR & Boundary & SHAPES & WS & Combined & Boundary & SHAPES & WS & Combined & Boundary & SHAPES & WS & Combined \\
\hline
GW170817            & -- & 0.1921 & -- & -- & -- & 0.1094 & -- & -- & -- & 0.1094 & -- & -- & -- \\
\hline
Blip                & 25 & 0.1001 & 22.4458 & 13.6502 & 23.3001 & 0.0938 & 23.0607 & 13.6482 & 23.3876 & 0.1250 & 22.0642 & 13.6502 & 23.3468 \\
\hline
Koi Fish  & 30 & 0.1125 & 29.2630 & 19.4828 & 29.4823 & 0.0938 & 29.8092 & 19.4886 & 29.7993 & 0.1094 & 29.6471 & 19.4828 & 29.7679 \\
\hline
Tomte               & 27 & 0.1118 & 24.2282 & 20.1158 & 25.7495 & 0.1094 & 24.9896 & 20.1136 & 25.8389 & 0.1094 & 24.2178 & 20.1158 & 25.7098 \\
\hline
Low frequency blip  & 25 & 0.1196 & 23.4842 & 15.7051 & 24.1364 & 0.1094 & 23.5905 & 15.7028 & 24.2804 & 0.1094 & 23.4624 & 15.7028 & 24.3804 \\
\hline

\end{tabular}%
}
\end{table*}

\begin{figure*}[htbp!]
\centering

\begin{subfigure}{0.235\textwidth}
\centering
\includegraphics[width=\textwidth]{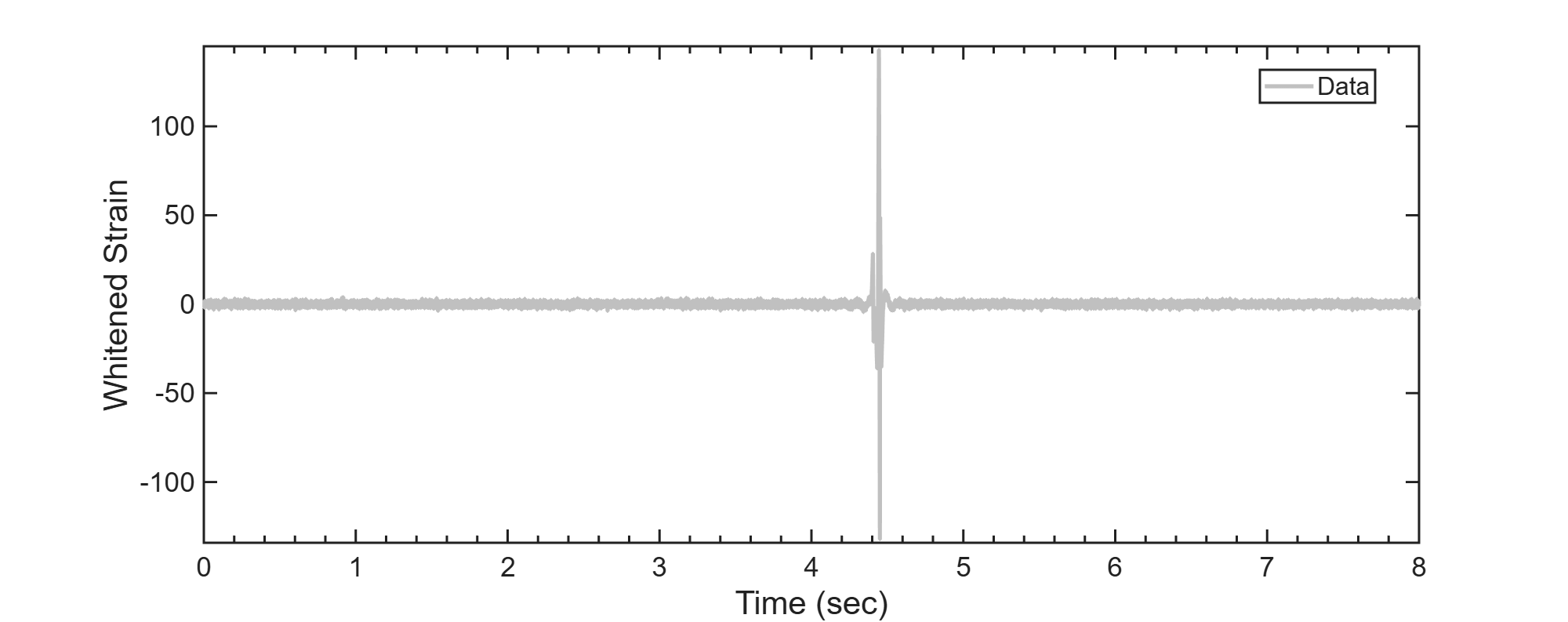}
\caption{}
\label{fig:amps_gw170817_ts}
\end{subfigure}
\hfill
\begin{subfigure}{0.235\textwidth}
\centering
\includegraphics[width=\textwidth, height = 0.65in]{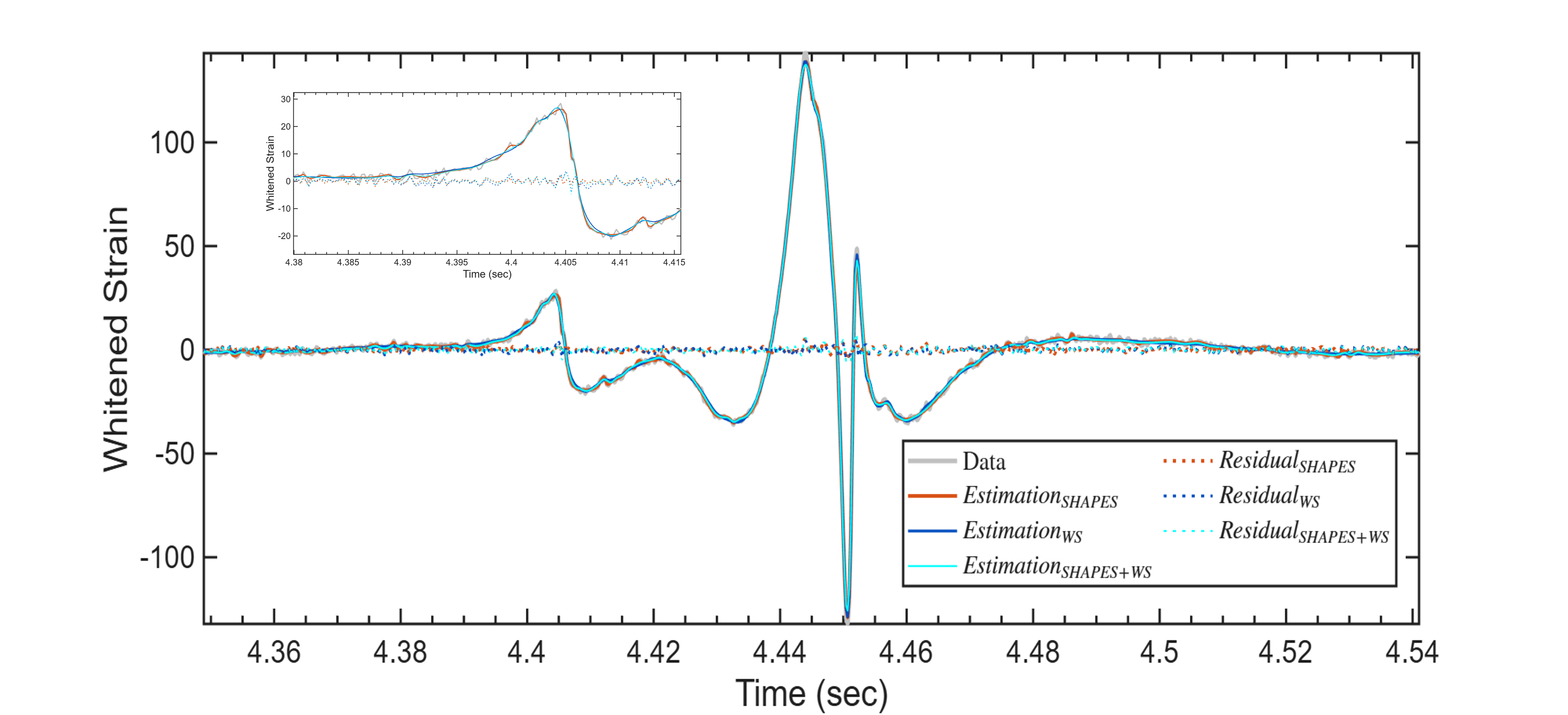}
\caption{}
\label{fig:amps_gw170817_zoom}
\end{subfigure}
\hfill
\begin{subfigure}{0.235\textwidth}
\centering
\includegraphics[width=\textwidth]{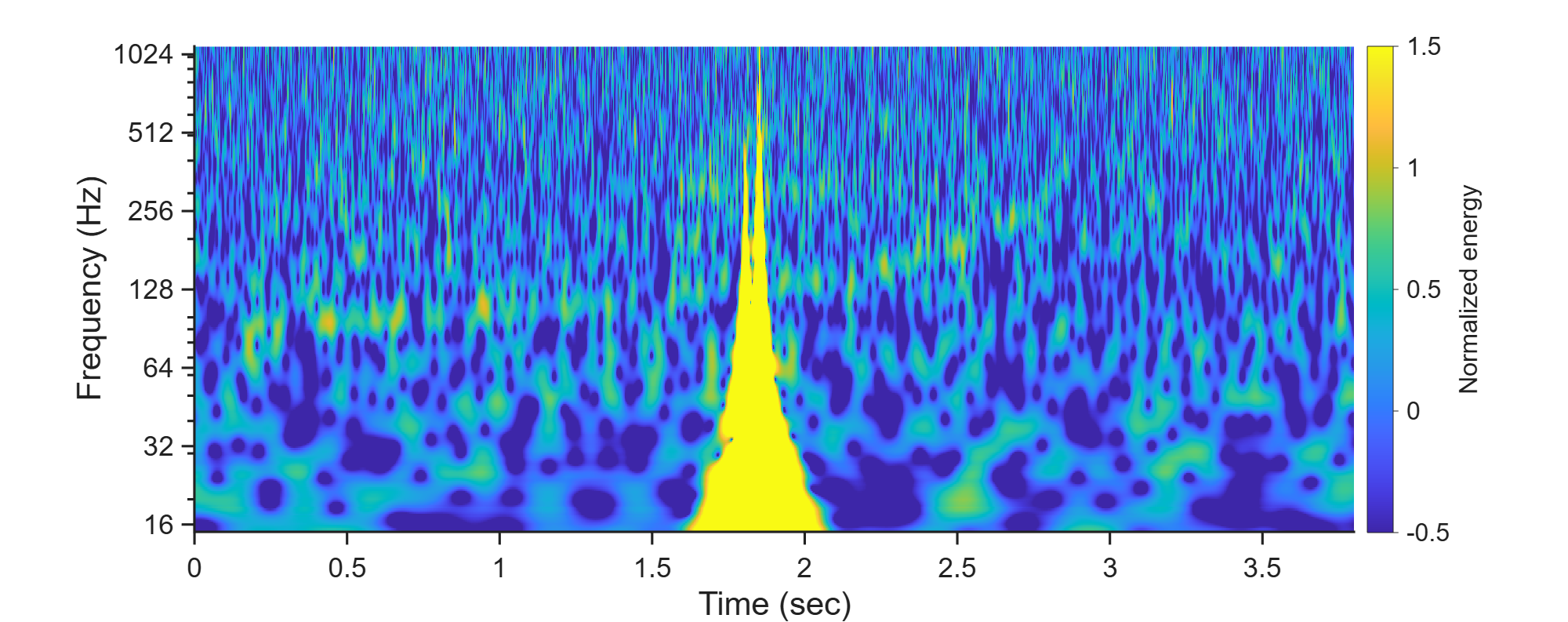}
\caption{}
\label{fig:amps_gw170817_cqt_before}
\end{subfigure}
\hfill
\begin{subfigure}{0.235\textwidth}
\centering
\includegraphics[width=\textwidth]{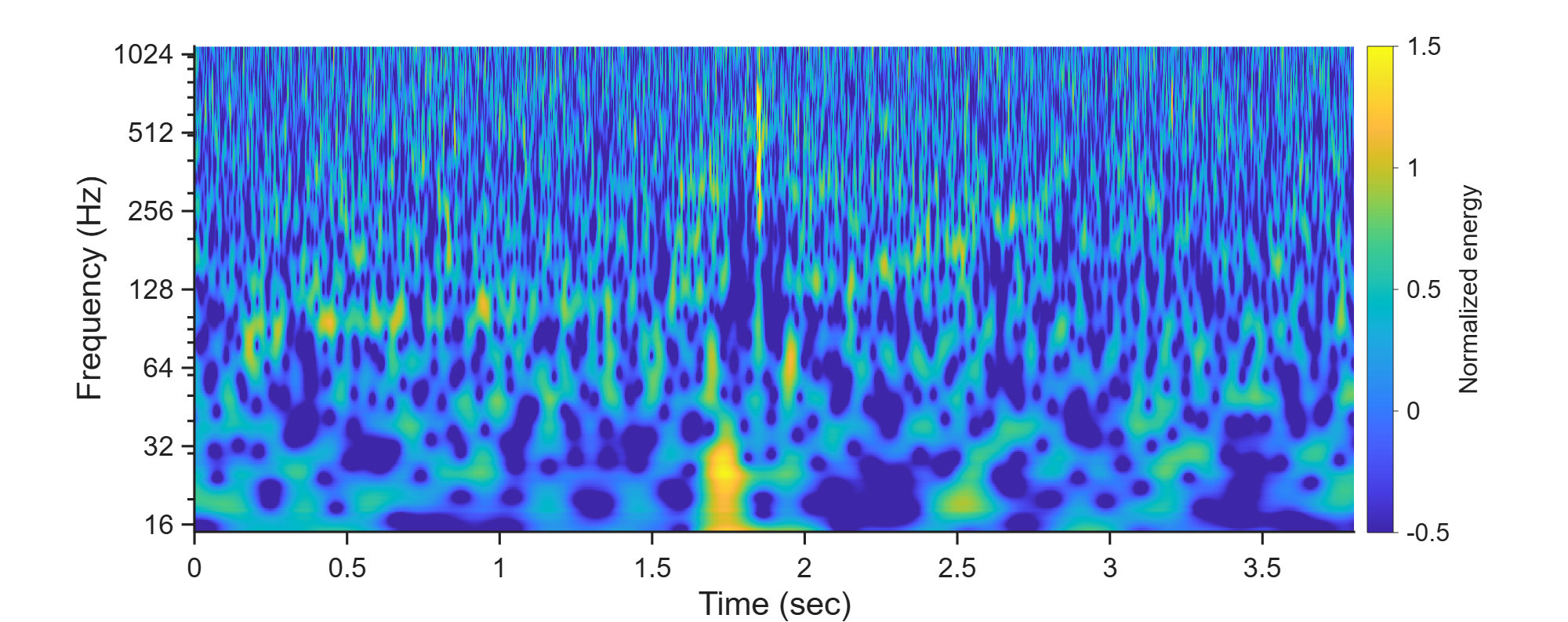}
\caption{}
\label{fig:amps_gw170817_cqt_after}
\end{subfigure}

\vspace{0.3cm}

\begin{subfigure}{0.235\textwidth}
\centering
\includegraphics[width=\textwidth]{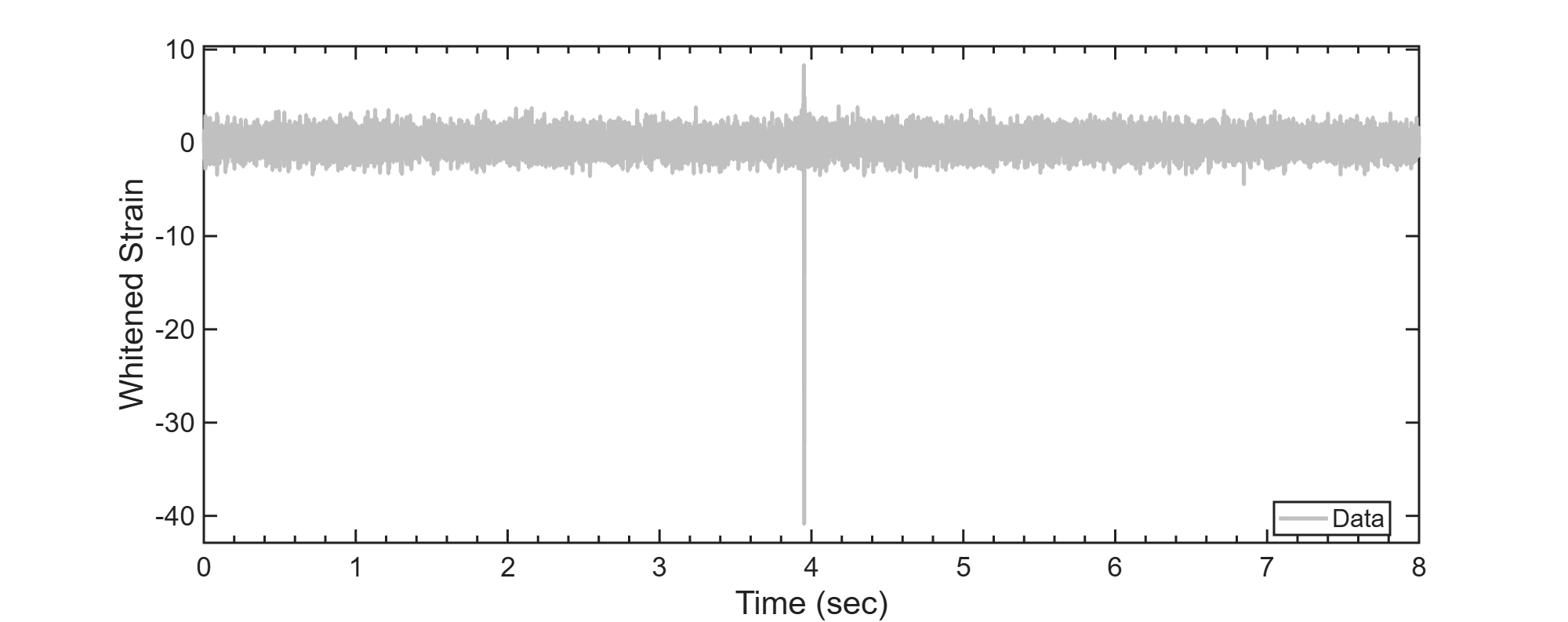}
\caption{}
\label{fig:amps_blip_ts}
\end{subfigure}
\hfill
\begin{subfigure}{0.235\textwidth}
\centering
\includegraphics[width=\textwidth]{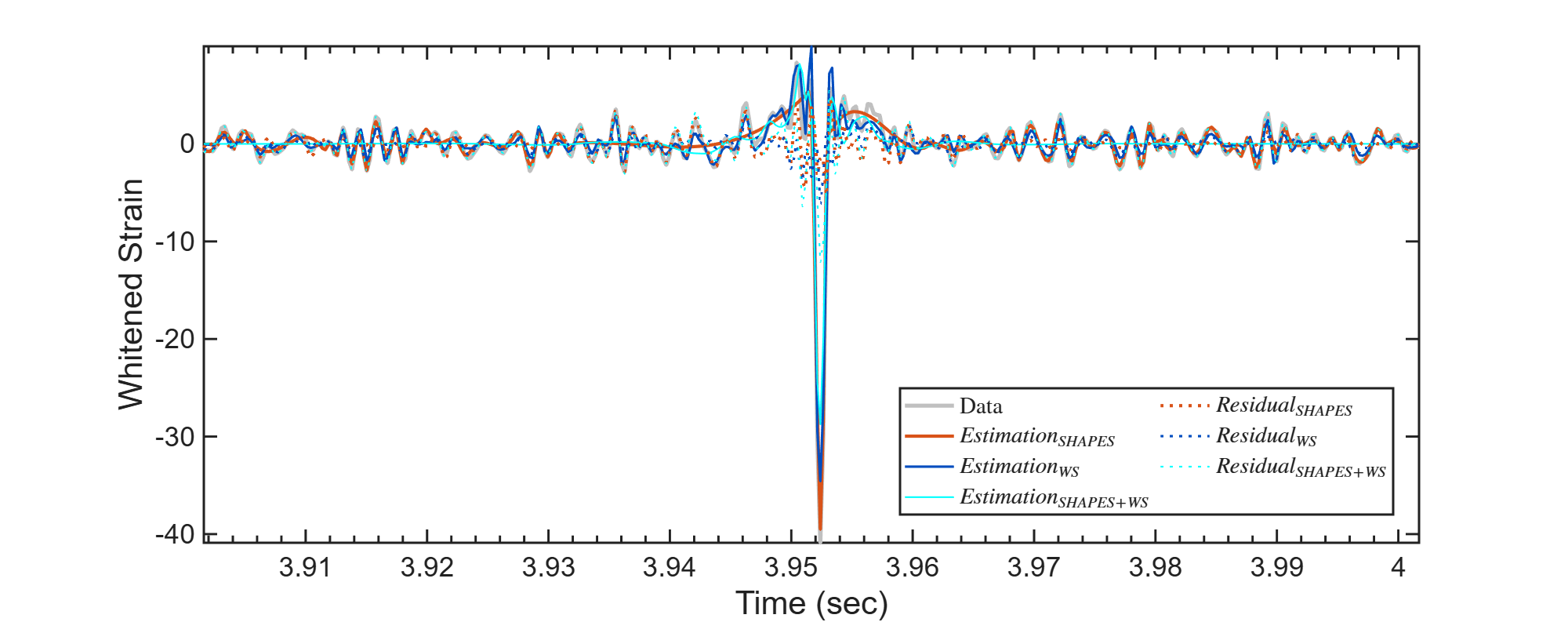}
\caption{}
\label{fig:amps_blip_zoom}
\end{subfigure}
\hfill
\begin{subfigure}{0.235\textwidth}
\centering
\includegraphics[width=\textwidth]{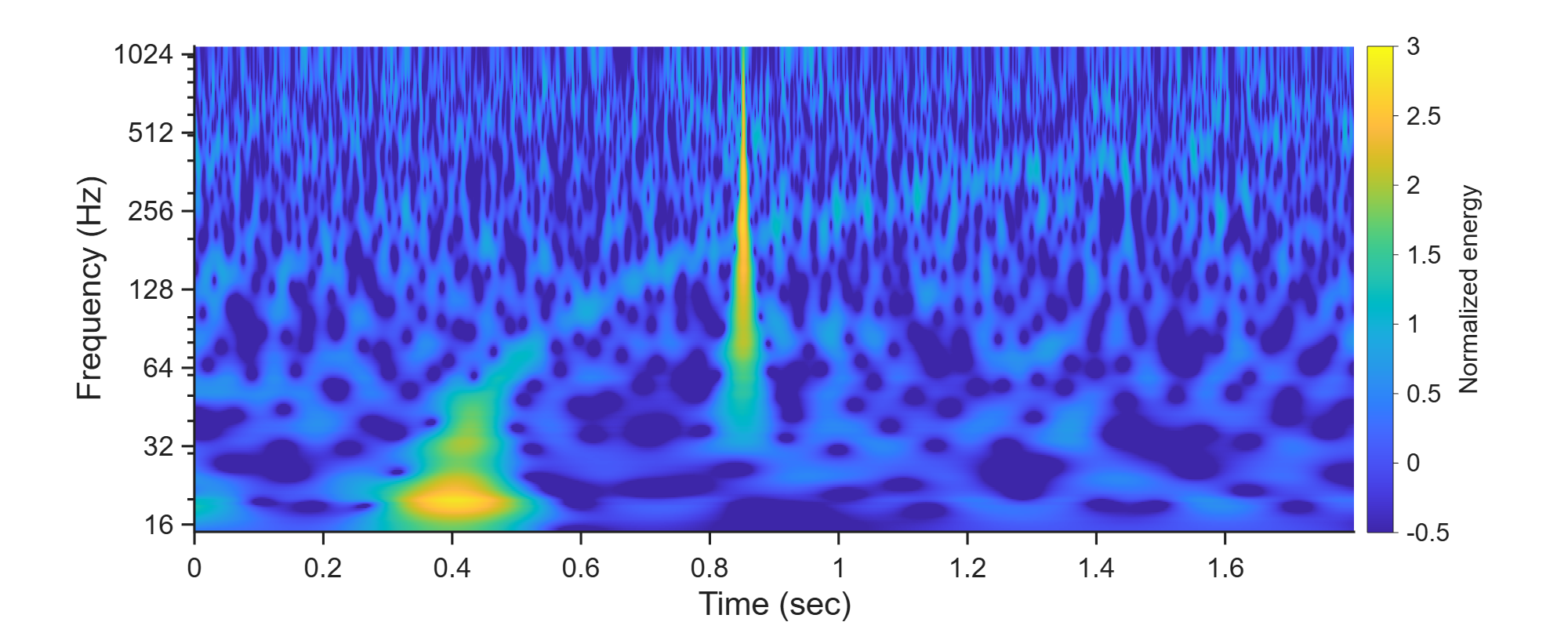}
\caption{}
\label{fig:amps_blip_cqt_before}
\end{subfigure}
\hfill
\begin{subfigure}{0.235\textwidth}
\centering
\includegraphics[width=\textwidth]{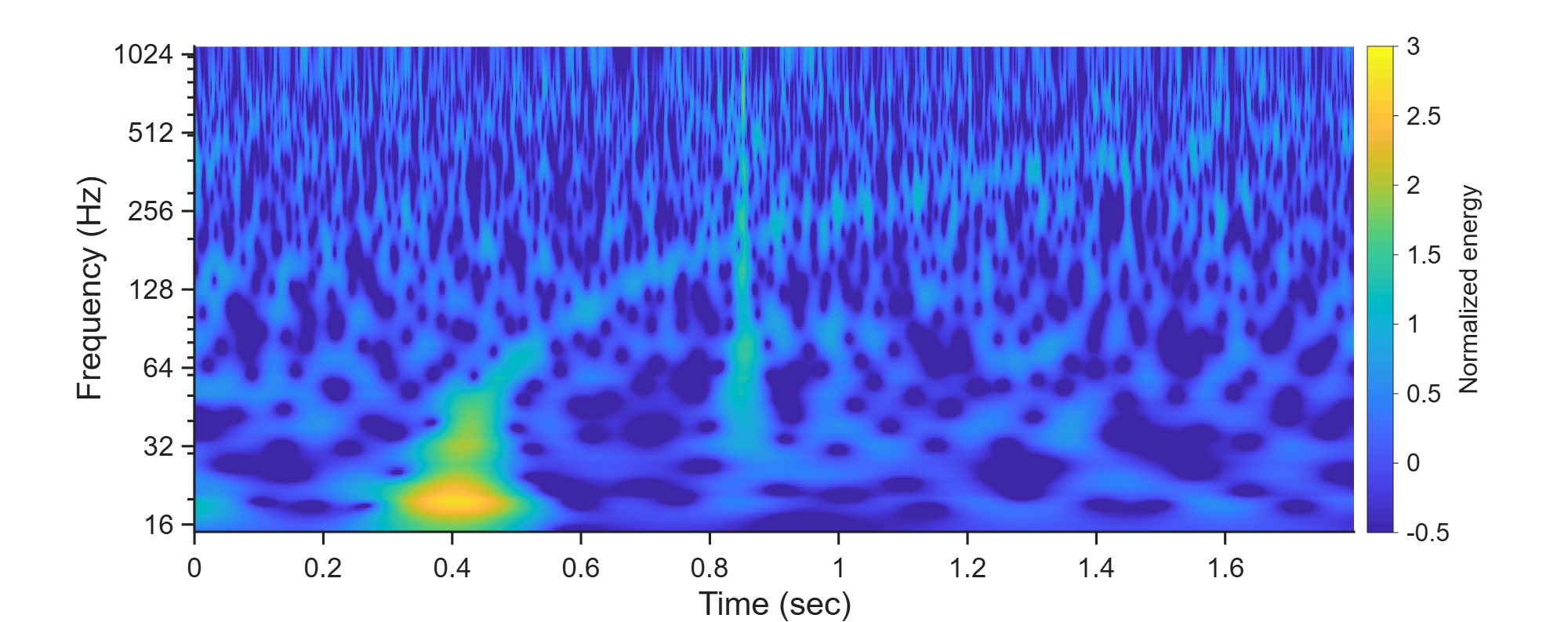}
\caption{}
\label{fig:amps_blip_cqt_after}
\end{subfigure}

\vspace{0.3cm}

\begin{subfigure}{0.235\textwidth}
\centering
\includegraphics[width=\textwidth]{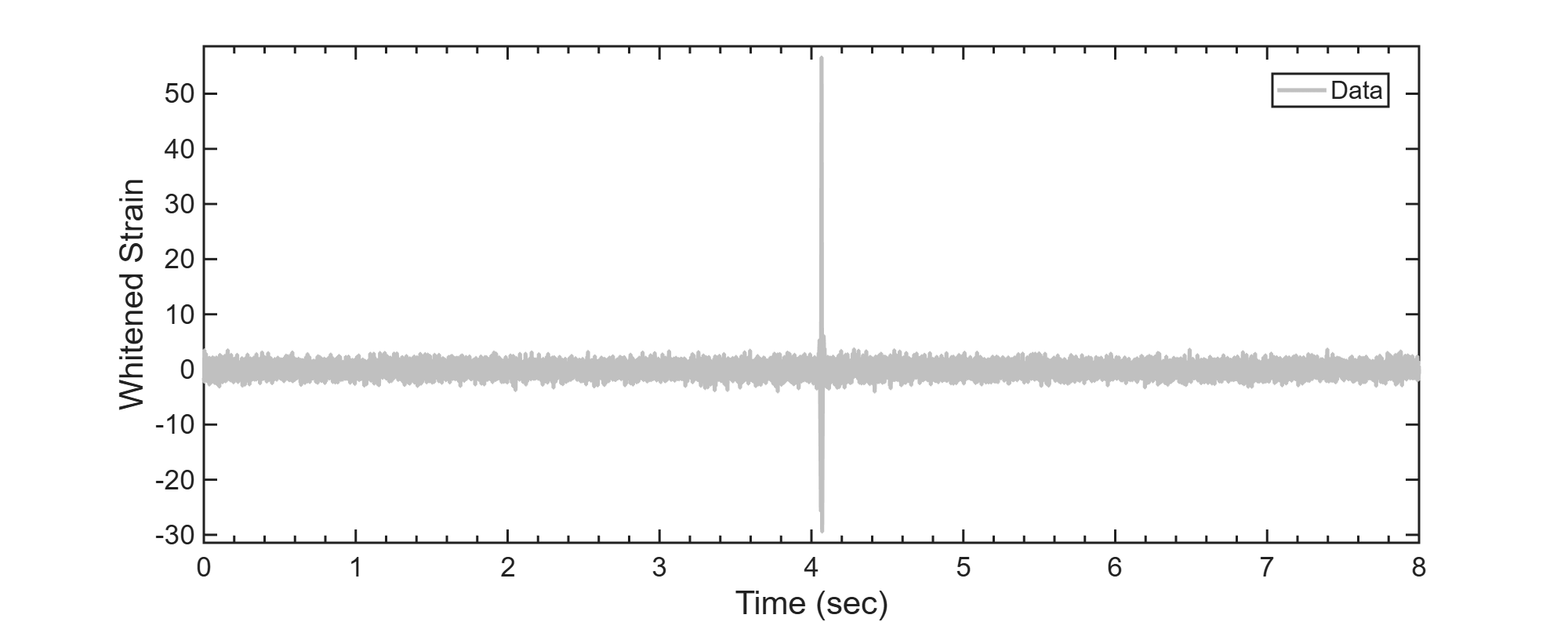}
\caption{}
\label{fig:amps_koifish_ts}
\end{subfigure}
\hfill
\begin{subfigure}{0.235\textwidth}
\centering
\includegraphics[width=\textwidth]{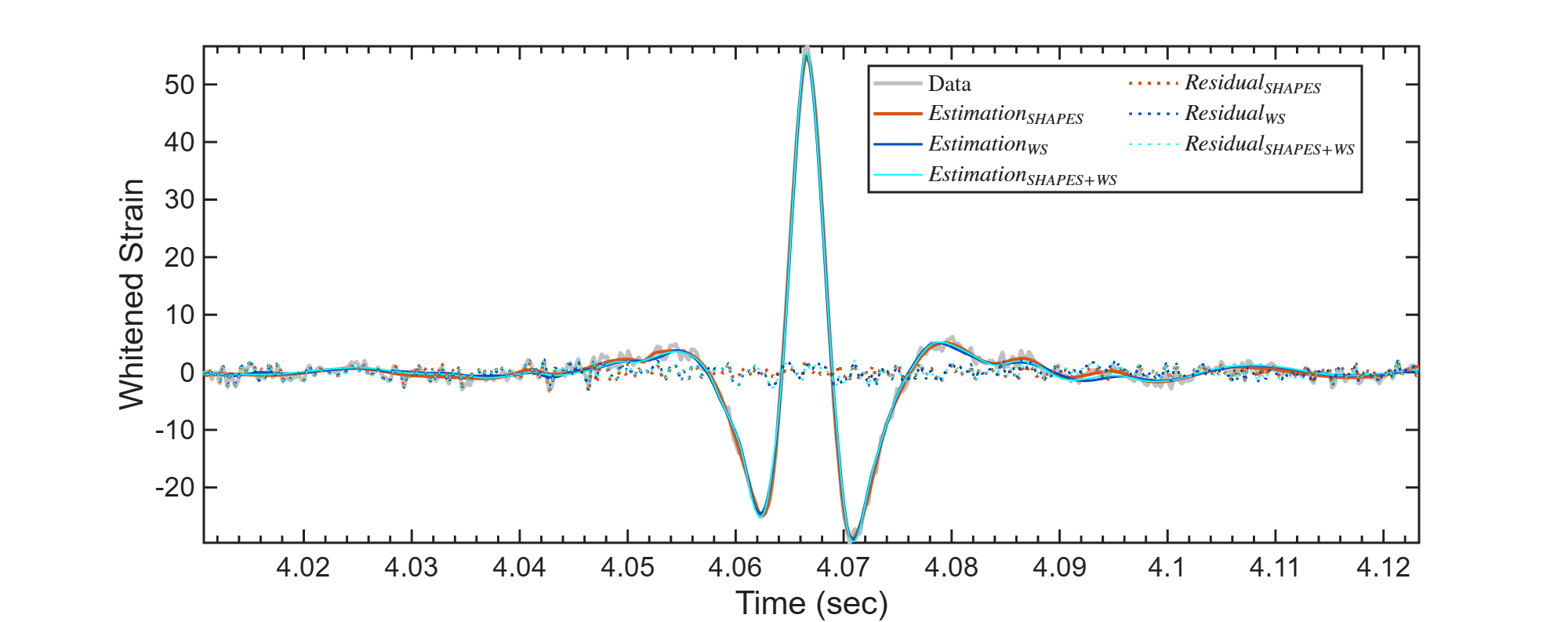}
\caption{}
\label{fig:amps_koifish_zoom}
\end{subfigure}
\hfill
\begin{subfigure}{0.235\textwidth}
\centering
\includegraphics[width=\textwidth]{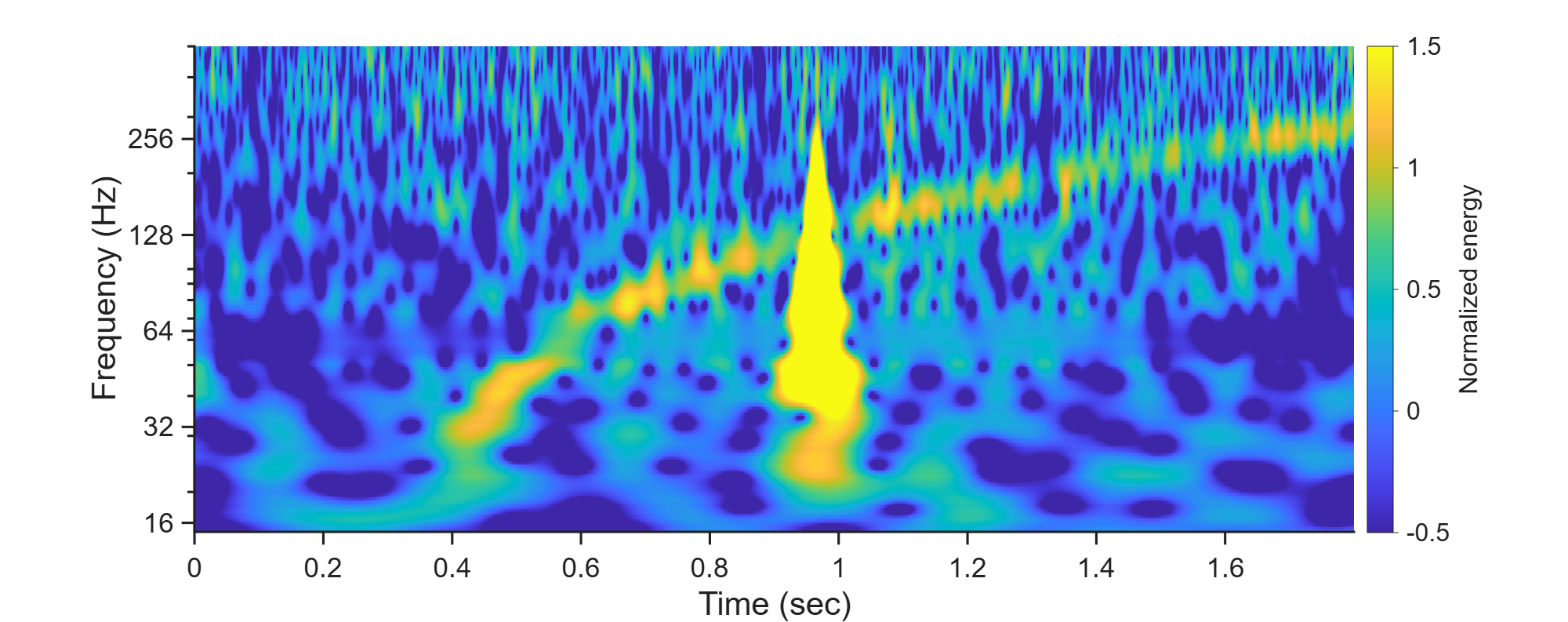}
\caption{}
\label{fig:amps_koifish_cqt_before}
\end{subfigure}
\hfill
\begin{subfigure}{0.235\textwidth}
\centering
\includegraphics[width=\textwidth]{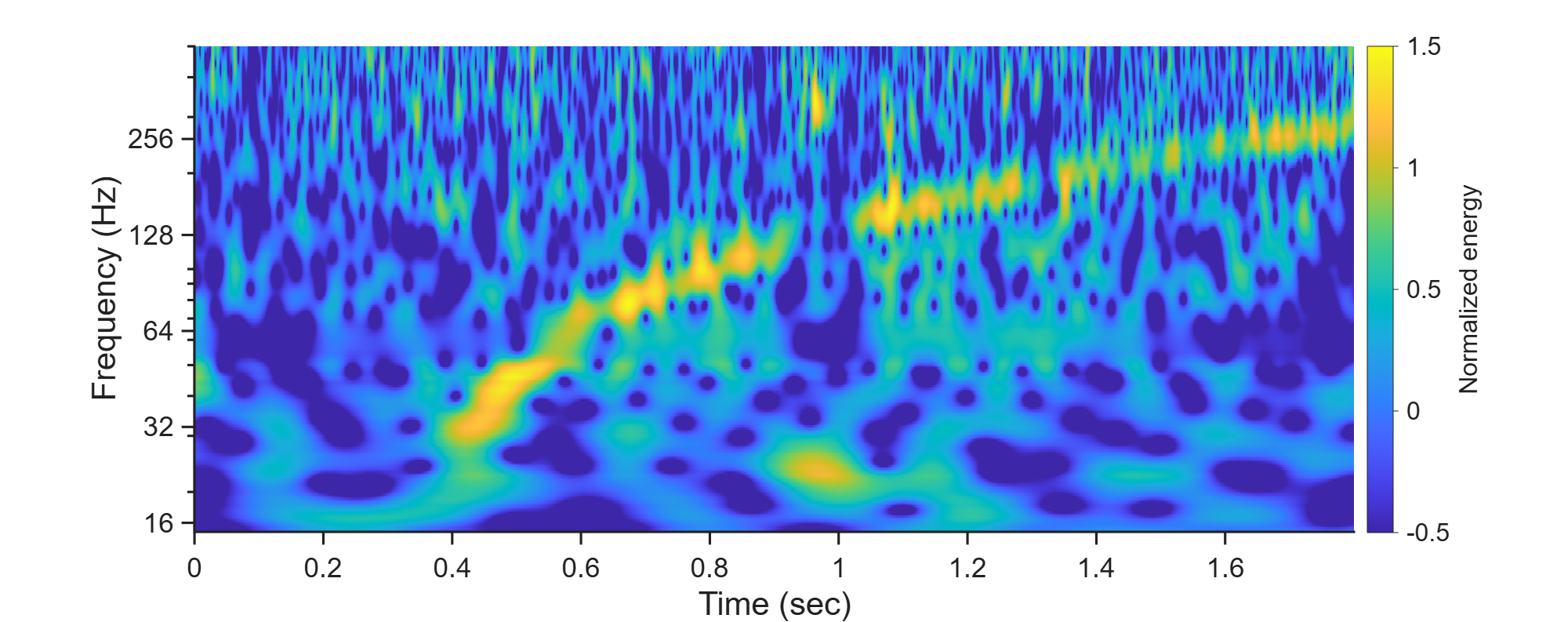}
\caption{}
\label{fig:amps_koifish_cqt_after}
\end{subfigure}

\vspace{0.3cm}

\begin{subfigure}{0.235\textwidth}
\centering
\includegraphics[width=\textwidth]{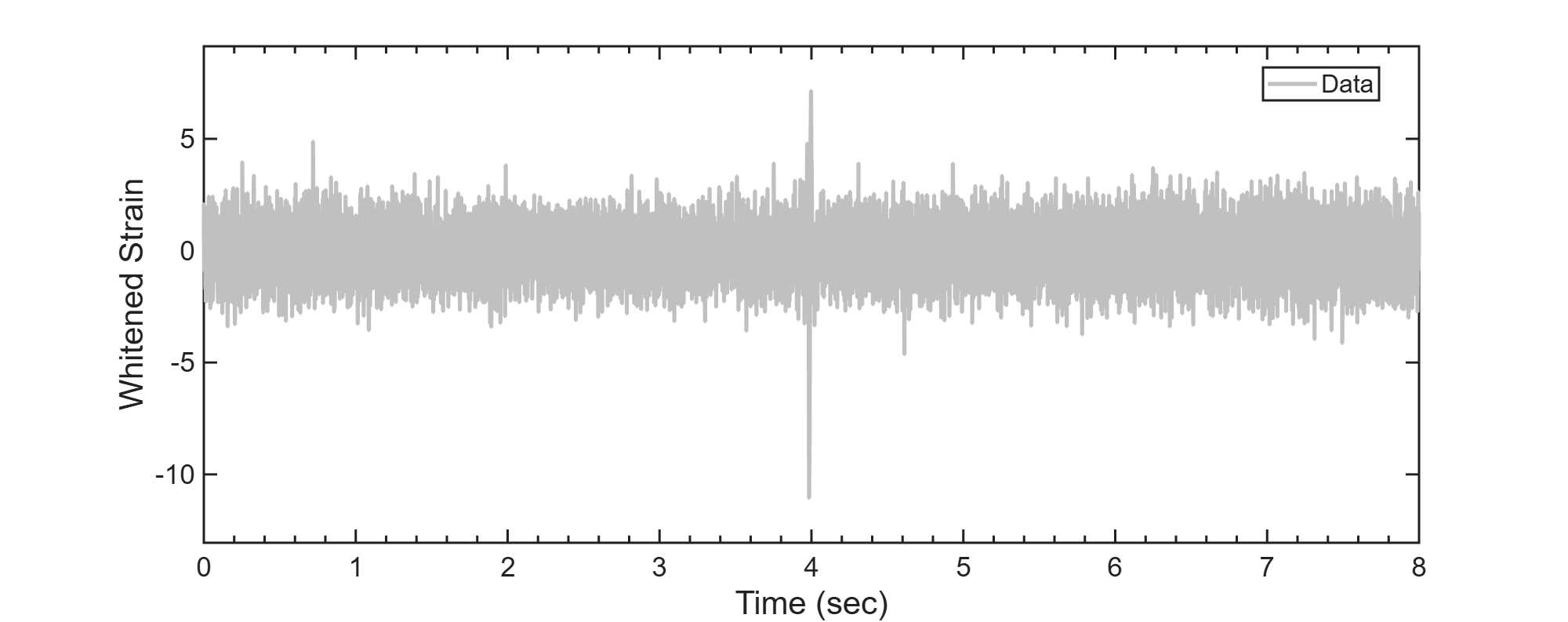}
\caption{}
\label{fig:amps_tomte_ts}
\end{subfigure}
\hfill
\begin{subfigure}{0.235\textwidth}
\centering
\includegraphics[width=\textwidth]{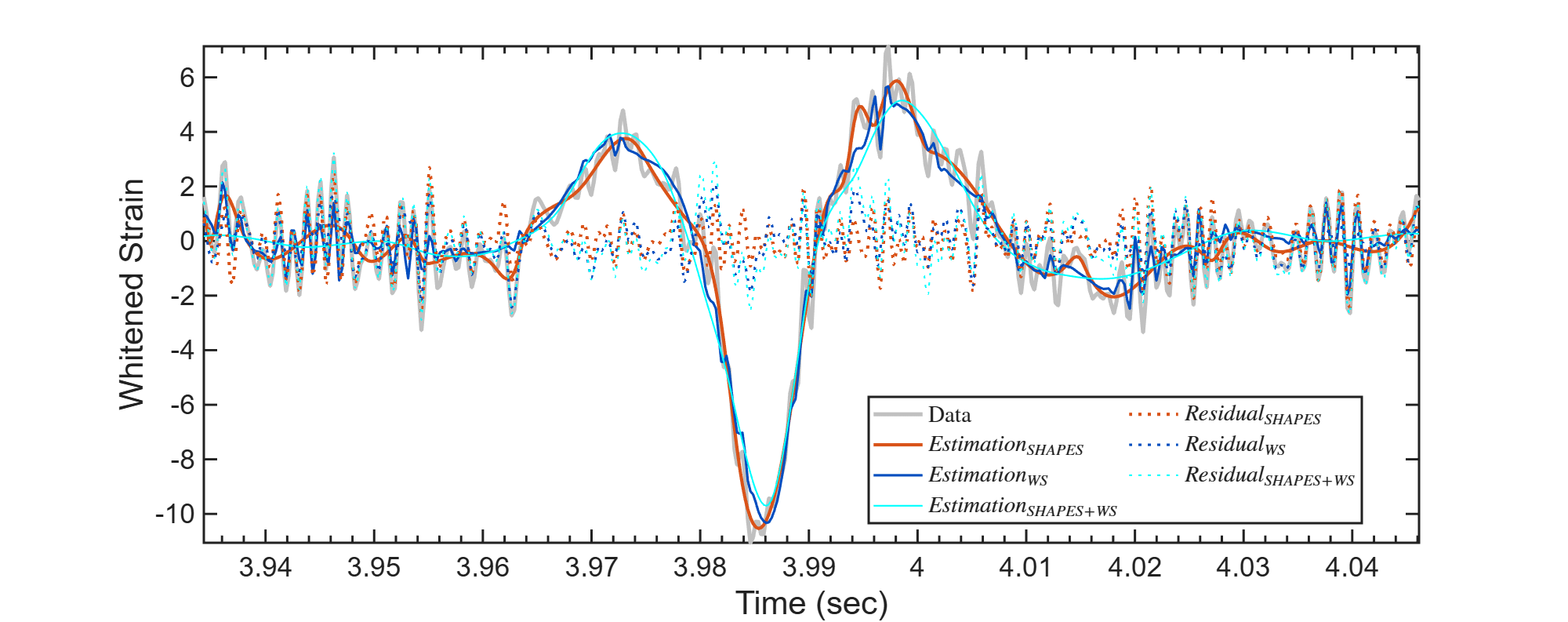}
\caption{}
\label{fig:amps_tomte_zoom}
\end{subfigure}
\hfill
\begin{subfigure}{0.235\textwidth}
\centering
\includegraphics[width=\textwidth]{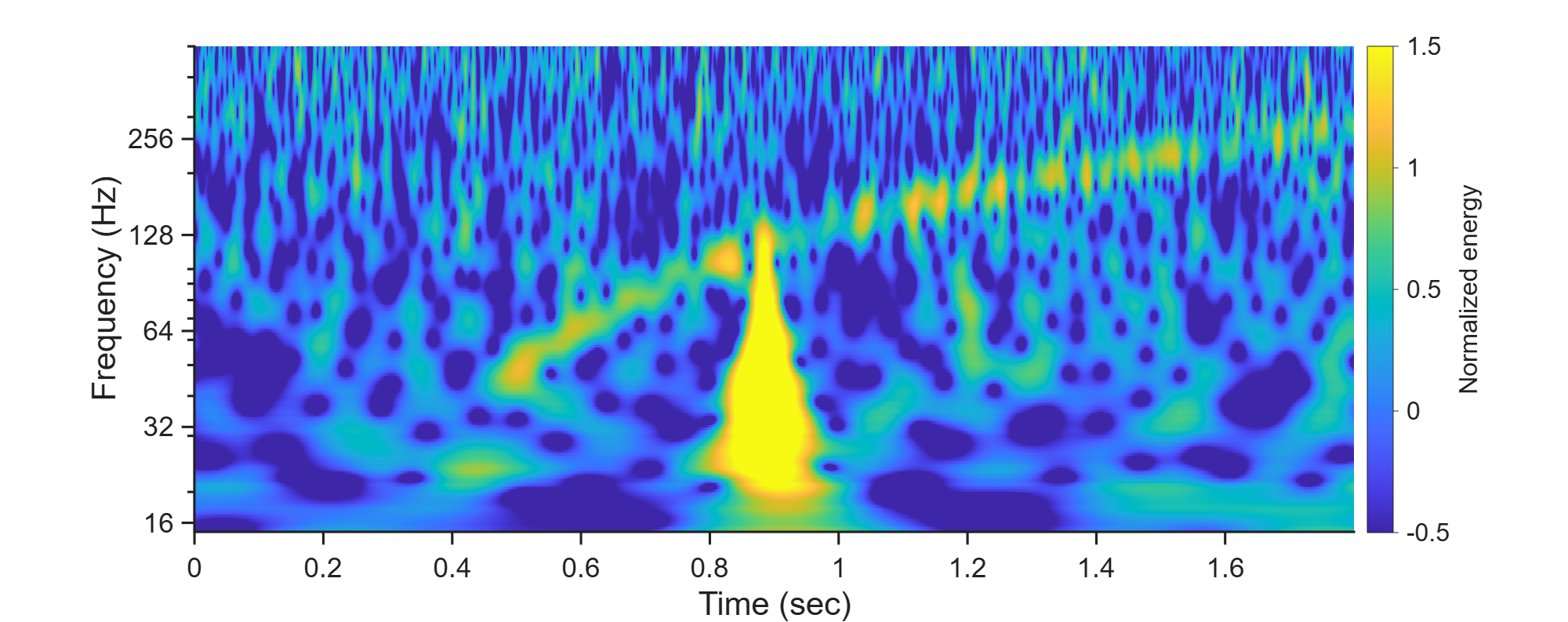}
\caption{}
\label{fig:amps_tomte_cqt_before}
\end{subfigure}
\hfill
\begin{subfigure}{0.235\textwidth}
\centering
\includegraphics[width=\textwidth]{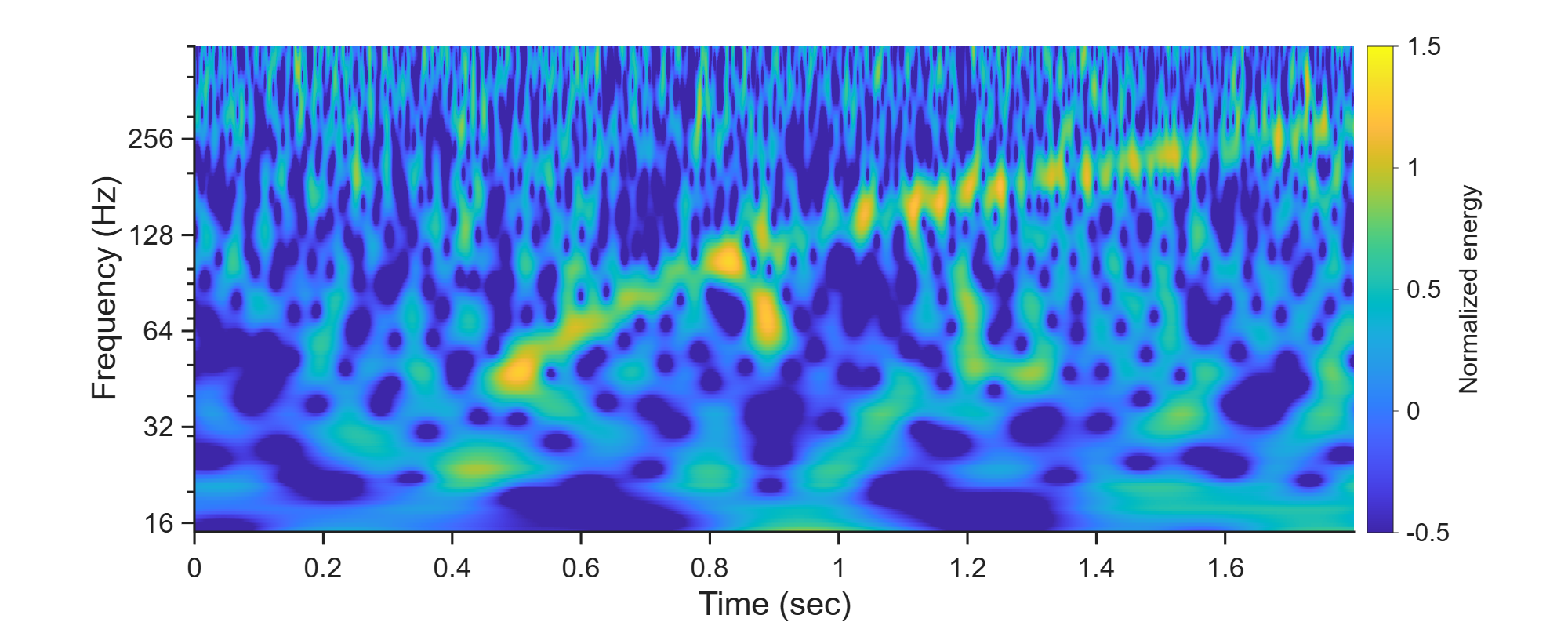}
\caption{}
\label{fig:amps_tomte_cqt_after}
\end{subfigure}

\vspace{0.3cm}

\begin{subfigure}{0.235\textwidth}
\centering
\includegraphics[width=\textwidth]{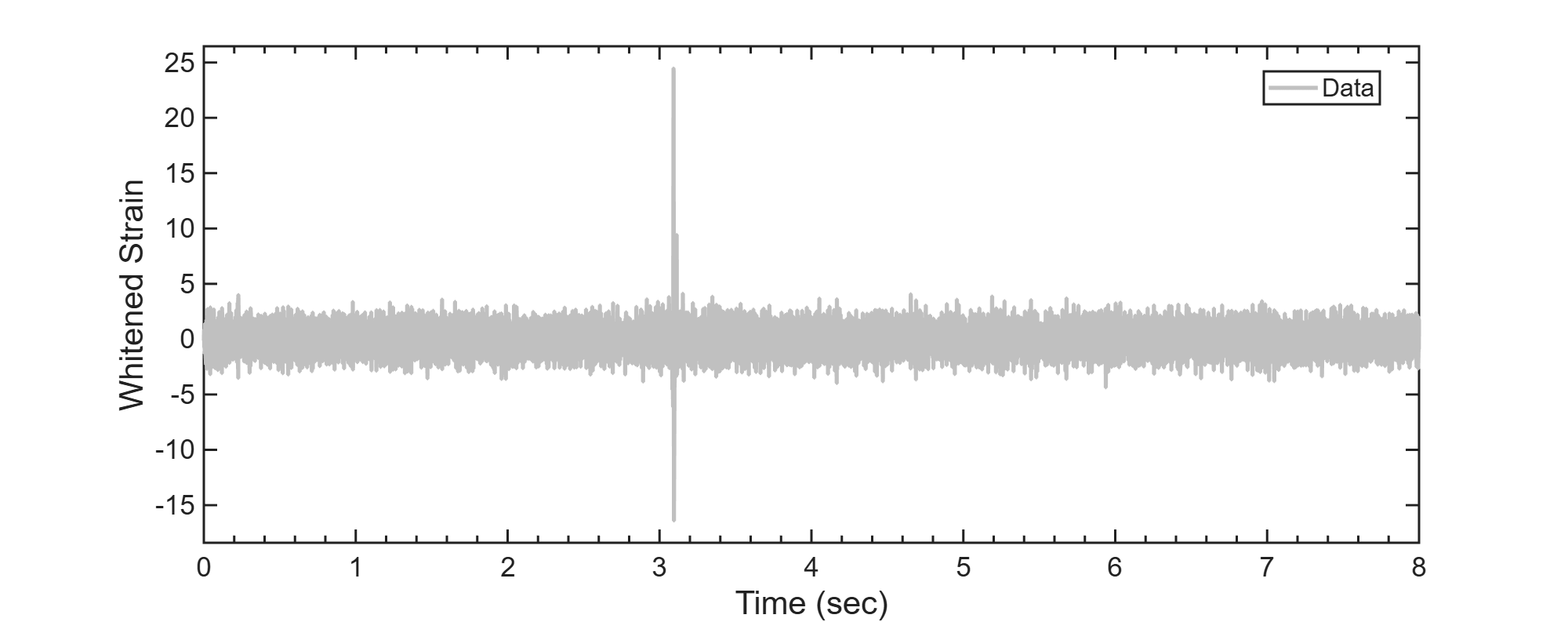}
\caption{}
\label{fig:amps_lfb_ts}
\end{subfigure}
\hfill
\begin{subfigure}{0.235\textwidth}
\centering
\includegraphics[width=\textwidth]{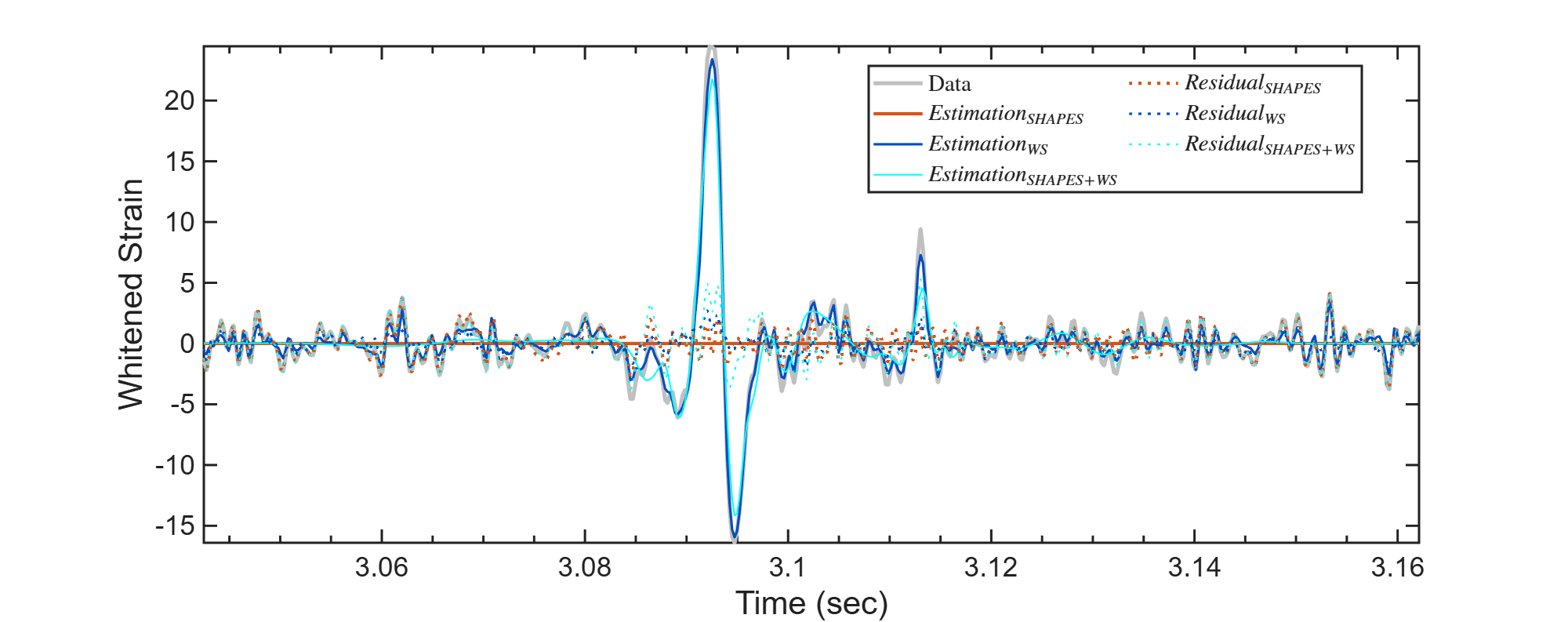}
\caption{}
\label{fig:amps_lfb_zoom}
\end{subfigure}
\hfill
\begin{subfigure}{0.235\textwidth}
\centering
\includegraphics[width=\textwidth]{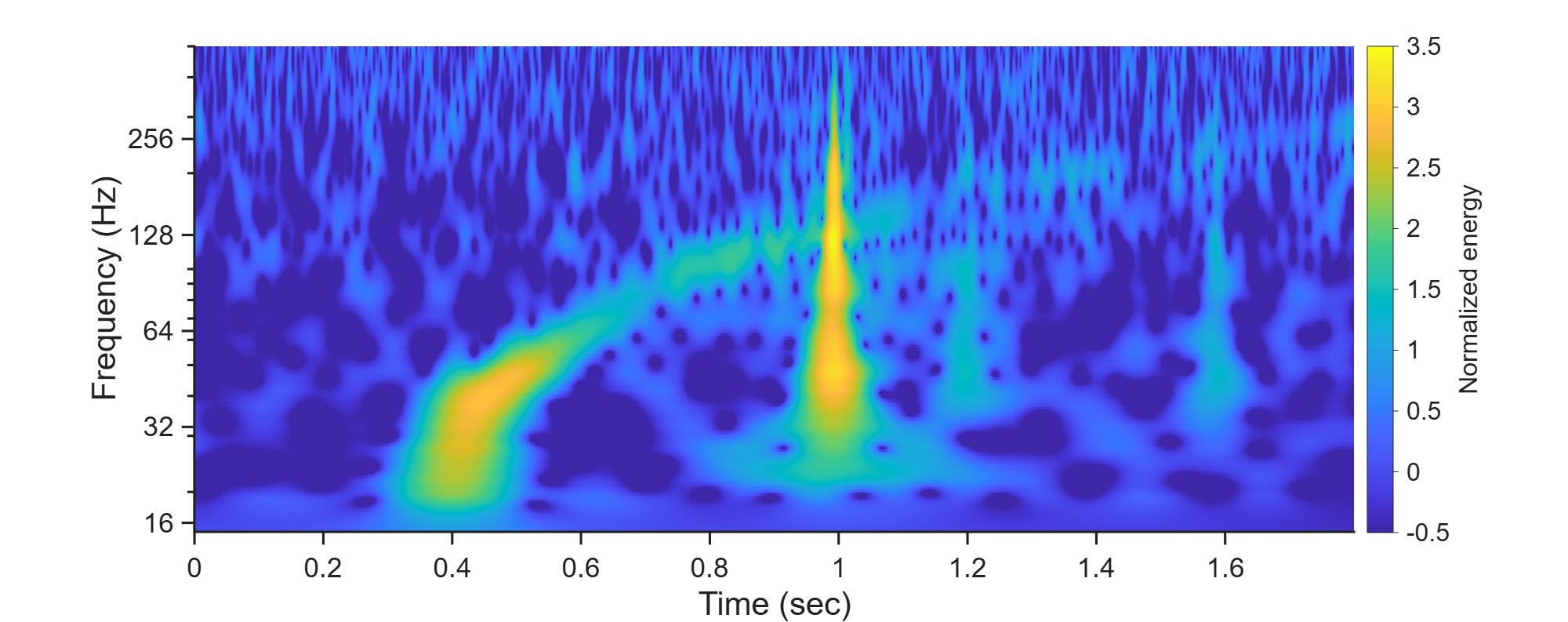}
\caption{}
\label{fig:amps_lfb_cqt_before}
\end{subfigure}
\hfill
\begin{subfigure}{0.235\textwidth}
\centering
\includegraphics[width=\textwidth]{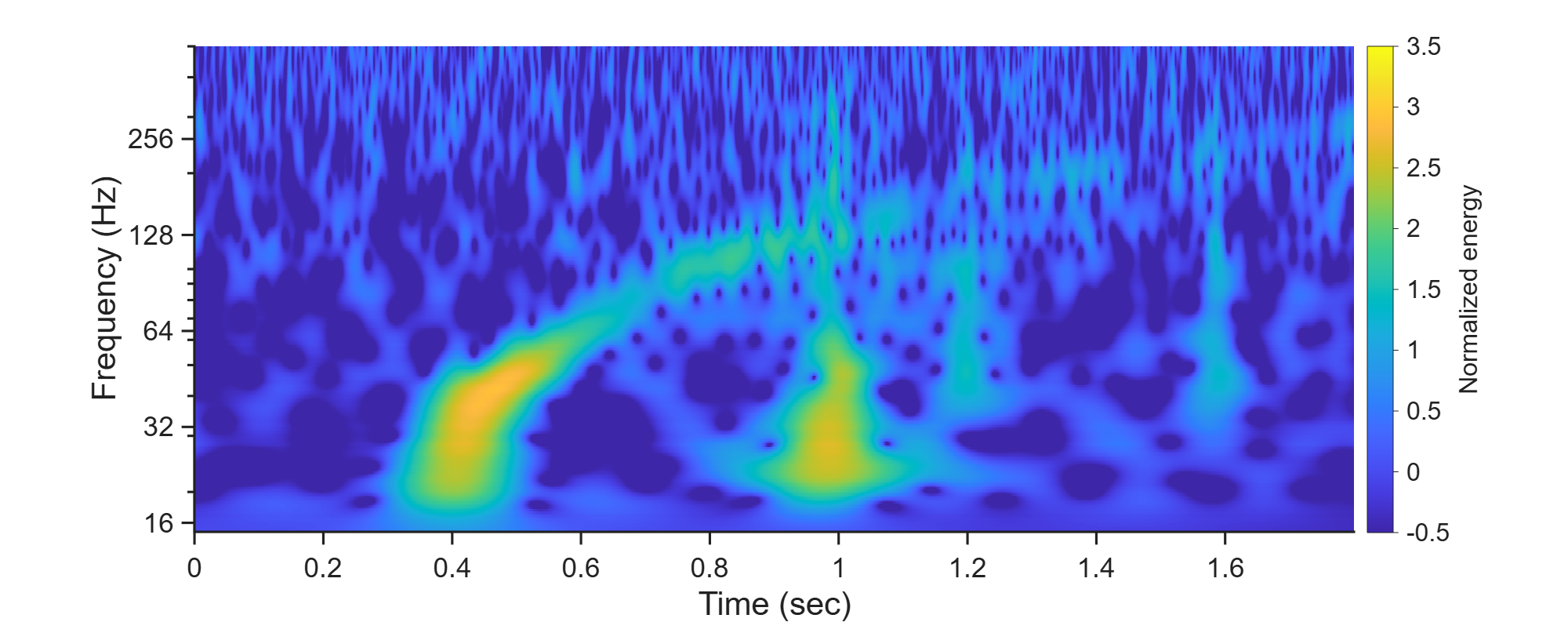}
\caption{}
\label{fig:amps_lfb_cqt_after}
\end{subfigure}

\caption{Identification and subtraction of the five glitch instances using AMPS, from top to bottom: GW170817, Blip, Koi Fish, Tomte, and the low frequency blip. First column: whitened strain time series around the glitch against time in seconds; this panel is shown once per glitch type and is not repeated for FLARE or CRISP later in this section. Second column: zoomed view of the identified glitch segment, whitened strain against time in seconds, with the SHAPES, WS, and SHAPES+WS estimates drawn as solid lines and their residuals as dotted lines of matching color. Third column: Q-transform of the segment before subtraction, time in seconds against frequency in Hz, with the colorbar giving normalized energy. Fourth column: Q-transform of the same segment after subtraction, using the same time window, frequency range, and color scale as the third column, so that any leftover glitch power or surviving chirp track can be read off by direct comparison between the two panels.}
\label{fig:amps_all}
\end{figure*}

\subsection{AMPS}
\label{sec:results_amps}

AMPS was applied to the five glitch instances, with the outcome shown in Figure~\ref{fig:amps_all}. The first column of the whitened time series, shown once per glitch type in Figure~\ref{fig:amps_all}, is not repeated in the FLARE and CRISP figures that follow, since the same underlying time series is shared across all three identification methods. The second column gives a closer view of the identified segment, with the SHAPES, WS, and Combined estimates and their residuals. The third and fourth columns give the Q-transform before and after subtraction, with the fourth column built from the Combined residual in every case so the five glitches can be compared on equal footing.

For Blip, Koi Fish, Tomte, and the low frequency blip, the aim of subtraction is to remove the glitch while keeping the injected chirp intact. The chirp track survives subtraction and stays clearly visible in all four cases, seen across the fourth column of rows two through five.

Table~\ref{tab:snr_comparison} lists the AMPS boundary and recovered SNR for each glitch. AMPS returns a wider boundary for GW170817 than for any of the four injected glitches, matching the longer duration of that glitch relative to the others. Consistent with its qualitative treatment in Section~\ref{sec:demo_data}, Table~\ref{tab:snr_comparison} reports only GW170817's boundary. Figure~\ref{fig:amps_gw170817_cqt_after} shows the underlying astrophysical signal remains visible after subtraction, with the glitch power removed and a faint trace consistent with the signal left in its place.

Koi Fish recovers the highest fraction of injected SNR among the four glitches under SHAPES and Combined. Blip and Tomte give up the largest fraction of injected SNR among the four glitches under both SHAPES and Combined, while the low frequency blip retains a higher recovered fraction than either, second only to Koi Fish.

Averaged over the four injected glitches, SHAPES recovers about 92.7\% of the injected SNR, WS alone recovers about 64.2\%, and Combined recovers about 95.8\%, each value capped at 100\% per glitch before averaging, since recovered SNR cannot exceed complete signal recovery. Combined gives the best recovered SNR for every glitch under AMPS boundaries, with the largest margin over SHAPES for Blip and Tomte. WS alone gives the lowest recovered SNR for every glitch by a clear margin. SHAPES and Combined both preserve the injected signal better than WS alone under AMPS boundaries, and Combined holds a small, consistent edge over SHAPES on its own.

\begin{figure*}[htbp!]
\centering

\begin{subfigure}{0.325\textwidth}
\centering
\includegraphics[width=\textwidth, height = 0.90in]{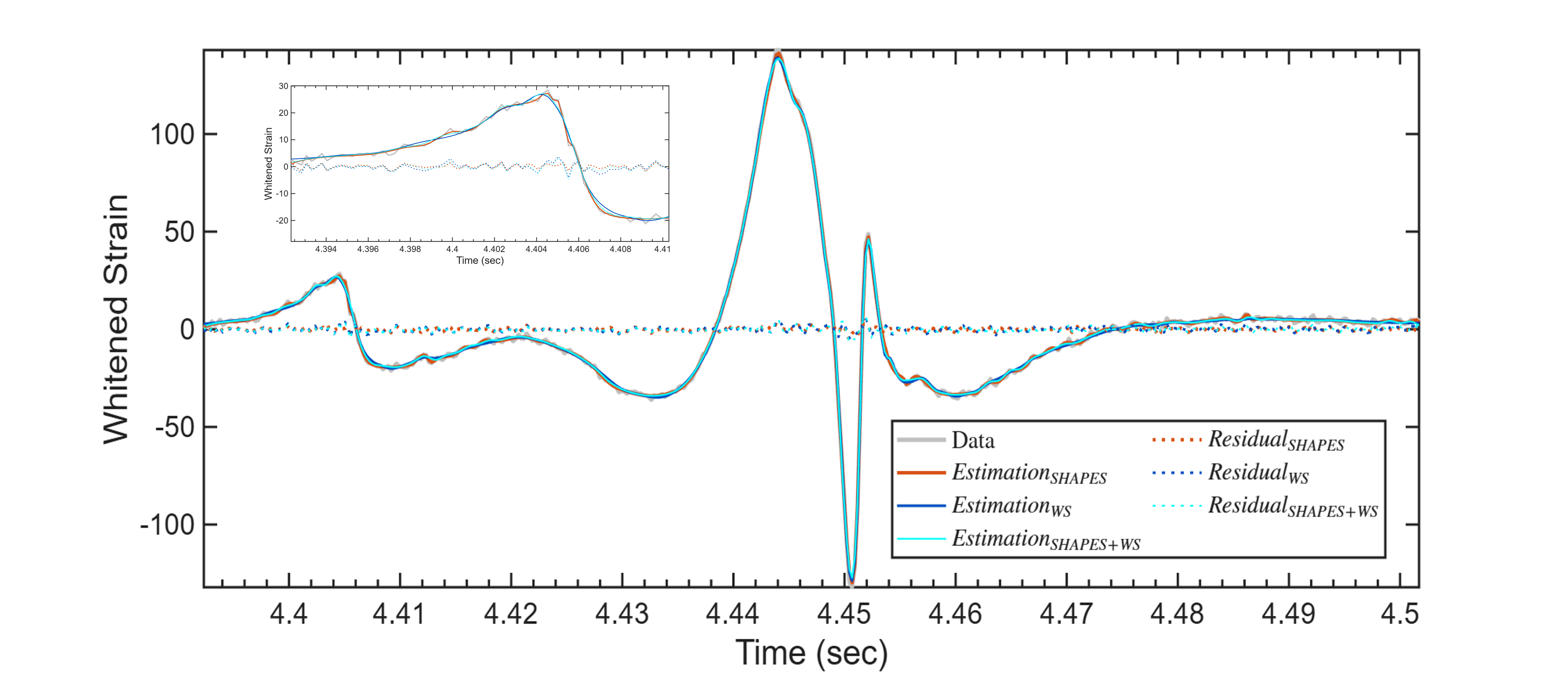}
\caption{}
\label{fig:FLARE_gw170817_zoom}
\end{subfigure}
\hfill
\begin{subfigure}{0.325\textwidth}
\centering
\includegraphics[width=\textwidth]{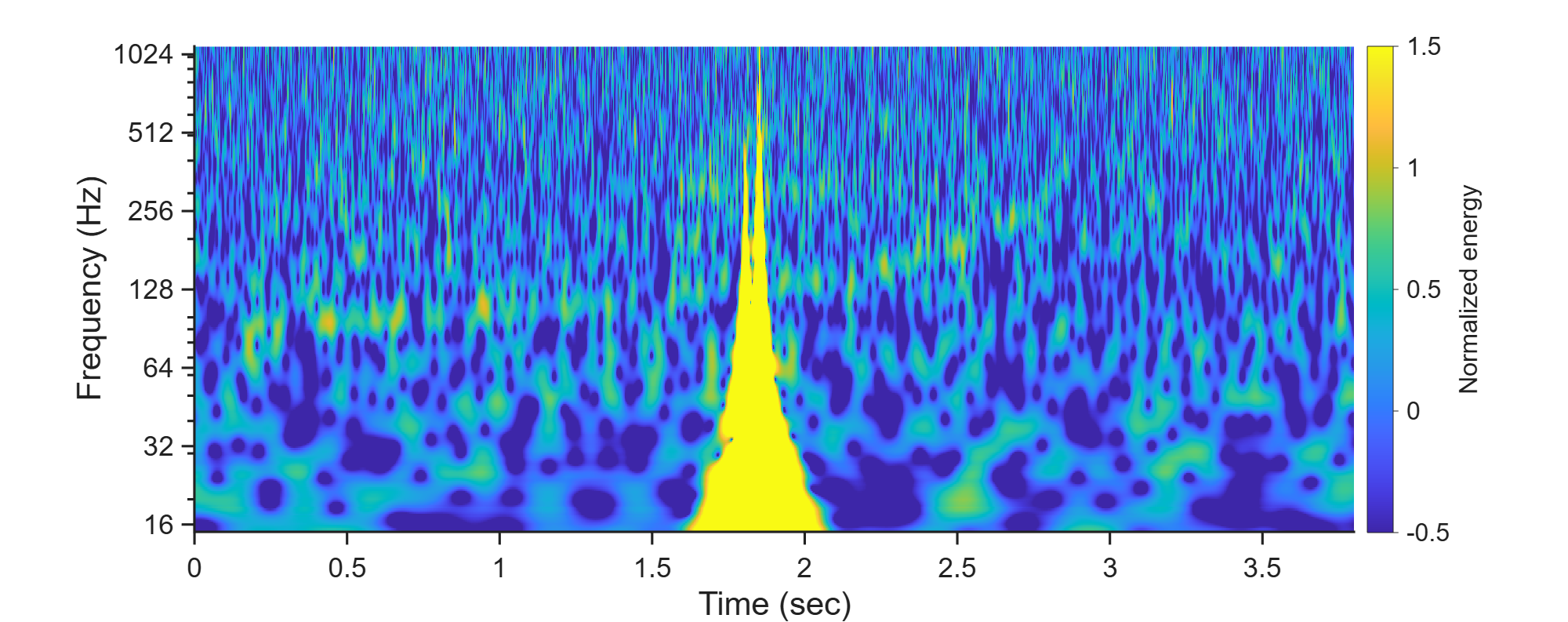}
\caption{}
\label{fig:FLARE_gw170817_cqt_before}
\end{subfigure}
\hfill
\begin{subfigure}{0.325\textwidth}
\centering
\includegraphics[width=\textwidth]{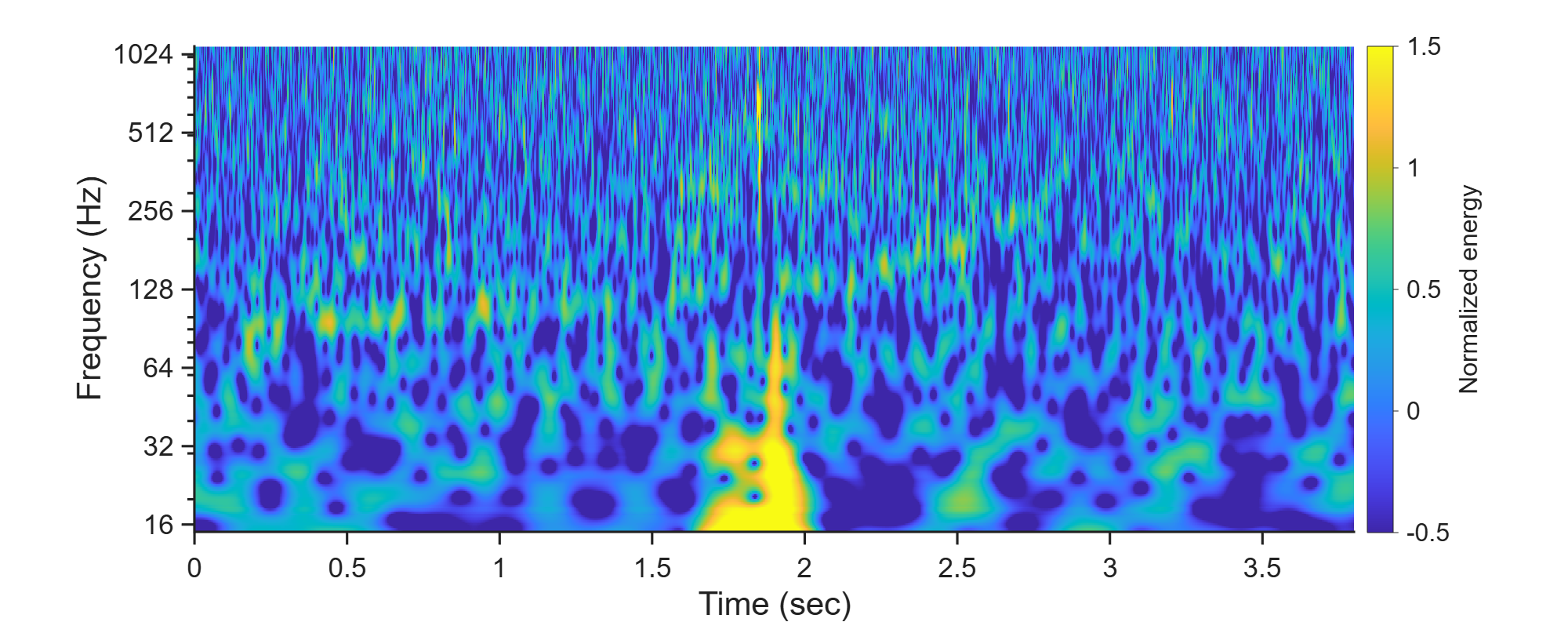}
\caption{}
\label{fig:FLARE_gw170817_cqt_after}
\end{subfigure}

\vspace{0.3cm}

\begin{subfigure}{0.325\textwidth}
\centering
\includegraphics[width=\textwidth]{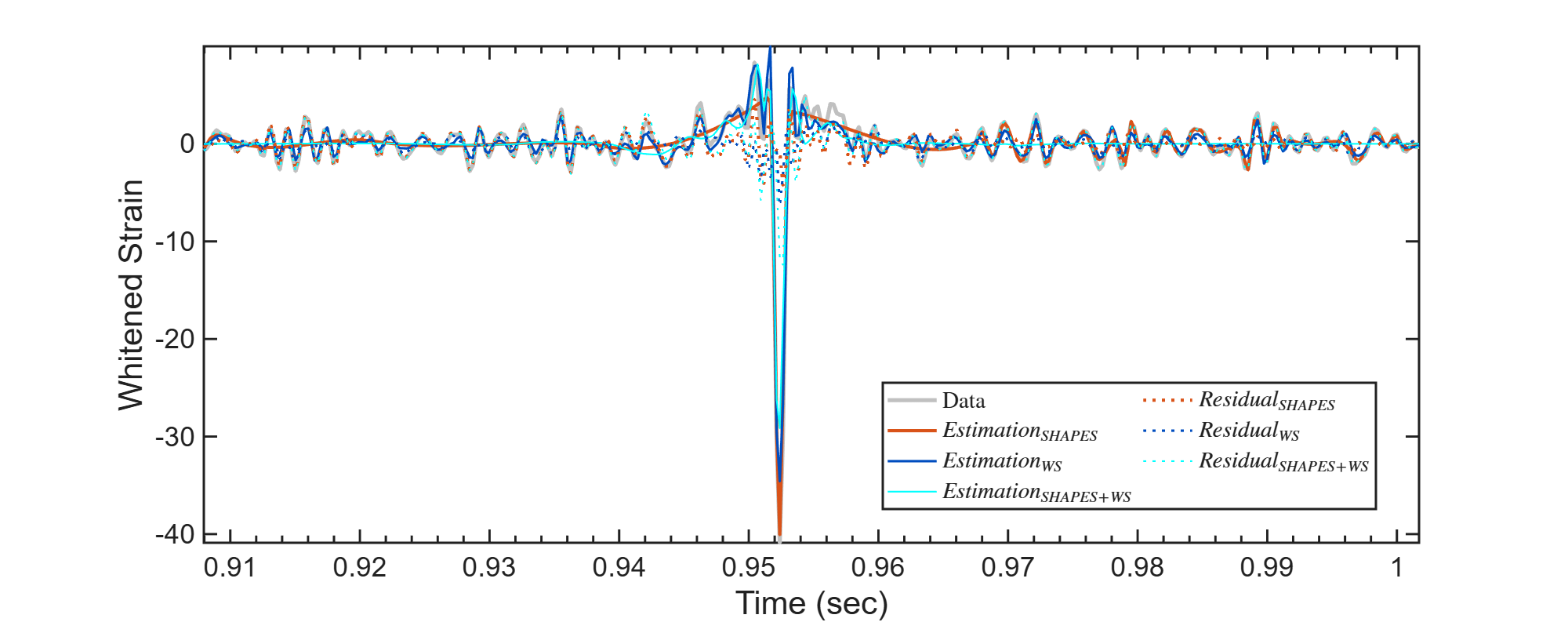}
\caption{}
\label{fig:FLARE_blip_zoom}
\end{subfigure}
\hfill
\begin{subfigure}{0.325\textwidth}
\centering
\includegraphics[width=\textwidth]{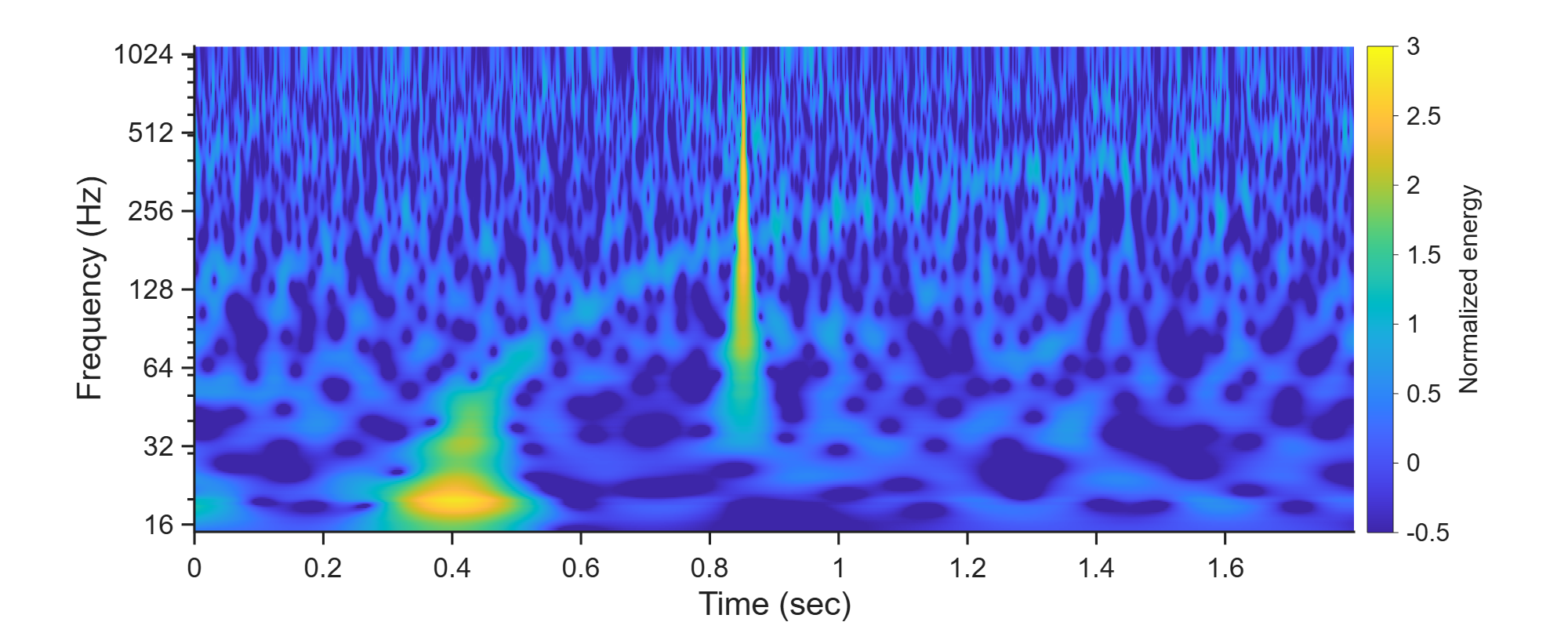}
\caption{}
\label{fig:FLARE_blip_cqt_before}
\end{subfigure}
\hfill
\begin{subfigure}{0.325\textwidth}
\centering
\includegraphics[width=\textwidth]{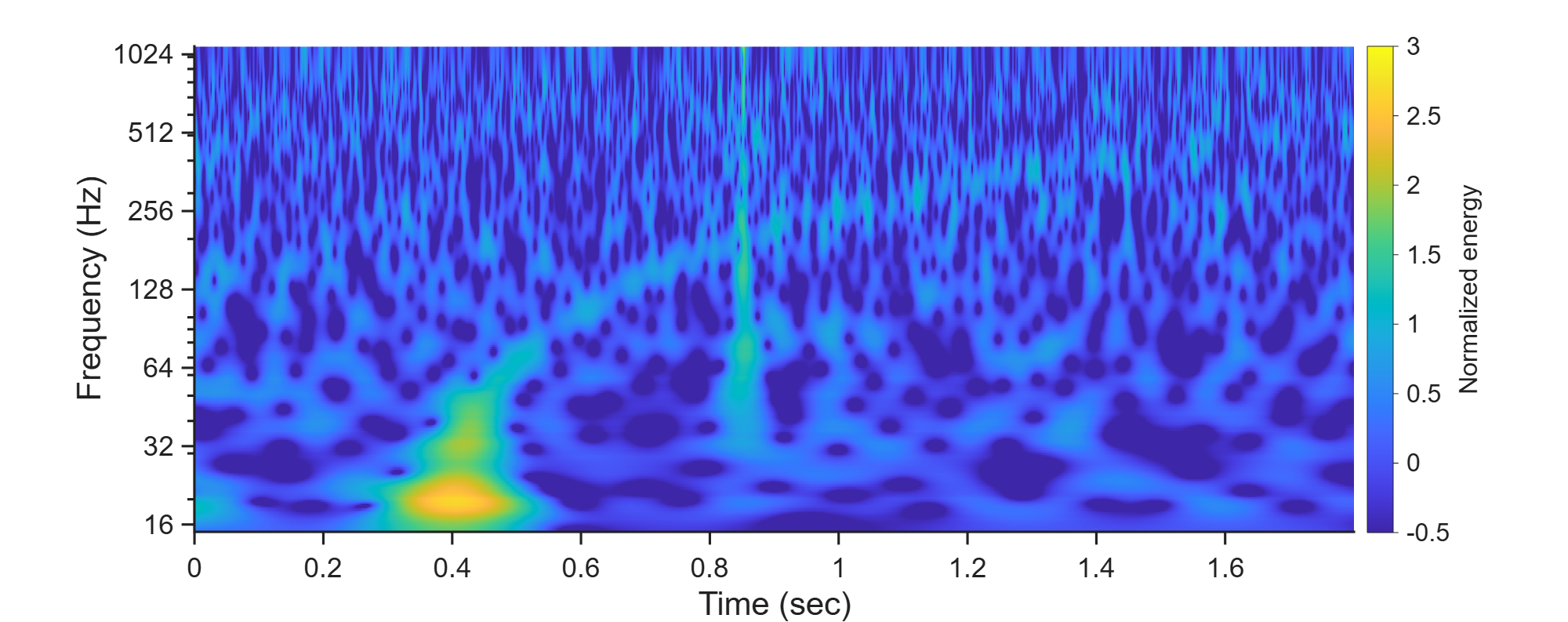}
\caption{}
\label{fig:FLARE_blip_cqt_after}
\end{subfigure}

\vspace{0.3cm}

\begin{subfigure}{0.325\textwidth}
\centering
\includegraphics[width=\textwidth]{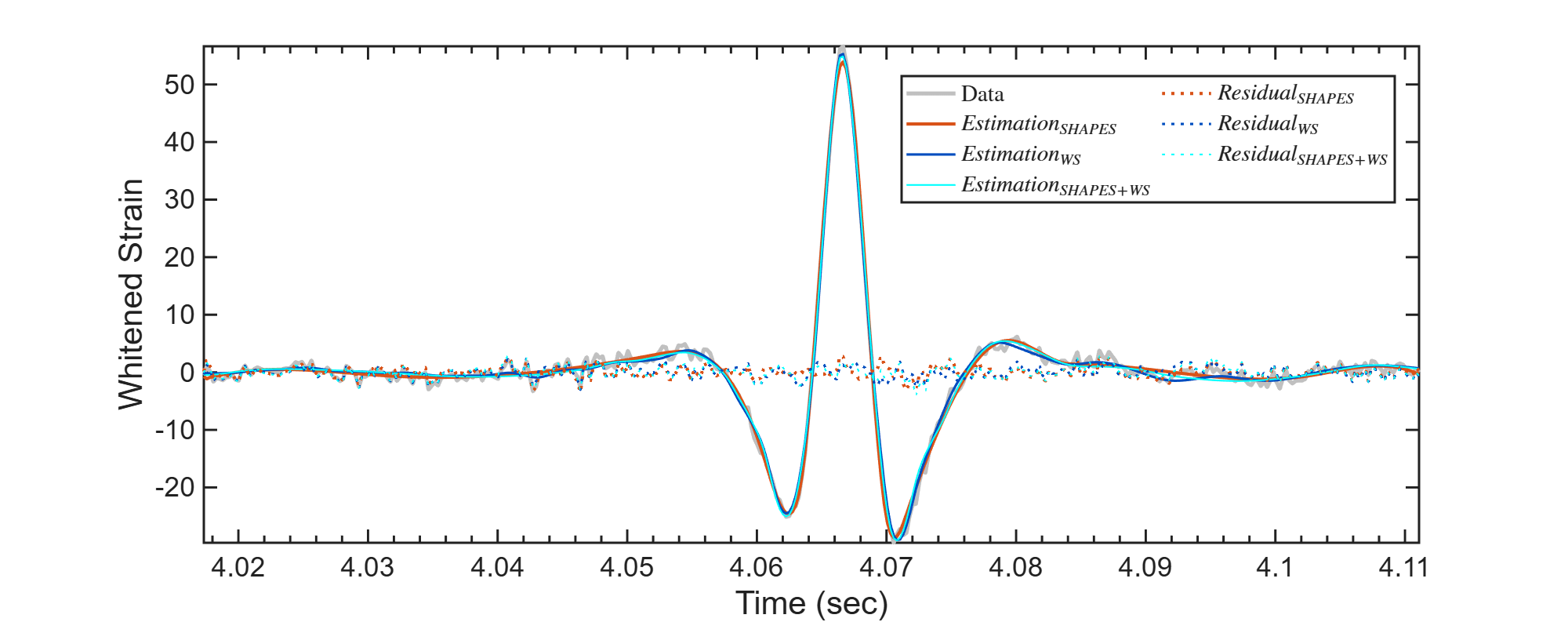}
\caption{}
\label{fig:FLARE_koifish_zoom}
\end{subfigure}
\hfill
\begin{subfigure}{0.325\textwidth}
\centering
\includegraphics[width=\textwidth]{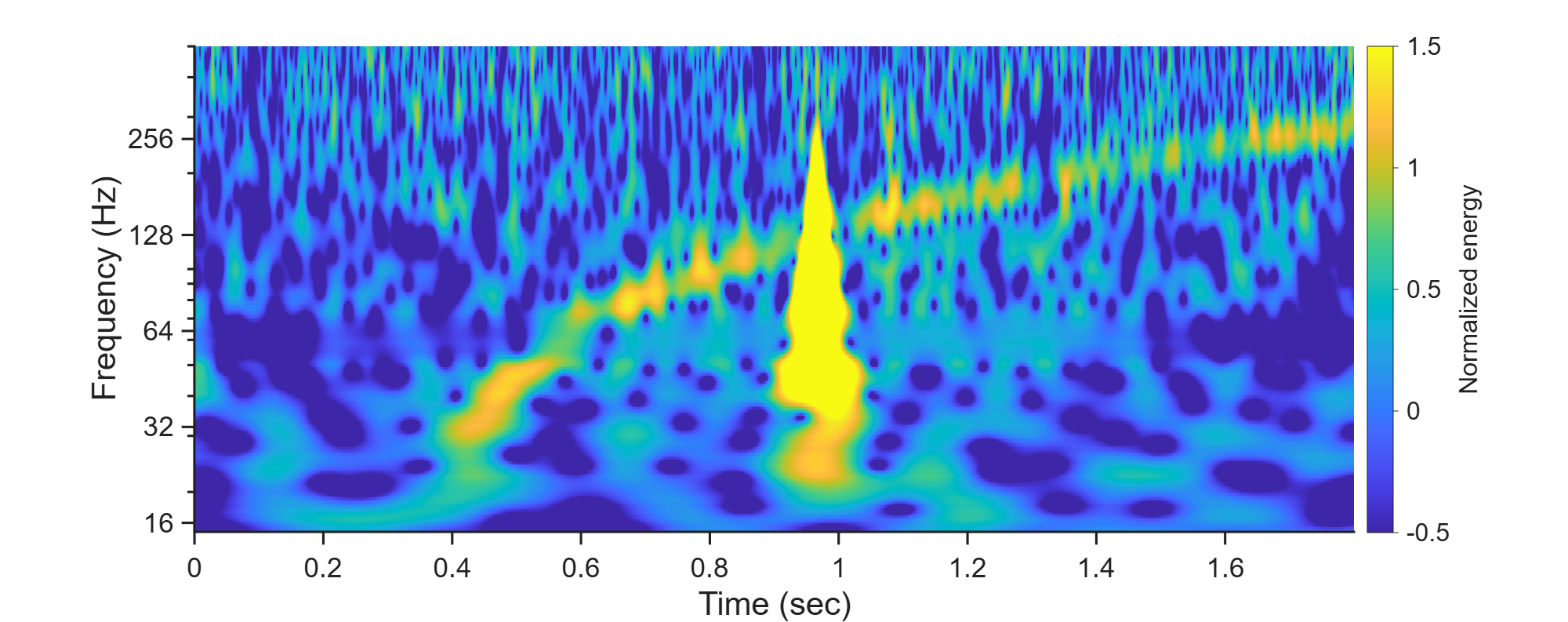}
\caption{}
\label{fig:FLARE_koifish_cqt_before}
\end{subfigure}
\hfill
\begin{subfigure}{0.325\textwidth}
\centering
\includegraphics[width=\textwidth]{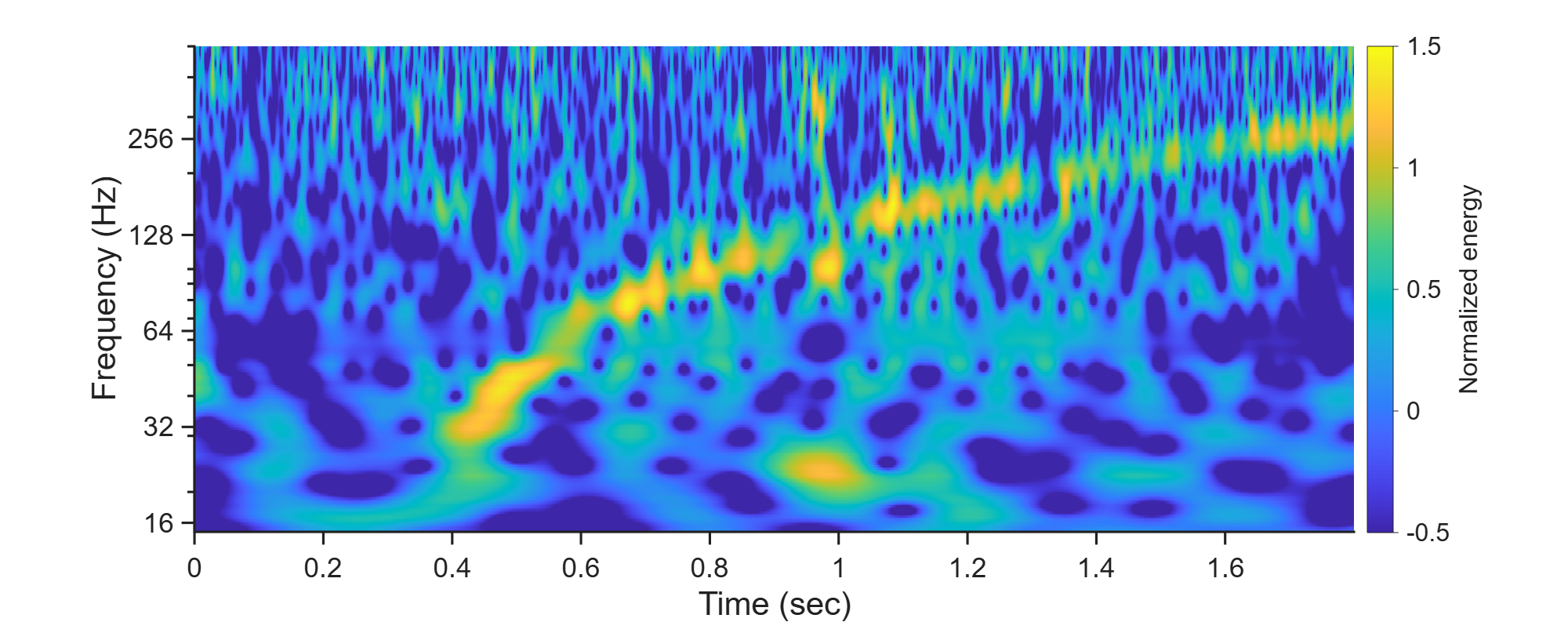}
\caption{}
\label{fig:FLARE_koifish_cqt_after}
\end{subfigure}

\vspace{0.3cm}

\begin{subfigure}{0.325\textwidth}
\centering
\includegraphics[width=\textwidth]{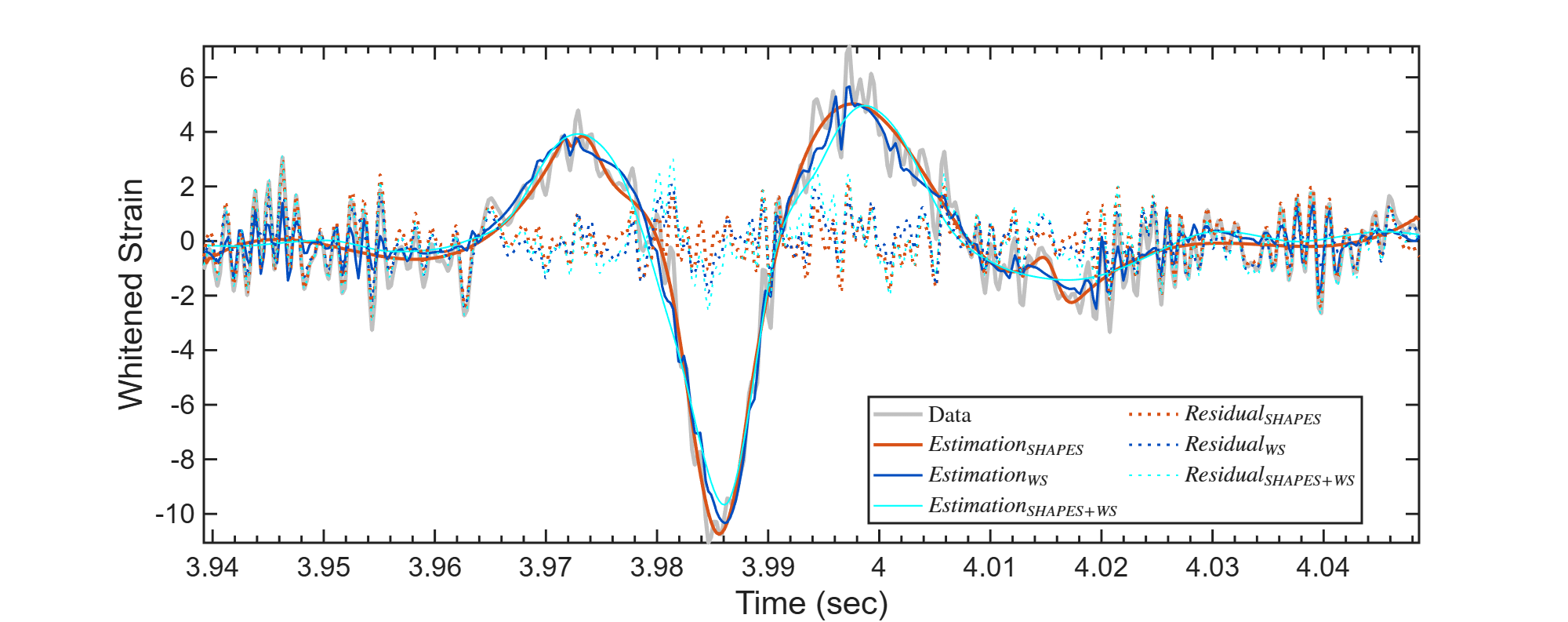}
\caption{}
\label{fig:FLARE_tomte_zoom}
\end{subfigure}
\hfill
\begin{subfigure}{0.325\textwidth}
\centering
\includegraphics[width=\textwidth]{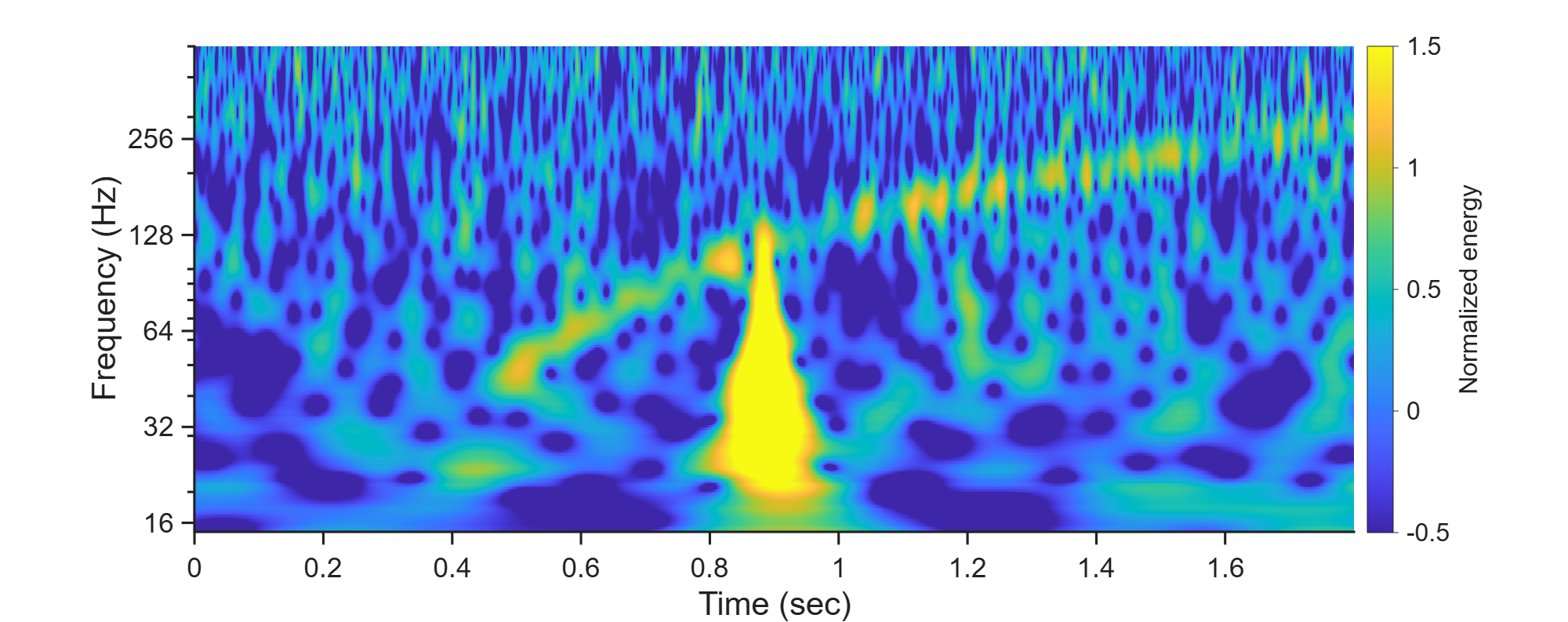}
\caption{}
\label{fig:FLARE_tomte_cqt_before}
\end{subfigure}
\hfill
\begin{subfigure}{0.325\textwidth}
\centering
\includegraphics[width=\textwidth]{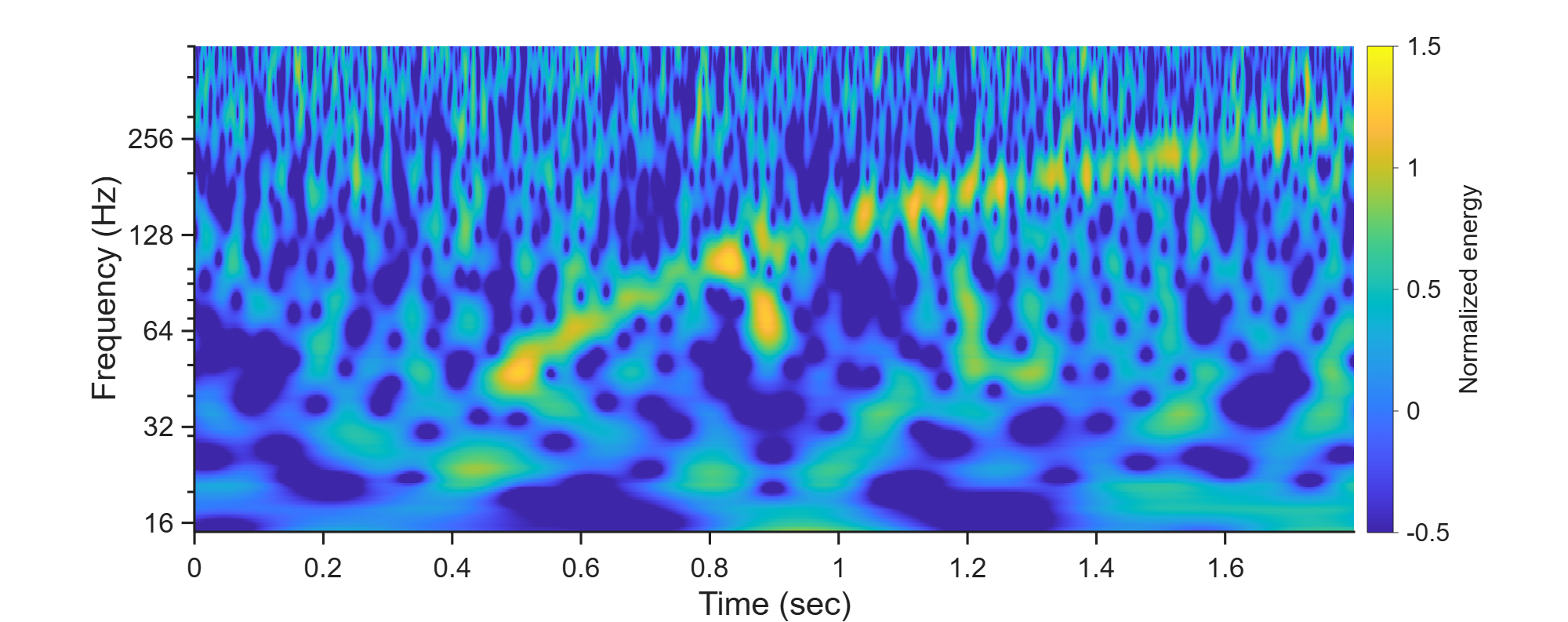}
\caption{}
\label{fig:FLARE_tomte_cqt_after}
\end{subfigure}

\vspace{0.3cm}

\begin{subfigure}{0.325\textwidth}
\centering
\includegraphics[width=\textwidth]{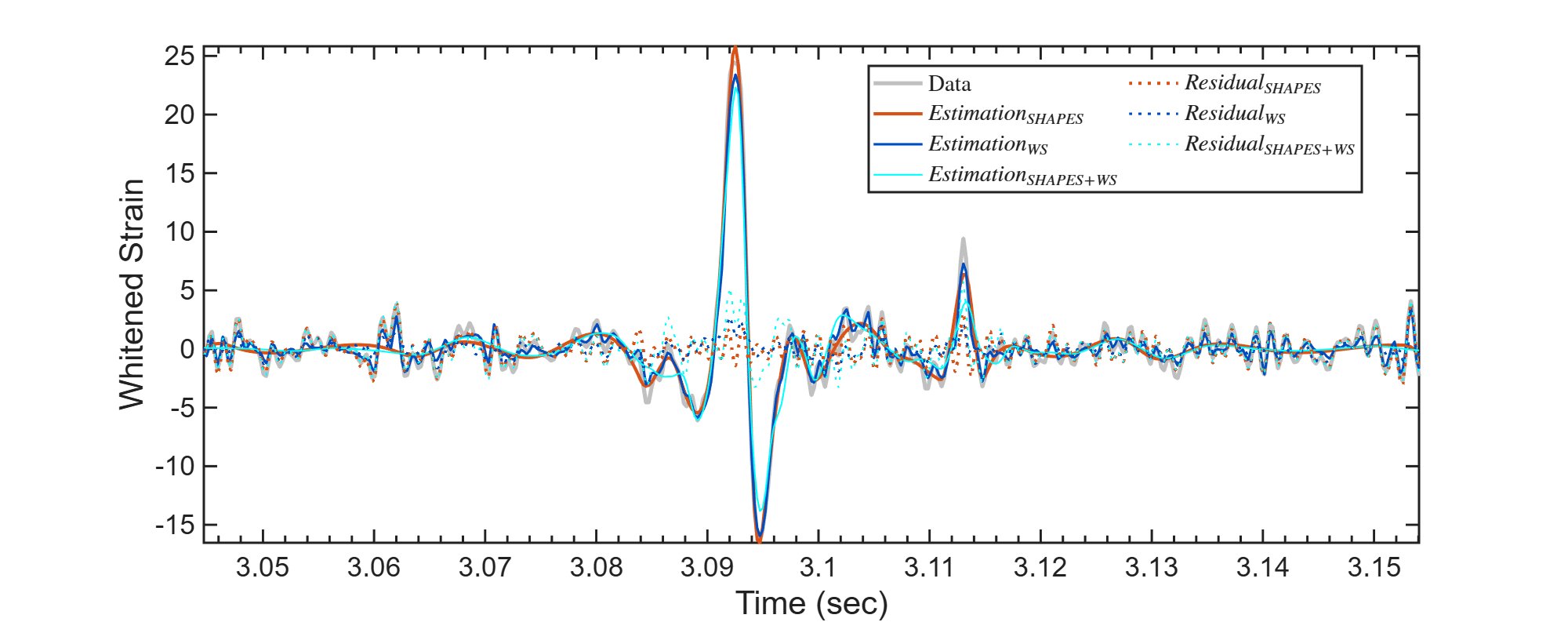}
\caption{}
\label{fig:FLARE_lfb_zoom}
\end{subfigure}
\hfill
\begin{subfigure}{0.325\textwidth}
\centering
\includegraphics[width=\textwidth]{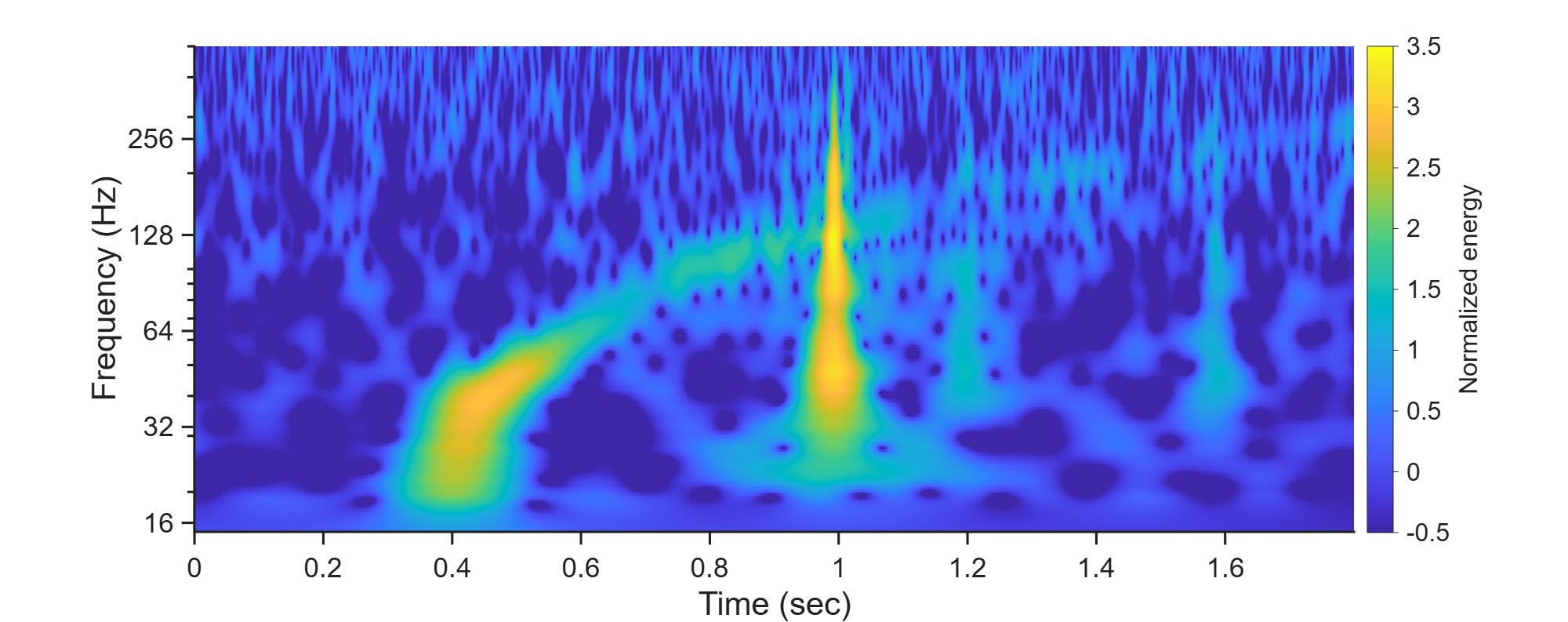}
\caption{}
\label{fig:FLARE_lfb_cqt_before}
\end{subfigure}
\hfill
\begin{subfigure}{0.325\textwidth}
\centering
\includegraphics[width=\textwidth]{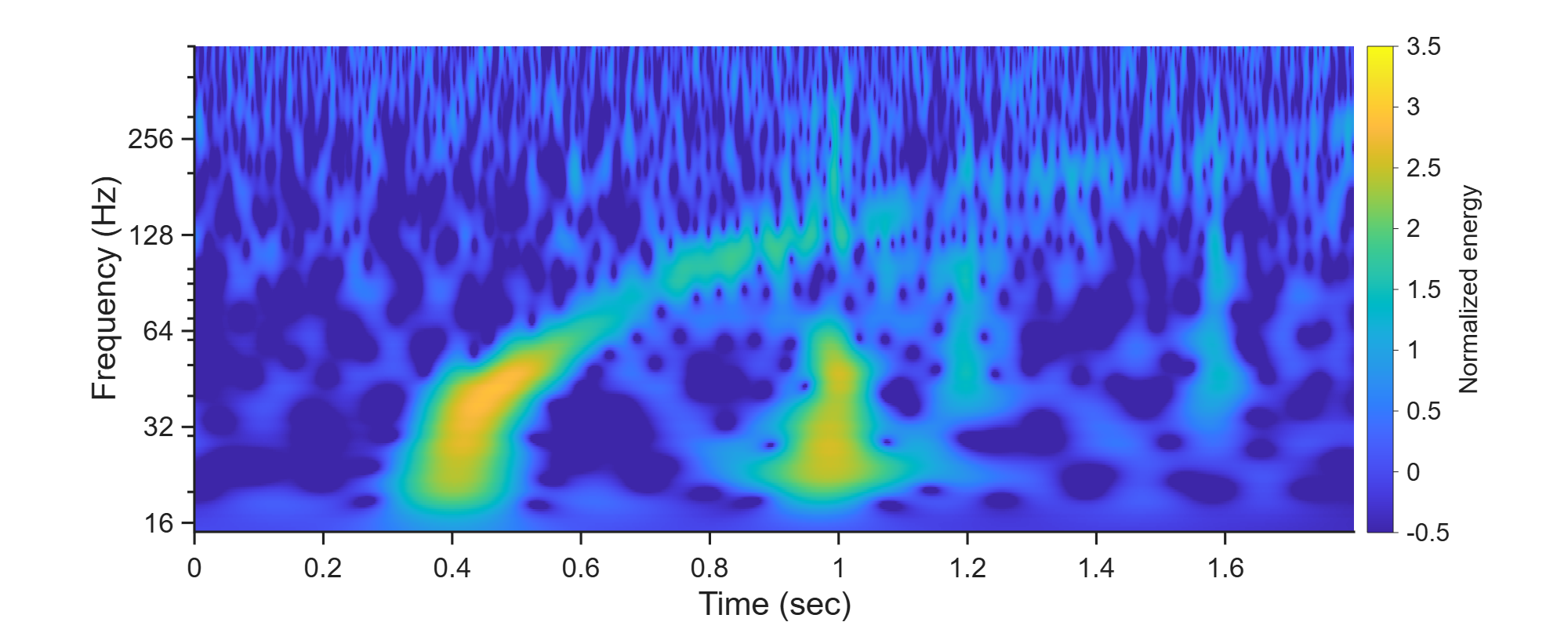}
\caption{}
\label{fig:flare_lfb_cqt_after}
\end{subfigure}

\caption{Identification and subtraction of the five glitch instances using FLARE, from top to bottom: GW170817, Blip, Koi Fish, Tomte, and the low frequency blip. First column: zoomed view of the segment FLARE identifies, whitened strain against time in seconds, with the SHAPES, WS, and Combined estimates drawn as solid lines and their residuals as dotted lines of matching color. Second column: Q-transform of the segment before subtraction, time in seconds against frequency in Hz, with the colorbar giving normalized energy. Third column: Q-transform of the same segment after subtraction, using the same time window, frequency range, and color scale as the second column, built from the Combined residual in every case.}
\label{fig:flare_all}
\end{figure*}

\subsection{FLARE}
\label{sec:flare_results}

FLARE was applied to the same five glitch instances used for AMPS, shown in Figure~\ref{fig:flare_all}. Each row gives the zoomed segment with the SHAPES, WS, and Combined estimates, followed by the Q-transform before and after subtraction, the latter built from the Combined residual.

Blip crosses the chirp track only briefly, at a single instant where the glitch spans nearly the full plotted frequency range, up to 1024~Hz. Koi Fish shows a similarly brief crossing, but confined to a lower band, reaching only about 300~Hz. Tomte and the low frequency blip each meet the rising chirp near 100 to 130~Hz, shortly before the glitch itself begins, the same brief, narrow character seen for Blip and Koi Fish, differing only in where the crossing falls.

GW170817 takes the longest to fit among the five cases, at 70.34~s, consistent with its greater duration and complexity. The low frequency blip is the next slowest, at 48.27~s against a 19.2~s average for Koi Fish, Tomte, and Blip. The cause of this extended SHAPES stage is not yet clear from the data alone, since Koi Fish and Blip also cross their chirp tracks without the same slowdown. 

Table~\ref{tab:snr_comparison} shows Blip and Tomte again give up the largest fraction of injected SNR among the four glitches, consistent with the pattern established under AMPS.Koi Fish gives the highest recovered fraction among the four glitches under both SHAPES and Combined, with SHAPES marginally ahead this time. Tomte and Blip behave as expected, with Combined ahead of SHAPES by a modest margin in both.

For GW170817, the Q-transform after subtraction still shows a low-frequency patch. This patch sits on the astrophysical signal's own track before subtraction; once removed, it no longer overlaps the recovered track and does not take out a meaningful amount of signal power in the process. Appendix~\ref{app:boundary_sensitivity} traces this patch to the fixed threshold multiplier used to set the boundary and shows that it shrinks, without disappearing entirely, as the multiplier is lowered.

Averaged over the four injected glitches (capped at 100\% as in Section~\ref{sec:results_amps}), Combined comes out ahead overall at about 96.4\%, with SHAPES close behind at 94.6\%. WS alone trails both by a wide margin, at roughly 64.2\%. Combined gives the best recovered SNR for Blip, Tomte, and the low frequency blip; on Koi Fish, SHAPES edges out Combined by a narrow margin instead. WS alone stays lowest across every glitch, never approaching the other two. Under FLARE boundaries, then, SHAPES and Combined again preserve the injected signal more effectively than WS alone, with Combined keeping a modest edge over SHAPES rather than a decisive one.

\begin{figure*}[htbp!]
\centering

\begin{subfigure}{0.32\textwidth}
\centering
\includegraphics[width=\textwidth, height = 0.90in]{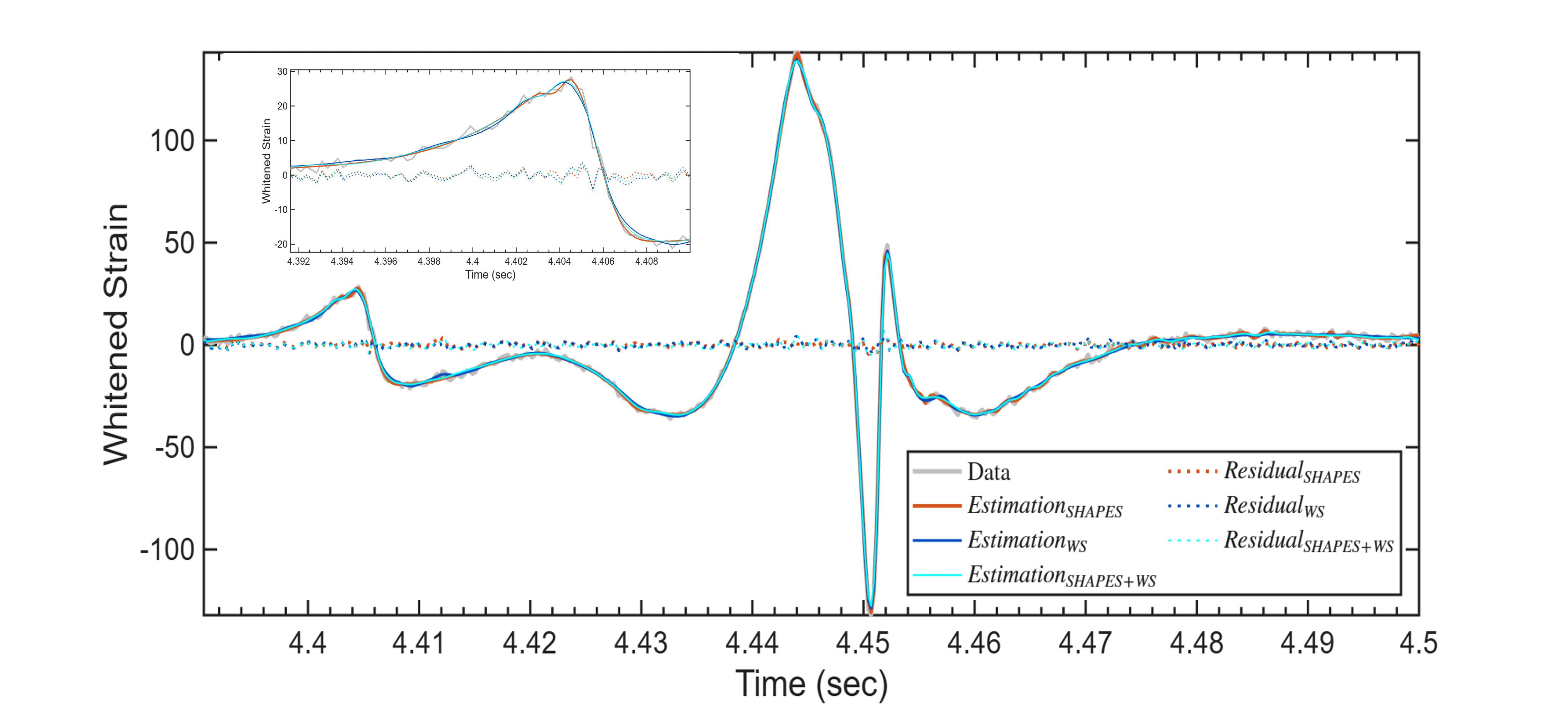}
\caption{}
\label{fig:CRISP_gw170817_zoom}
\end{subfigure}
\hfill
\begin{subfigure}{0.32\textwidth}
\centering
\includegraphics[width=\textwidth]{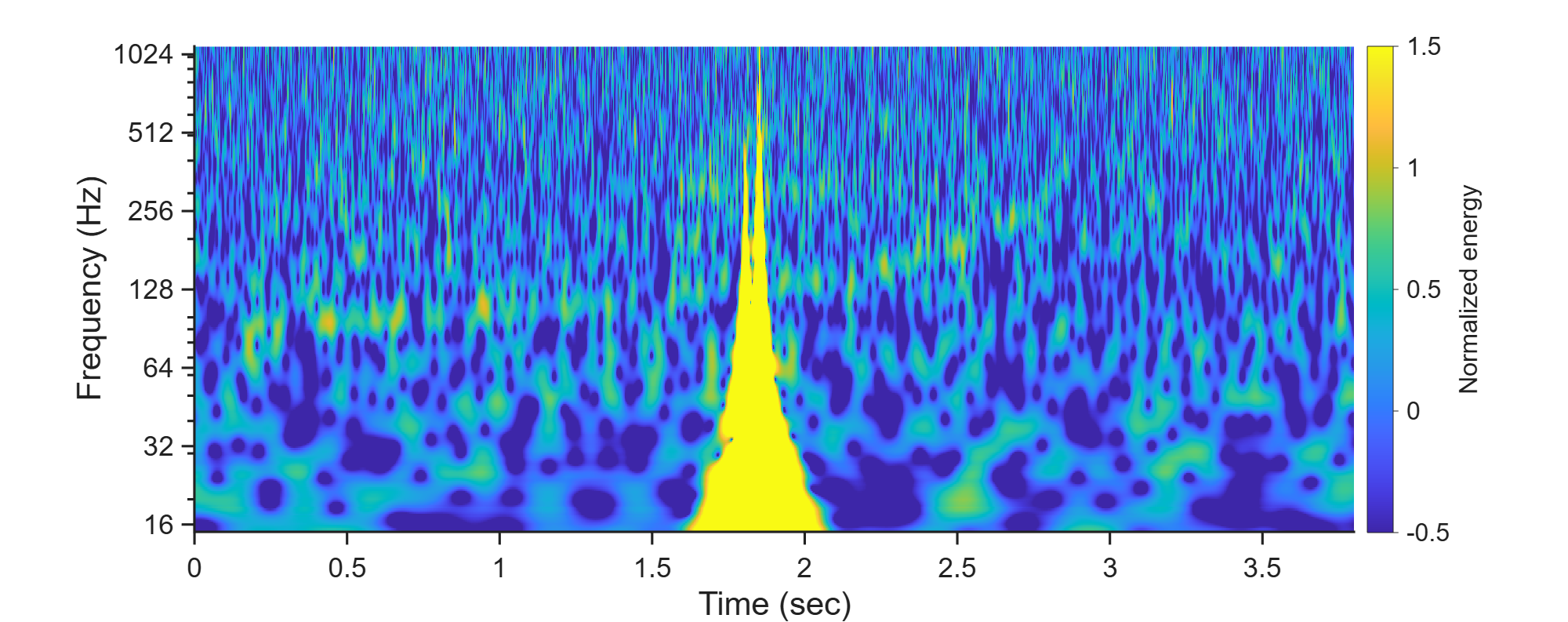}
\caption{}
\label{fig:CRISP_gw170817_cqt_before}
\end{subfigure}
\hfill
\begin{subfigure}{0.32\textwidth}
\centering
\includegraphics[width=\textwidth]{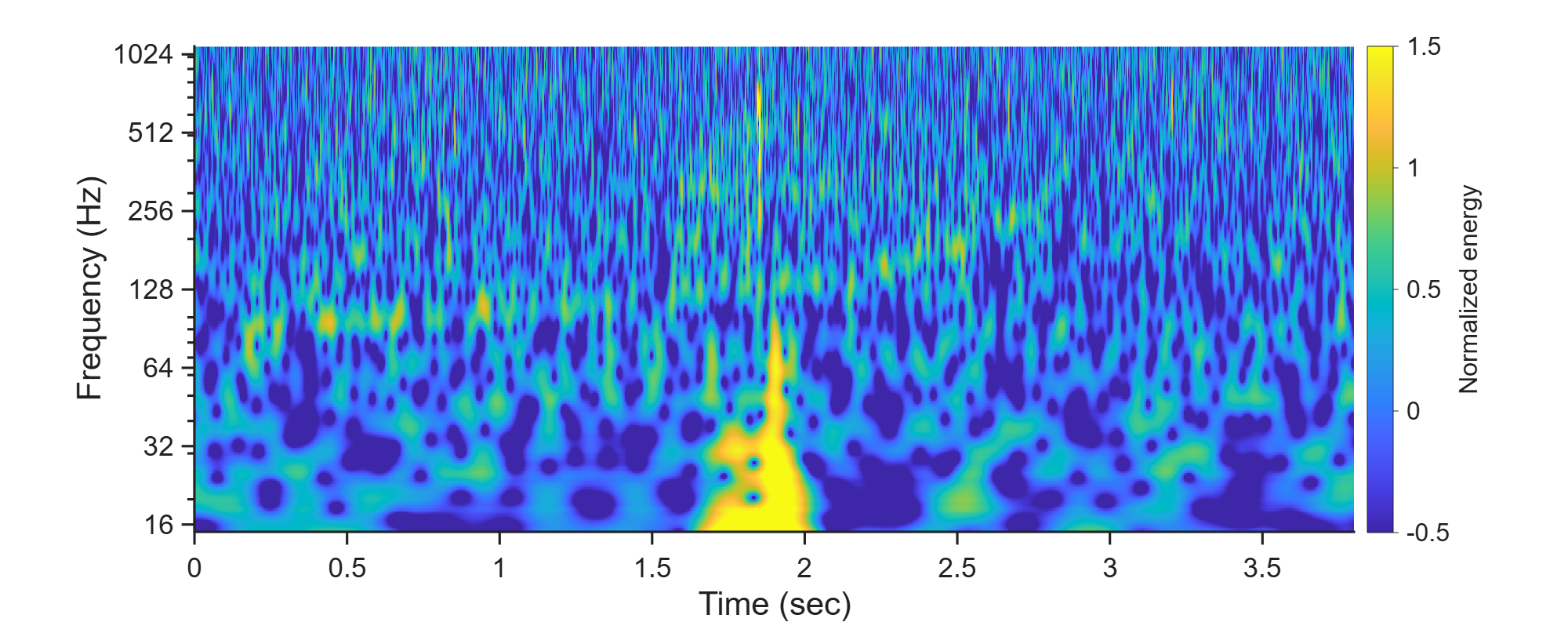}
\caption{}
\label{fig:CRISP_gw170817_cqt_after}
\end{subfigure}

\vspace{0.3cm}

\begin{subfigure}{0.32\textwidth}
\centering
\includegraphics[width=\textwidth]{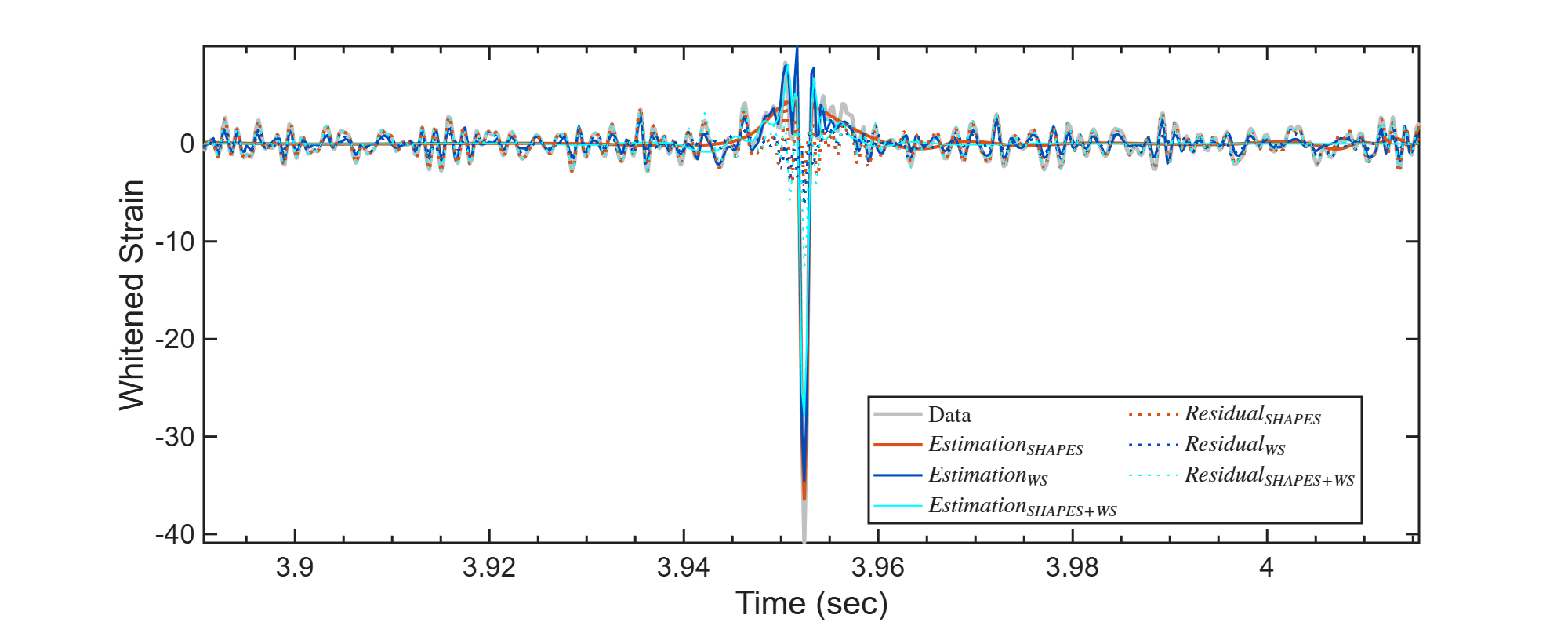}
\caption{}
\label{fig:CRISP_blip_zoom}
\end{subfigure}
\hfill
\begin{subfigure}{0.32\textwidth}
\centering
\includegraphics[width=\textwidth]{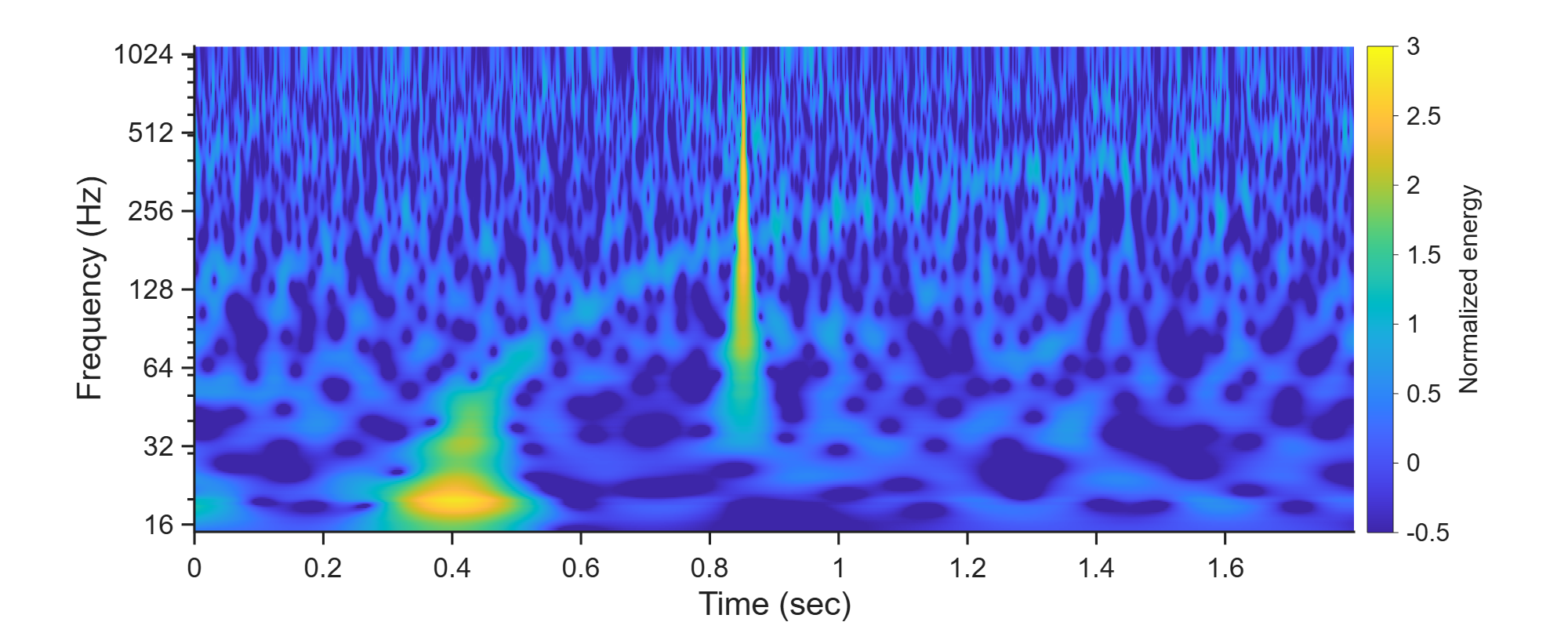}
\caption{}
\label{fig:CRISP_blip_cqt_before}
\end{subfigure}
\hfill
\begin{subfigure}{0.32\textwidth}
\centering
\includegraphics[width=\textwidth]{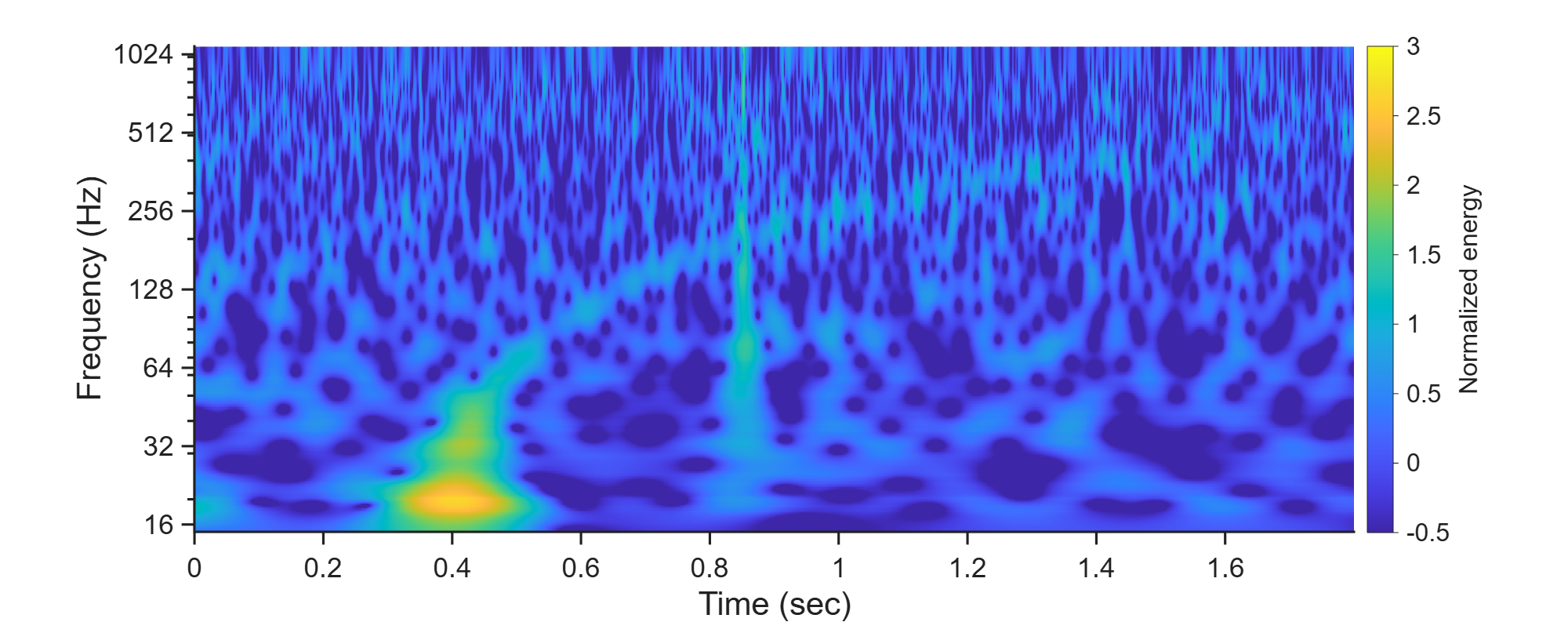}
\caption{}
\label{fig:CRISP_blip_cqt_after}
\end{subfigure}

\vspace{0.3cm}

\begin{subfigure}{0.32\textwidth}
\centering
\includegraphics[width=\textwidth]{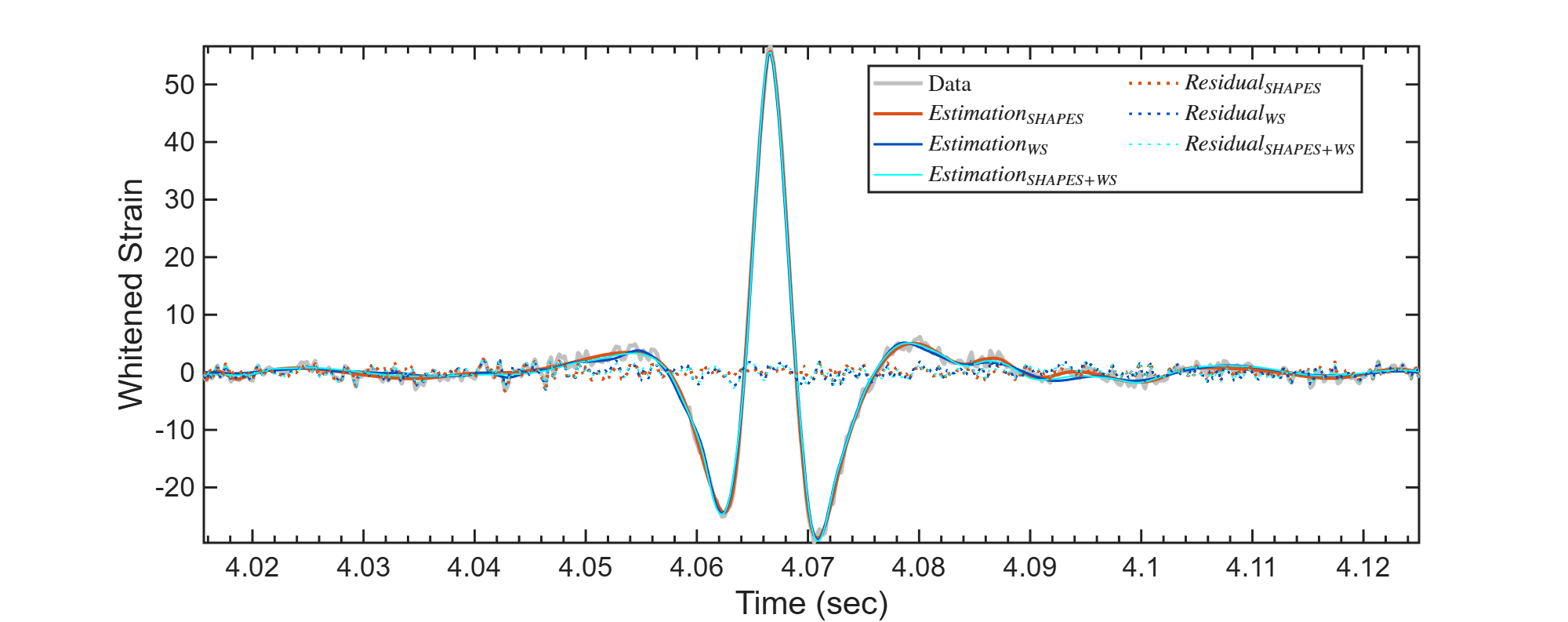}
\caption{}
\label{fig:CRISP_koifish_zoom}
\end{subfigure}
\hfill
\begin{subfigure}{0.32\textwidth}
\centering
\includegraphics[width=\textwidth]{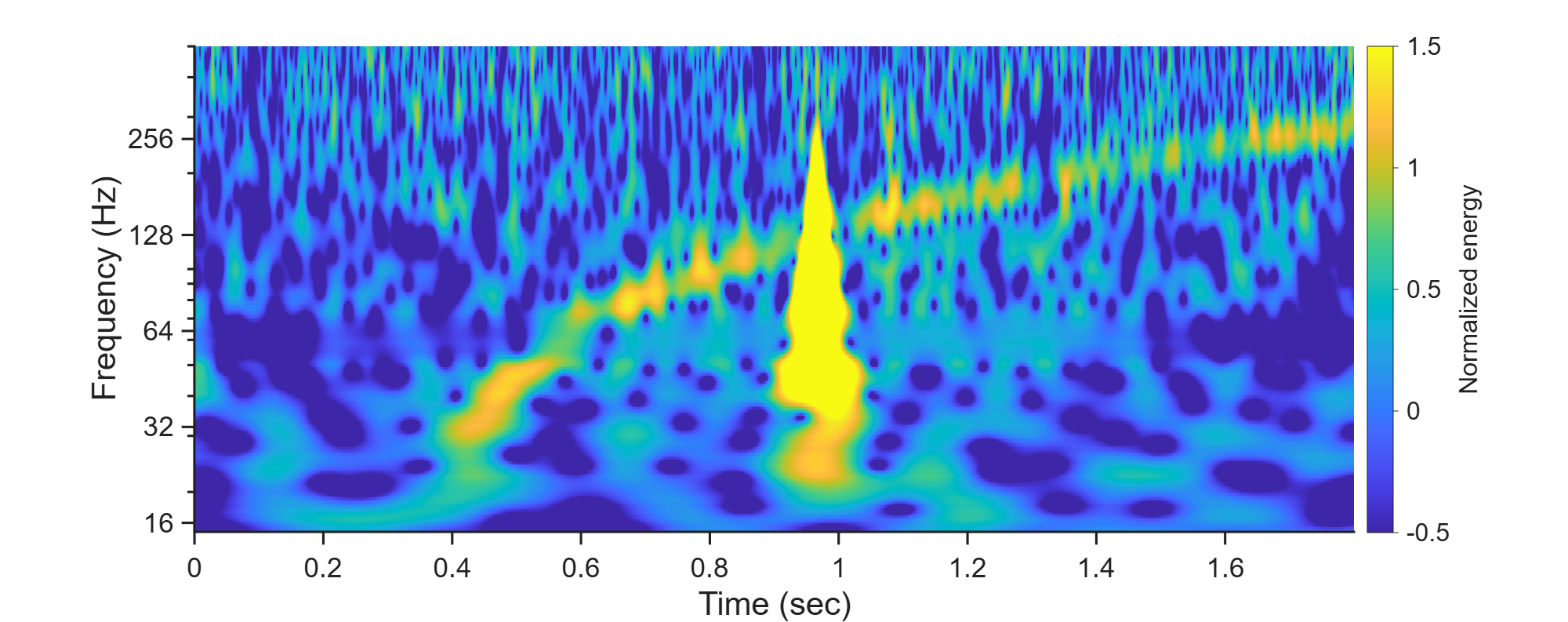}
\caption{}
\label{fig:CRISP_koifish_cqt_before}
\end{subfigure}
\hfill
\begin{subfigure}{0.32\textwidth}
\centering
\includegraphics[width=\textwidth]{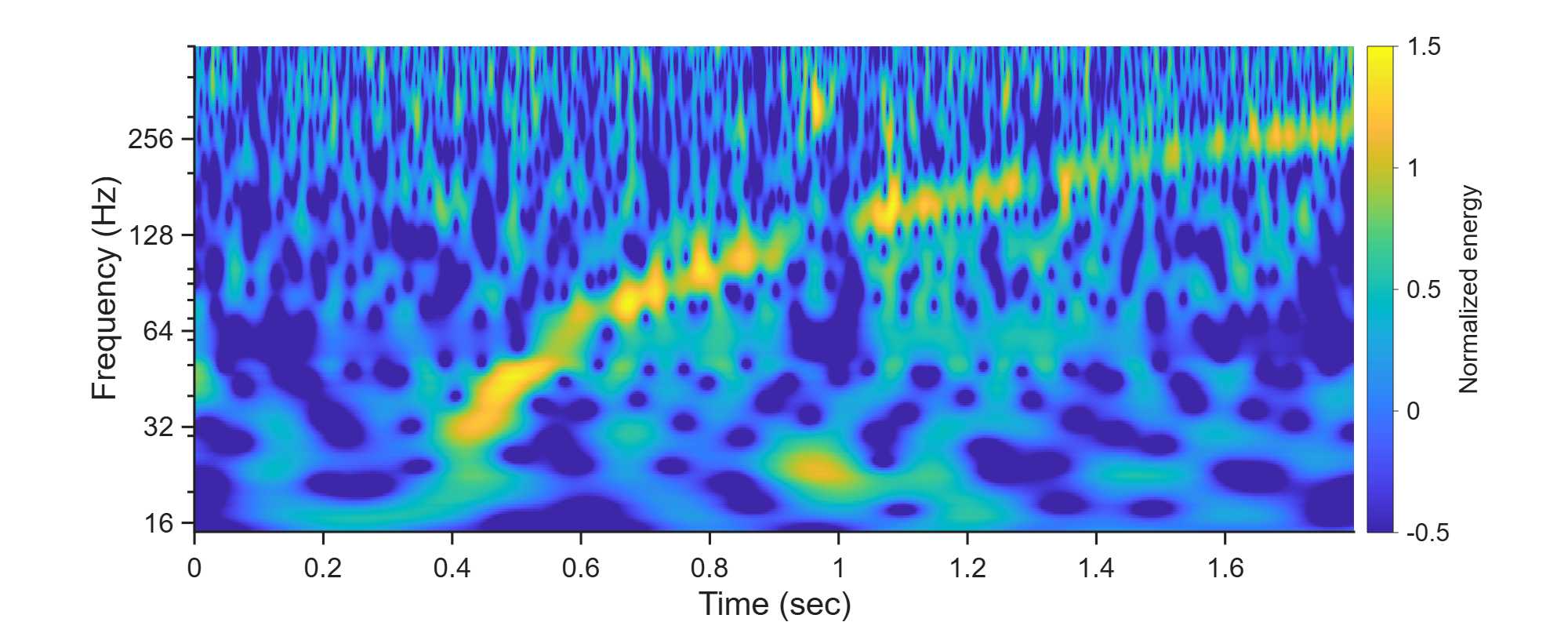}
\caption{}
\label{fig:CRISP_koifish_cqt_after}
\end{subfigure}

\vspace{0.3cm}

\begin{subfigure}{0.32\textwidth}
\centering
\includegraphics[width=\textwidth]{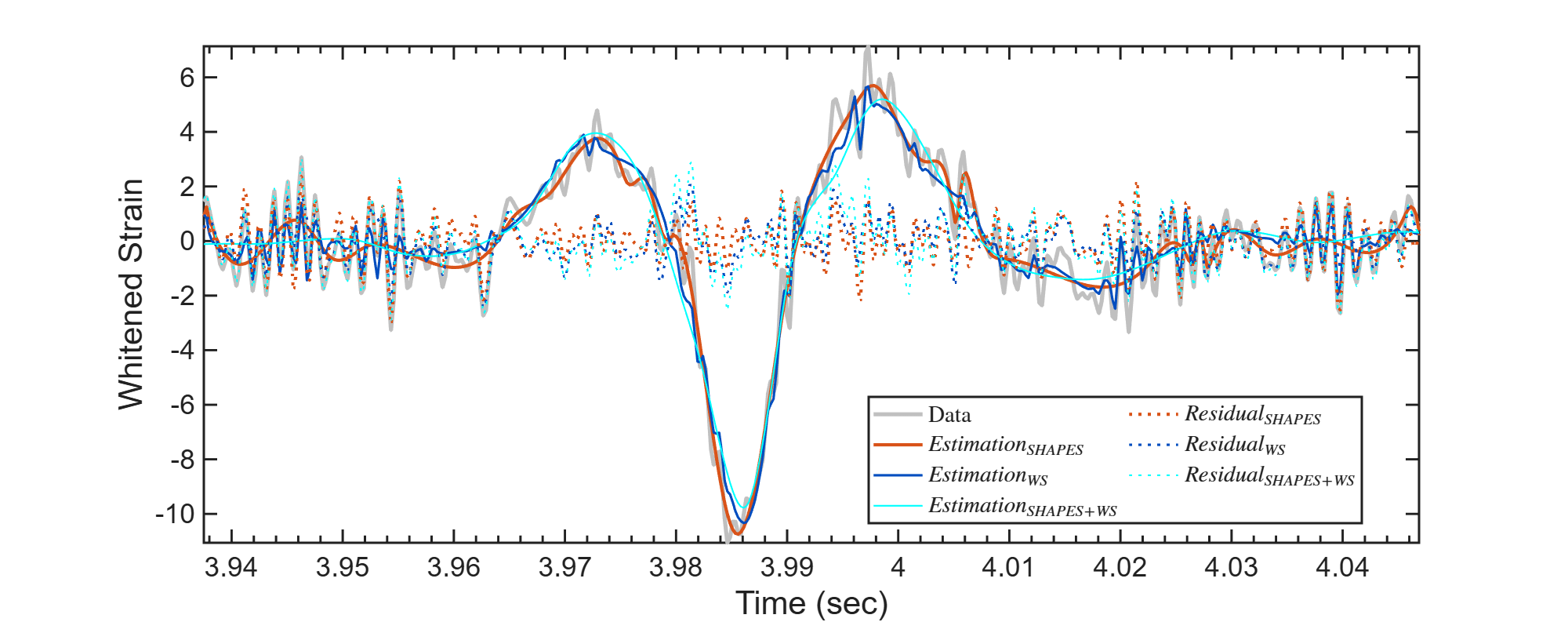}
\caption{}
\label{fig:CRISP_tomte_zoom}
\end{subfigure}
\hfill
\begin{subfigure}{0.32\textwidth}
\centering
\includegraphics[width=\textwidth]{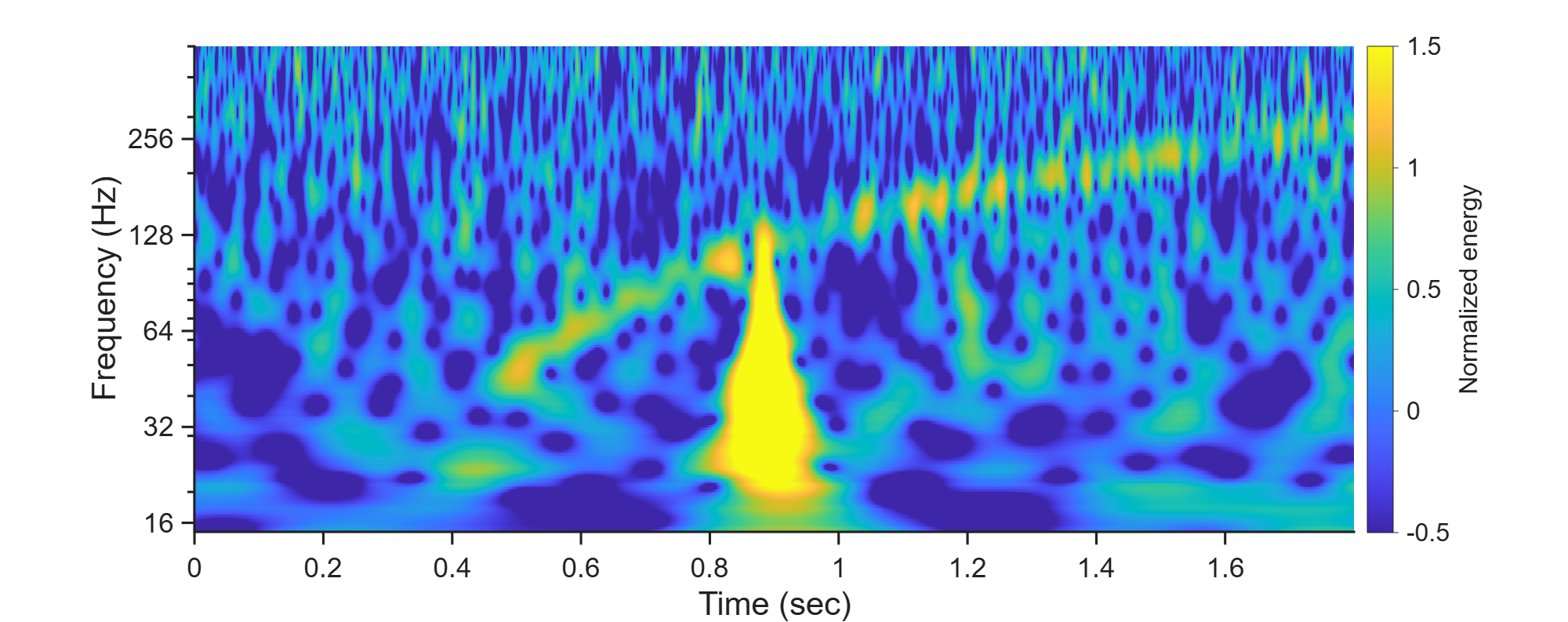}
\caption{}
\label{fig:CRISP_tomte_cqt_before}
\end{subfigure}
\hfill
\begin{subfigure}{0.32\textwidth}
\centering
\includegraphics[width=\textwidth]{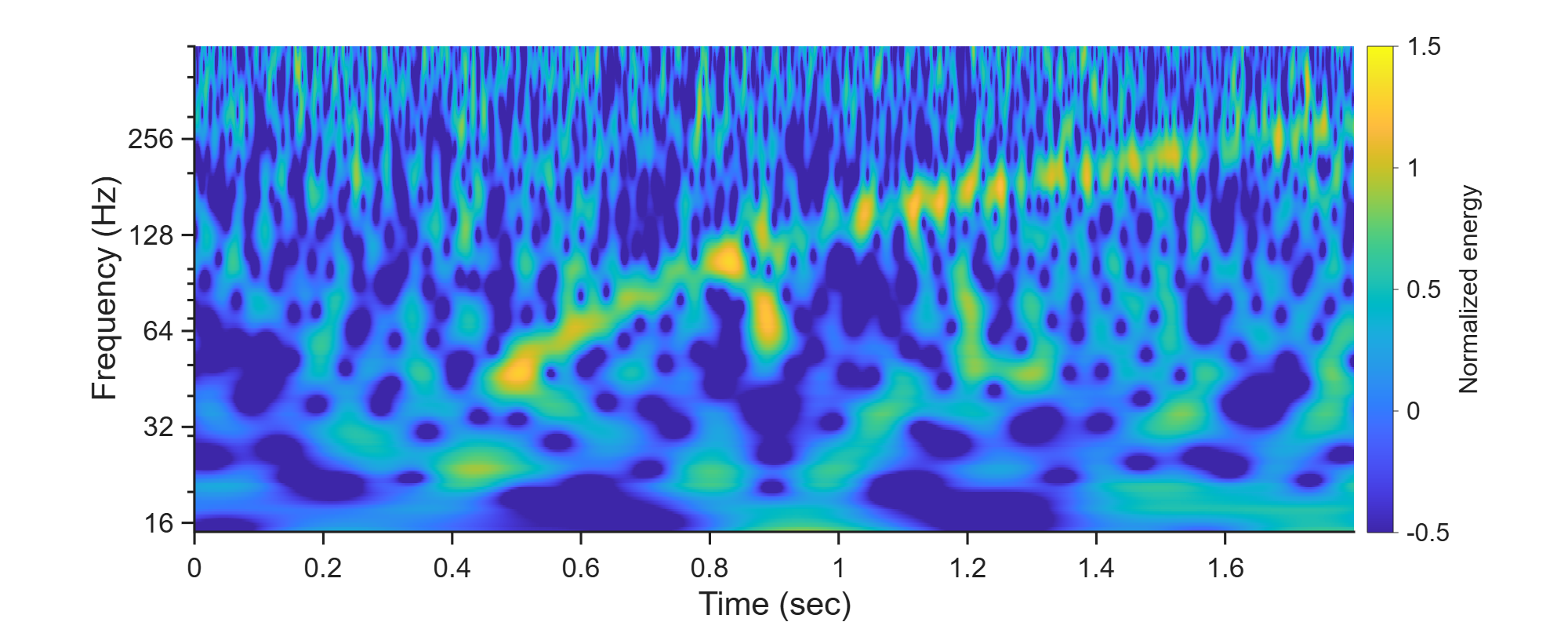}
\caption{}
\label{fig:CRISP_tomte_cqt_after}
\end{subfigure}

\vspace{0.3cm}

\begin{subfigure}{0.32\textwidth}
\centering
\includegraphics[width=\textwidth]{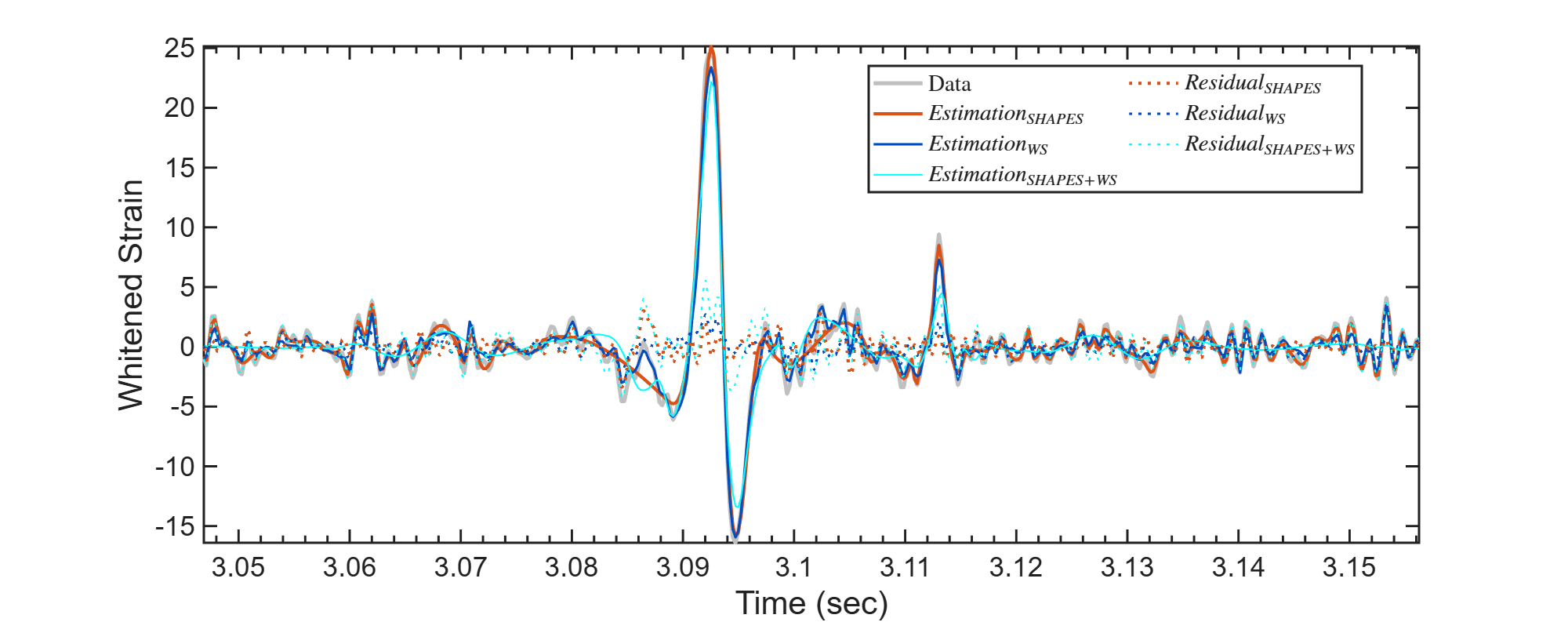}
\caption{}
\label{fig:CRISP_lfb_zoom}
\end{subfigure}
\hfill
\begin{subfigure}{0.32\textwidth}
\centering
\includegraphics[width=\textwidth]{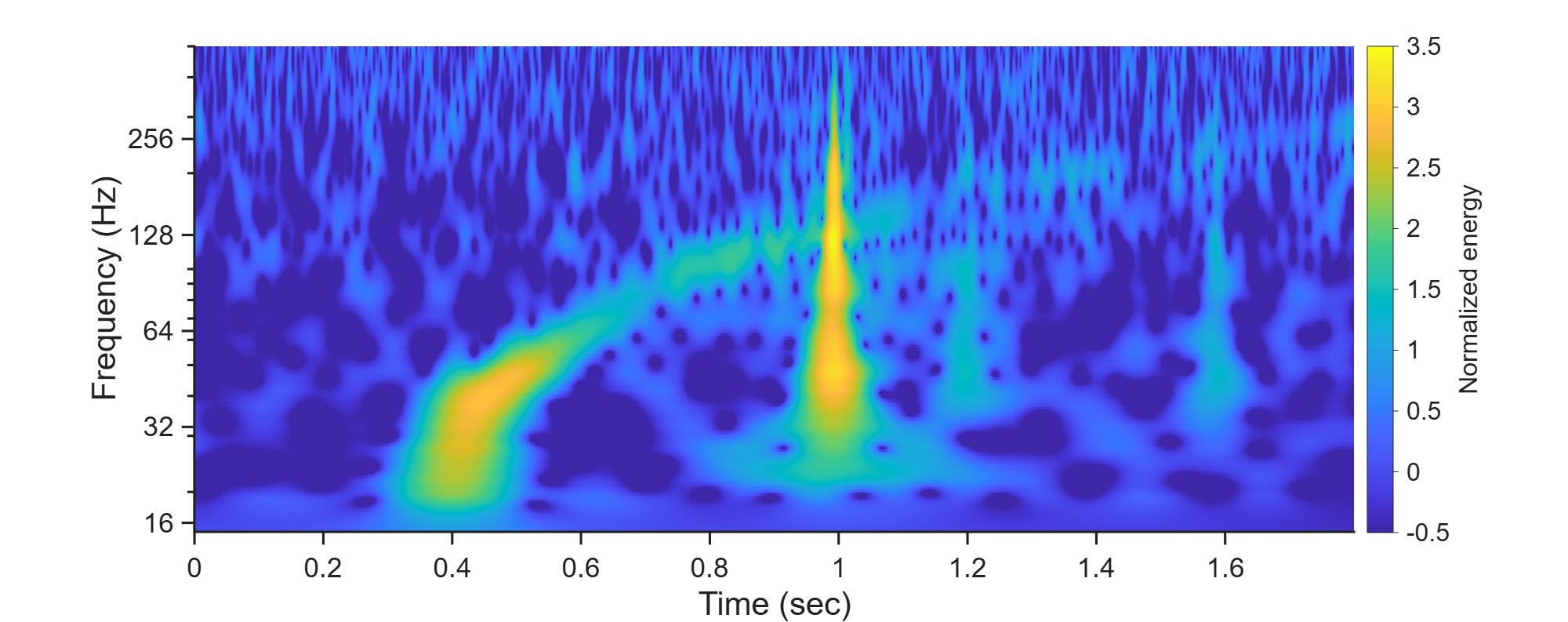}
\caption{}
\label{fig:CRISP_lfb_cqt_before}
\end{subfigure}
\hfill
\begin{subfigure}{0.32\textwidth}
\centering
\includegraphics[width=\textwidth]{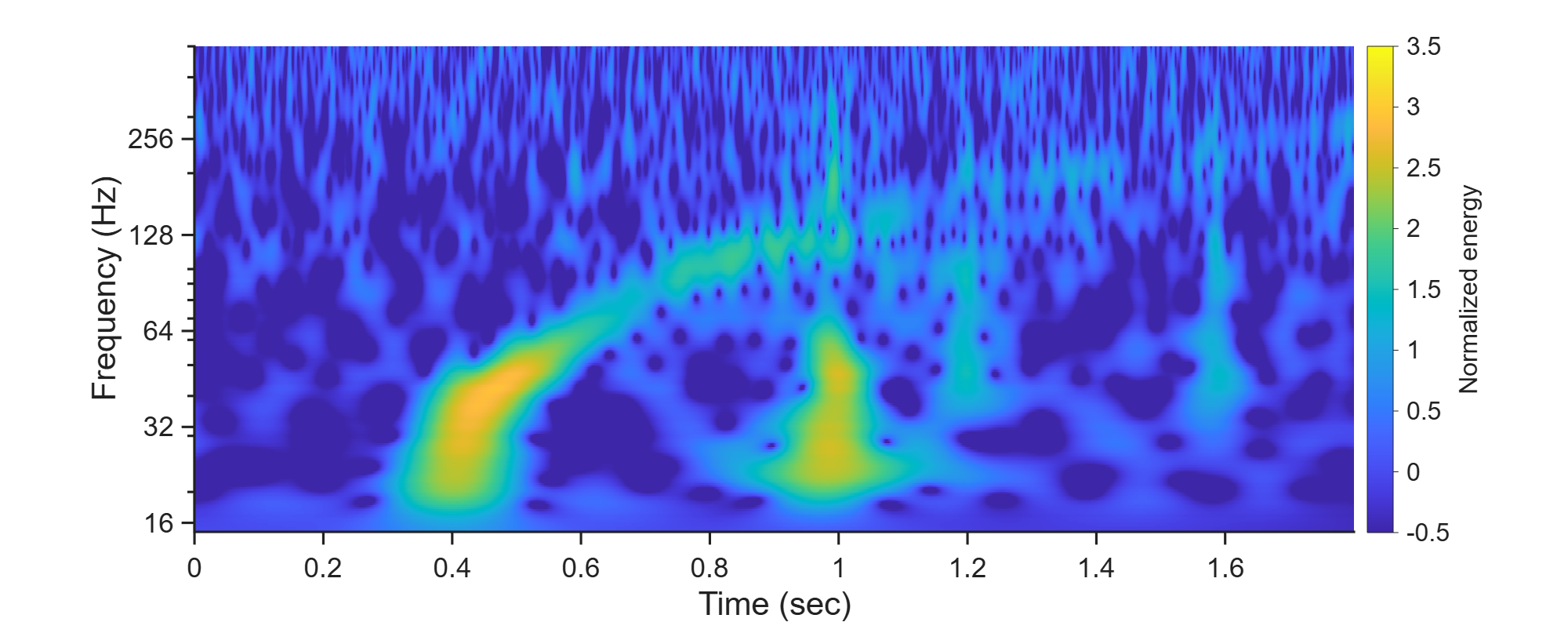}
\caption{}
\label{fig:CRISP_lfb_cqt_after}
\end{subfigure}

\caption{Identification and subtraction of five glitch instances using CRISP: GW170817, Blip, Koi Fish, Tomte, and the low frequency blip, top to bottom. The first column (whitened strain versus time, in seconds) is shared with Figure~\ref{fig:amps_all} and is not repeated here. First column of each row: zoomed view of the boundary CRISP identifies, with the SHAPES, WS, and Combined estimates as solid lines and their residuals as dotted lines of matching color, whitened strain versus time in seconds. Second column: Q-transform of the segment before subtraction, time in seconds versus frequency in Hz, colorbar giving normalized energy. Third column: Q-transform of the same segment after subtraction, built from the Combined residual, using the same time window, frequency range, and color scale as the second column.}
\label{fig:CRISP_all}
\end{figure*}

\subsection{CRISP}
\label{sec:results_crisp}

CRISP was run on the same five instances as AMPS and FLARE, shown in Figure~\ref{fig:CRISP_all}. Each row gives the zoomed boundary region with the three subtraction estimates, then the Q-transform before and after subtraction.

The injected chirp survives subtraction in all four glitch cases and stays visible to the eye directly, with no need to search for it against the background. GW170817 shows the same low-frequency patch seen under FLARE, expected since CRISP and FLARE share the 0.1094 s boundary for this glitch (Section~\ref{sec:flare_results}).

CRISP settles on a 448-sample boundary (0.1094~s) for four of the five glitches, GW170817, Koi Fish, Tomte, and the low frequency blip, and widens slightly to 512 samples (0.1250~s) for Blip. That near-uniform width follows from how the boundary is set, a spectrogram power threshold applied independently of any particular fitting technique, rather than from anything AMPS's fixed-pad rule would produce on the same five cases. Estimation time, covering the subsequent SHAPES fit, is longer and more varied: 30.5~s for the low frequency blip up to 66.7~s for Blip, with GW170817, Koi Fish, and Tomte landing between 49 and 51~s.

Koi Fish again gives the highest recovered fraction among the four glitches under CRISP boundaries, consistent with the pattern seen under AMPS and FLARE. Blip and Tomte again give up the largest fraction of injected SNR under CRISP, matching the pattern seen under AMPS and FLARE. For Tomte and Blip, Combined again gives the highest recovered SNR, SHAPES sits a moderate step behind it, and WS alone trails both by a wide margin, most visibly for Blip.

Averaged over the four injected glitches and capping each at 100\% before averaging, SHAPES reaches roughly 92.7\%, WS alone about 64.2\%, and Combined about 96.3\%. Combined leads on every glitch, with the narrowest margin over SHAPES on Koi Fish. WS alone stays well below the other two for every glitch, with the largest shortfall on Blip. CRISP's recovery numbers land close to FLARE's on this same set, while its boundary widths cluster far more tightly around a single value than either AMPS or FLARE manage, a point taken up directly in the cross-method comparison that follows.

\subsection{Comparison of identification methods}
\label{sec:comparison_identification}

Table~\ref{tab:snr_comparison} brings together the boundary and residual SNR values obtained from AMPS, FLARE, and CRISP for each glitch, and Table~\ref{tab:timing_comparison} lists the time each method took to identify a glitch in an 8~s data stretch.

\begin{table}[htbp!]
\centering
\caption{Glitch identification time for AMPS, FLARE, and CRISP on an 8~s data segment.}
\label{tab:timing_comparison}
\begin{tabular}{|l|c|c|c|}
\hline
Glitch type & AMPS (s) & FLARE (s) & CRISP (s) \\
\hline
GW170817           & 0.0221 & 45.05 & 0.1020 \\
Blip                & 0.0261 & 11.88 & 0.1479 \\
Koi Fish            & 0.0364 & 15.12 & 0.0903 \\
Tomte               & 0.0139 & 18.91 & 0.0708 \\
Low frequency blip  & 0.0444 & 13.03 & 0.1027 \\
\hline
\end{tabular}
\end{table}

AMPS is the fastest of the three by a wide margin against FLARE, two to three orders of magnitude, identifying every glitch in well under a tenth of a second. Against CRISP the gap is smaller, roughly two to six times depending on the glitch, since both methods rely on a threshold scan rather than a fitted model. This speed comes at the cost of boundary precision: AMPS locates a glitch by an amplitude threshold and pads a fixed number of samples on each side, rather than resolving the actual extent of the excess power. CRISP still performs a coarse power scan of the full segment before refining a candidate, which accounts for its larger runtime relative to AMPS, but it remains a fraction of a second per glitch. FLARE is the slowest of the three, needing between about 12 and 45 seconds per glitch, since it fits a segmented shape model with particle swarm optimization across the full data stretch instead of localizing the glitch first. That cost buys FLARE the most direct link between its boundary and the underlying waveform shape, whereas CRISP's boundary comes from a spectrogram power criterion and AMPS's from padding around an amplitude threshold.

Boundary values in Table~\ref{tab:snr_comparison} follow this same logic. FLARE and CRISP agree closely on boundary duration for most glitches, since both resolve the excess-power region directly, while AMPS boundaries differ more, reflecting fixed-padding construction rather than a measured glitch duration. Residual SNR tracks this pattern only loosely: CRISP's recovered SNR stays within about 0.2 SNR of AMPS and FLARE for three of the four injected glitches, so a coarser, cheaper boundary is often enough to match a costlier, best-fit one at the subtraction stage.

The choice of injection position within the overlap window affects recovered SNR appreciably, and at some positions recovered SNR for Koi Fish exceeds the injected value under SHAPES and Combined. This is consistent with expected matched-filter SNR fluctuations under stationary, Gaussian noise~\cite{Gerosa_2024, wang2014ligosurf}. Appendix~\ref{app:koi_fish_reposition} examines this sensitivity directly and traces the excess to a chance noise contribution in the matched-filter output at those positions, rather than to a property of the subtraction method.

Combined recovers less SNR than a more aggressive fit would, since the goal is avoiding removal of structure that belongs to the signal rather than the glitch. Wavelet shrinkage is built on this trade-off directly: it accepts more bias in exchange for a large reduction in variance, shrinking small, noise-dominated coefficients while leaving coefficients carrying real signal structure intact~\cite{David_1995}. Fitting the residual more tightly to erase every trace of the glitch would push the estimate toward the noise-fitting end of that trade-off, risking removal of genuine signal power along with the glitch. BayesWave's glitch-subtraction procedure reflects the same caution operationally: when a glitch overlaps an astrophysical signal, it relies on multi-detector coherence to separate signal power from glitch power, since an aggressive single-detector fit cannot make that distinction on its own~\cite{Hourihane_2022}. A small residual left after subtraction is therefore consistent with established practice, not a shortcoming specific to Combined.

For context against other detection tools already in operational use, Omicron, the excess-power trigger generator used throughout LIGO-Virgo-KAGRA, produces triggers on the order of one second per data stretch on standard computing infrastructure and needs no training, matching AMPS and CRISP for ease of deployment though without AMPS's raw-amplitude simplicity~\cite{Robinet_2020}. Coherent WaveBurst, which additionally tests coherence across a detector network to separate signals from glitches, is considerably more demanding than any of the three methods here, since it searches multiple time-frequency resolutions and detector combinations rather than a single spectrogram~\cite{Klimenko_2016}. Machine-learning classifiers built on Omicron triggers, such as Gravity Spy, add classification within a few seconds per glitch once trained, but need a labelled training set and substantial upfront training time that AMPS, FLARE, and CRISP do not~\cite{Zevin_2017}. Against this backdrop, AMPS and CRISP sit closer to Omicron's speed and simplicity, while FLARE trades speed for a boundary tied directly to a fitted glitch model, a trade-off it shares conceptually with BayesWave~\cite{Cornish_2015}, though BayesWave's runtime of roughly one hour per glitch is far higher still~\cite{Dooney_2025}.

\begin{figure*}[htbp!]
    \centering
    \captionsetup{justification=centering}
    \begin{subfigure}[b]{0.23\textwidth}
        \includegraphics[width=\linewidth]{Figures/FLARE/GW170817/glitch_qfig.png}
        \caption{}
        \label{fig:subcomp_gw_raw}
    \end{subfigure}
    \hfill
    \begin{subfigure}[b]{0.23\textwidth}
        \includegraphics[width=\linewidth]{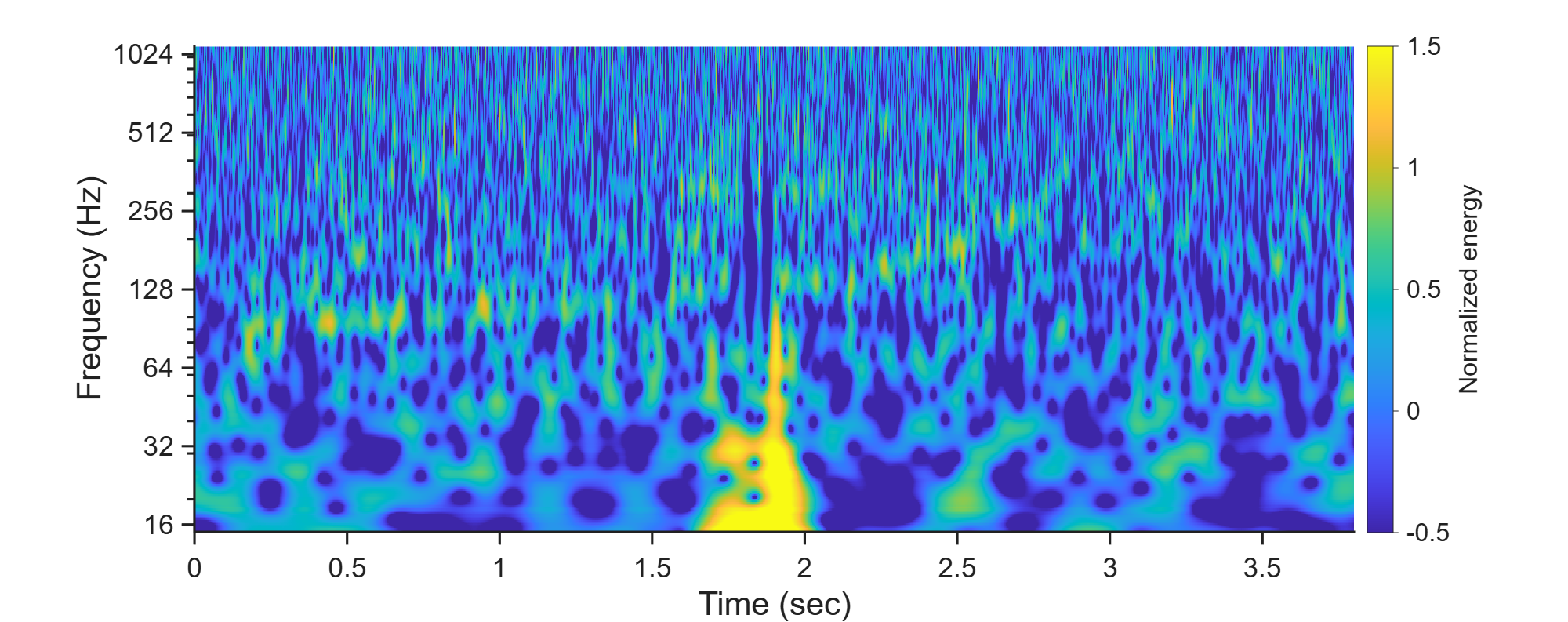}
        \caption{}
        \label{fig:subcomp_gw_shapes}
    \end{subfigure}
    \hfill
    \begin{subfigure}[b]{0.23\textwidth}
        \includegraphics[width=\linewidth]{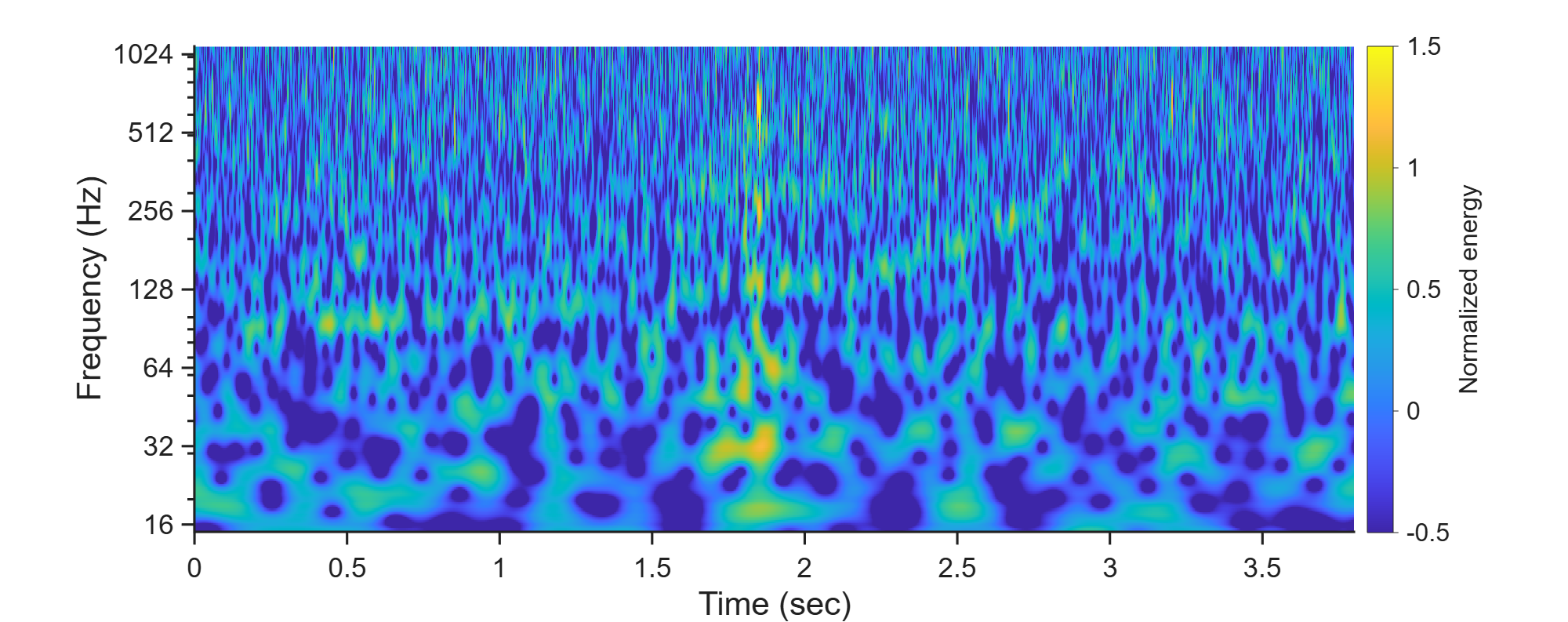}
        \caption{}
        \label{fig:subcomp_gw_ws}
    \end{subfigure}
    \hfill
    \begin{subfigure}[b]{0.23\textwidth}
        \includegraphics[width=\linewidth]{Figures/FLARE/GW170817/residual_qfig.png}
        \caption{}
        \label{fig:subcomp_gw_shapesws}
    \end{subfigure}

    \vspace{0.5em}
    \begin{subfigure}[b]{0.23\textwidth}
        \includegraphics[width=\linewidth]{Figures/FLARE/blip/glitch_qfig.png}
        \caption{}
        \label{fig:subcomp_blip_raw}
    \end{subfigure}
    \hfill
    \begin{subfigure}[b]{0.23\textwidth}
        \includegraphics[width=\linewidth]{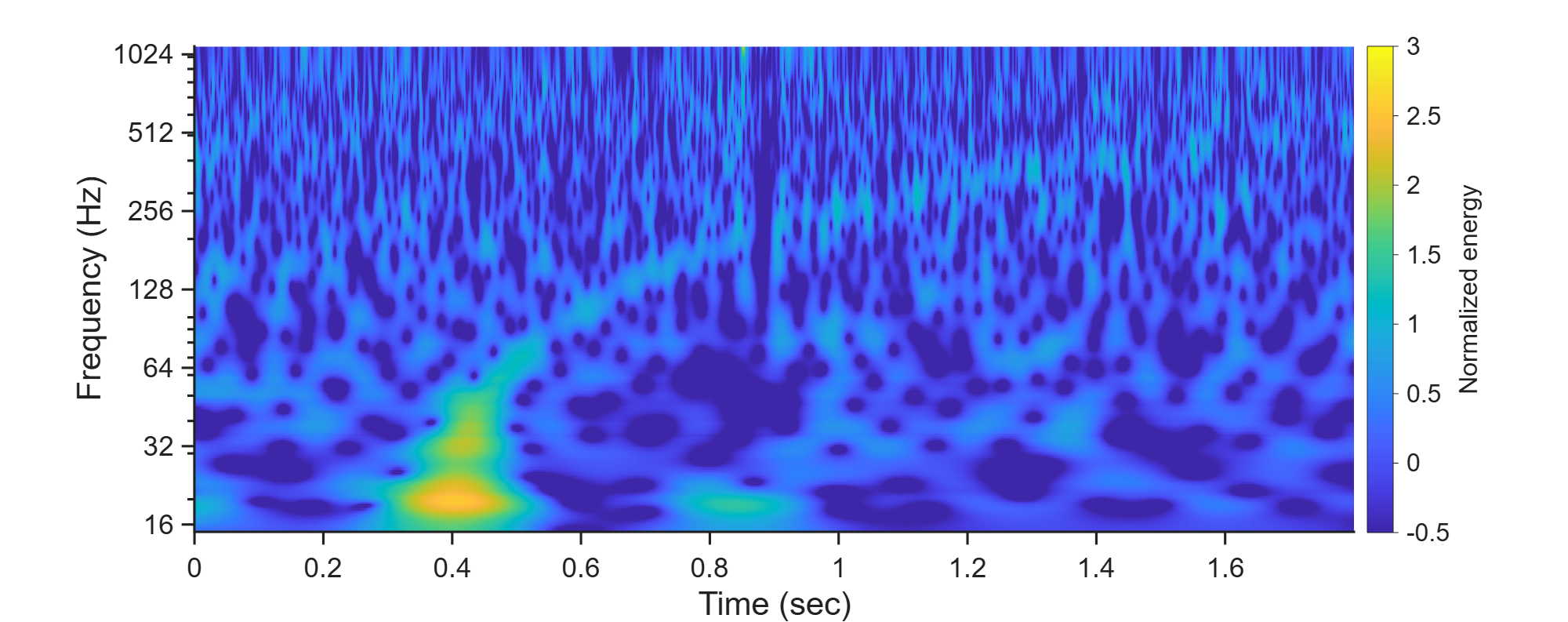}
        \caption{}
        \label{fig:subcomp_blip_shapes}
    \end{subfigure}
    \hfill
    \begin{subfigure}[b]{0.23\textwidth}
        \includegraphics[width=\linewidth]{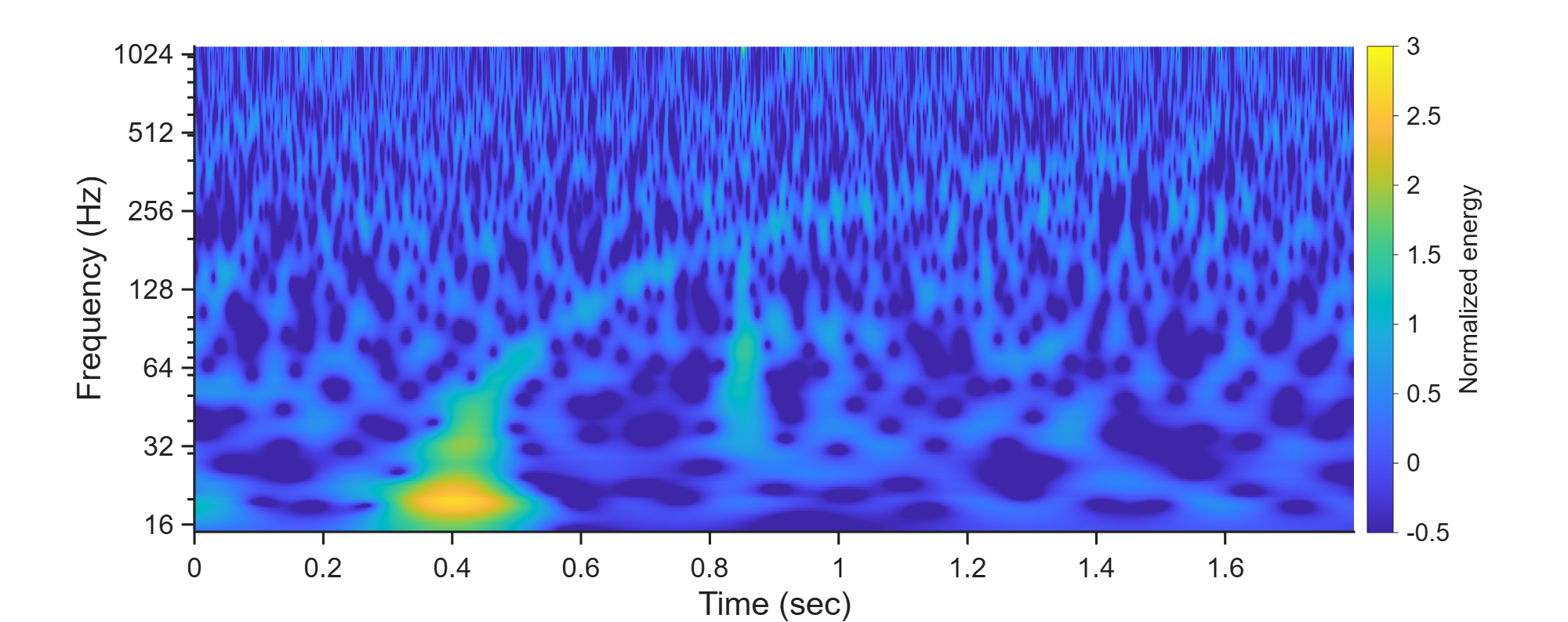}
        \caption{}
        \label{fig:subcomp_blip_ws}
    \end{subfigure}
    \hfill
    \begin{subfigure}[b]{0.23\textwidth}
        \includegraphics[width=\linewidth]{Figures/FLARE/blip/residual_qfig.png}
        \caption{}
        \label{fig:subcomp_blip_shapesws}
    \end{subfigure}

    \vspace{0.5em}
    \begin{subfigure}[b]{0.23\textwidth}
        \includegraphics[width=\linewidth]{Figures/FLARE/kf/glitch_qfig.png}
        \caption{}
        \label{fig:subcomp_kf_raw}
    \end{subfigure}
    \hfill
    \begin{subfigure}[b]{0.23\textwidth}
        \includegraphics[width=\linewidth]{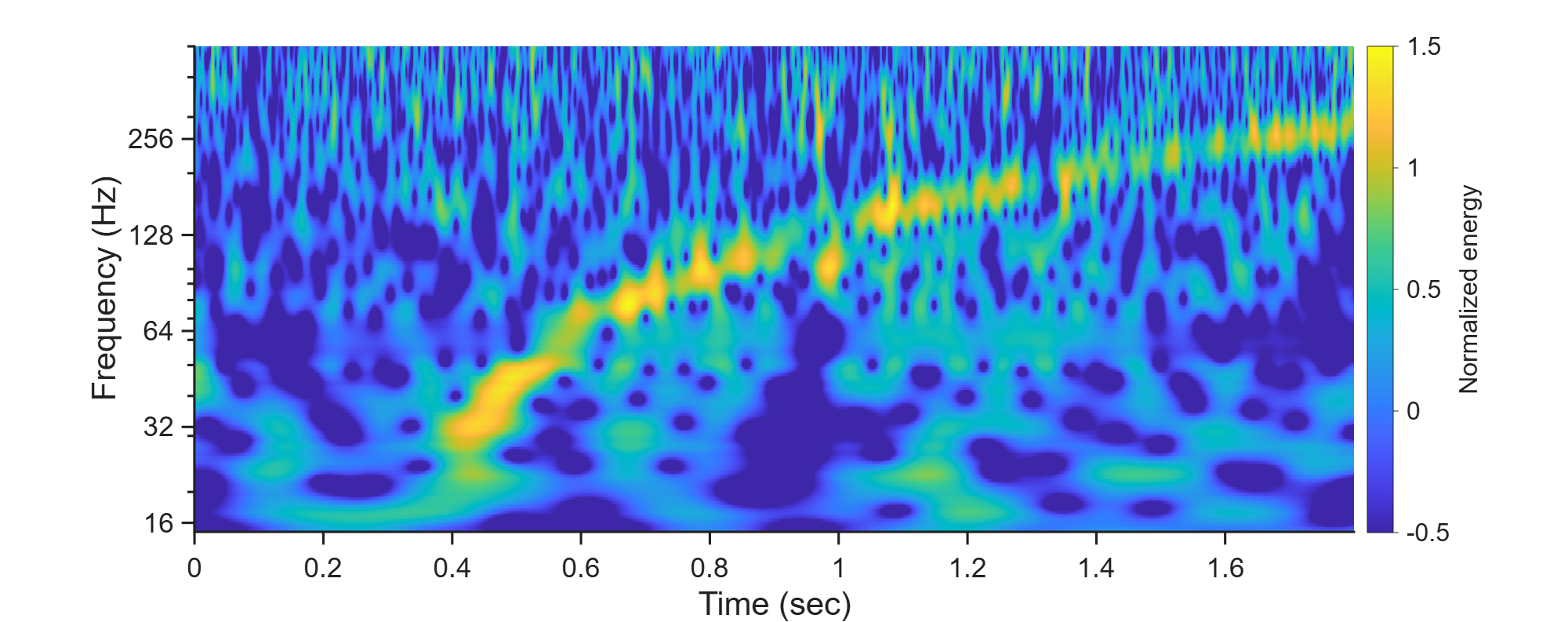}
        \caption{}
        \label{fig:subcomp_kf_shapes}
    \end{subfigure}
    \hfill
    \begin{subfigure}[b]{0.23\textwidth}
        \includegraphics[width=\linewidth]{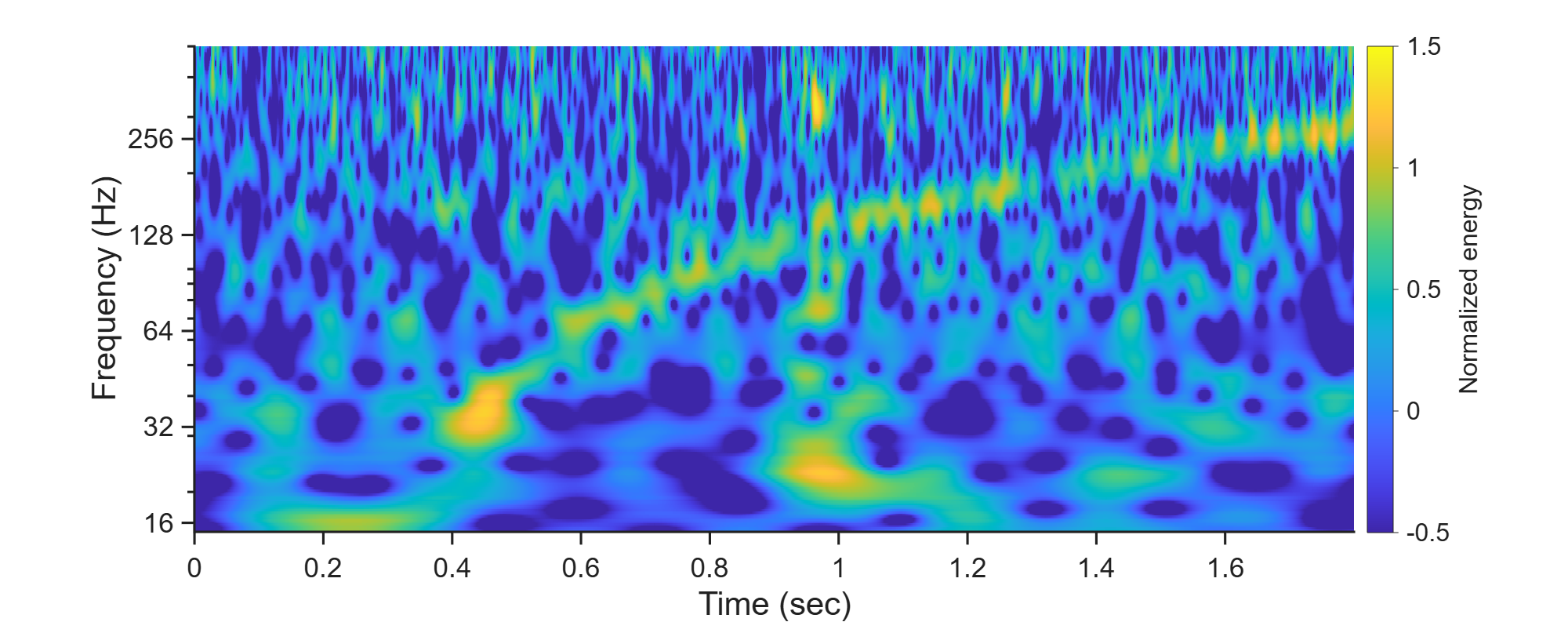}
        \caption{}
        \label{fig:subcomp_kf_ws}
    \end{subfigure}
    \hfill
    \begin{subfigure}[b]{0.23\textwidth}
        \includegraphics[width=\linewidth]{Figures/FLARE/kf/residual_qfig.png}
        \caption{}
        \label{fig:subcomp_kf_shapesws}
    \end{subfigure}

    \vspace{0.5em}
    \begin{subfigure}[b]{0.23\textwidth}
        \includegraphics[width=\linewidth]{Figures/FLARE/tomte/glitch_qfig.png}
        \caption{}
        \label{fig:subcomp_tomte_raw}
    \end{subfigure}
    \hfill
    \begin{subfigure}[b]{0.23\textwidth}
        \includegraphics[width=\linewidth]{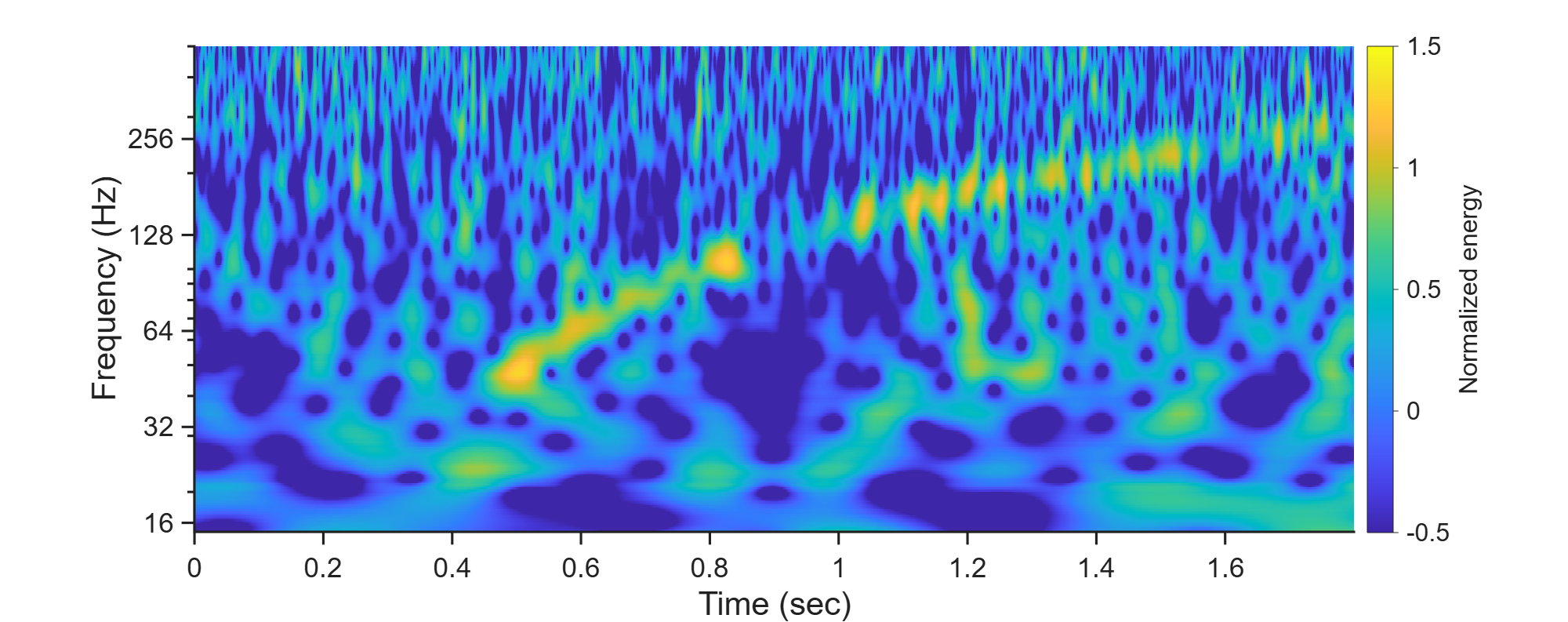}
        \caption{}
        \label{fig:subcomp_tomte_shapes}
    \end{subfigure}
    \hfill
    \begin{subfigure}[b]{0.23\textwidth}
        \includegraphics[width=\linewidth]{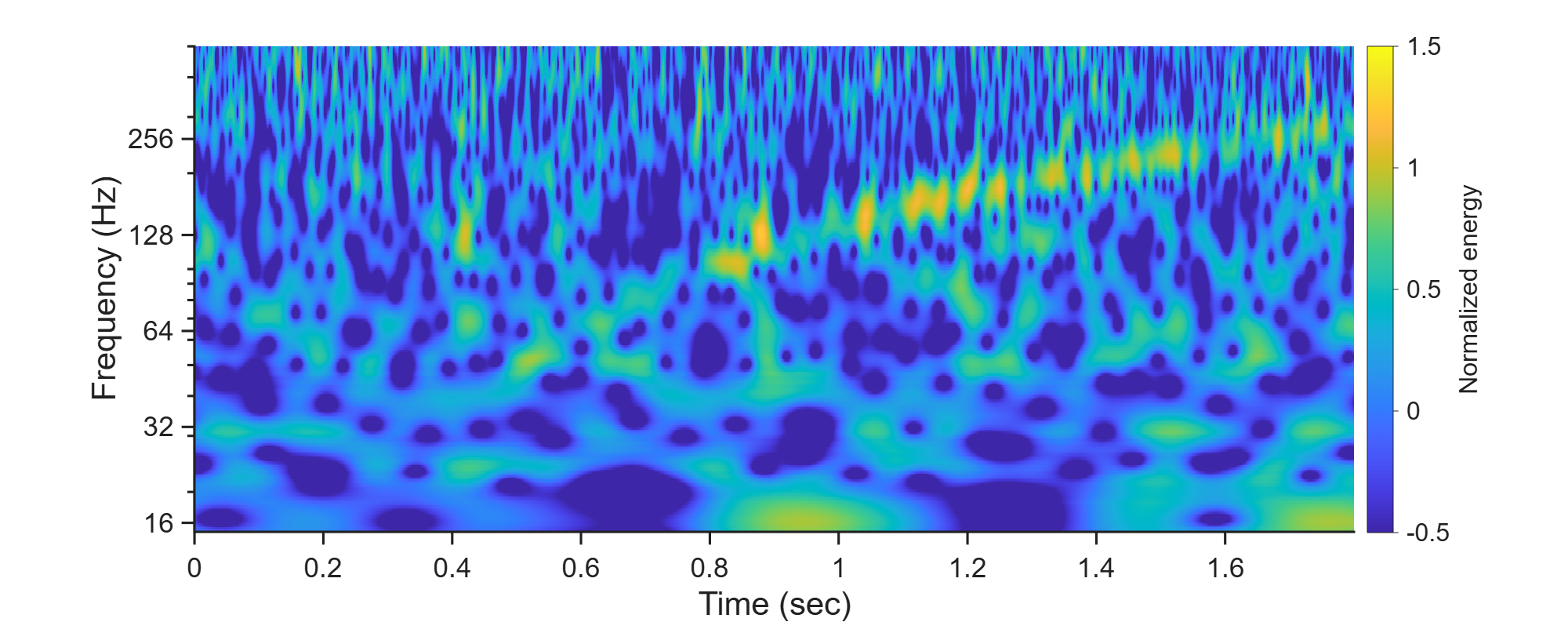}
        \caption{}
        \label{fig:subcomp_tomte_ws}
    \end{subfigure}
    \hfill
    \begin{subfigure}[b]{0.23\textwidth}
        \includegraphics[width=\linewidth]{Figures/FLARE/tomte/residual_qfig.png}
        \caption{}
        \label{fig:subcomp_tomte_shapesws}
    \end{subfigure}

    \vspace{0.5em}
    \begin{subfigure}[b]{0.23\textwidth}
        \includegraphics[width=\linewidth]{Figures/FLARE/lfb/glitch_qfig.png}
        \caption{}
        \label{fig:subcomp_lfb_raw}
    \end{subfigure}
    \hfill
    \begin{subfigure}[b]{0.23\textwidth}
        \includegraphics[width=\linewidth]{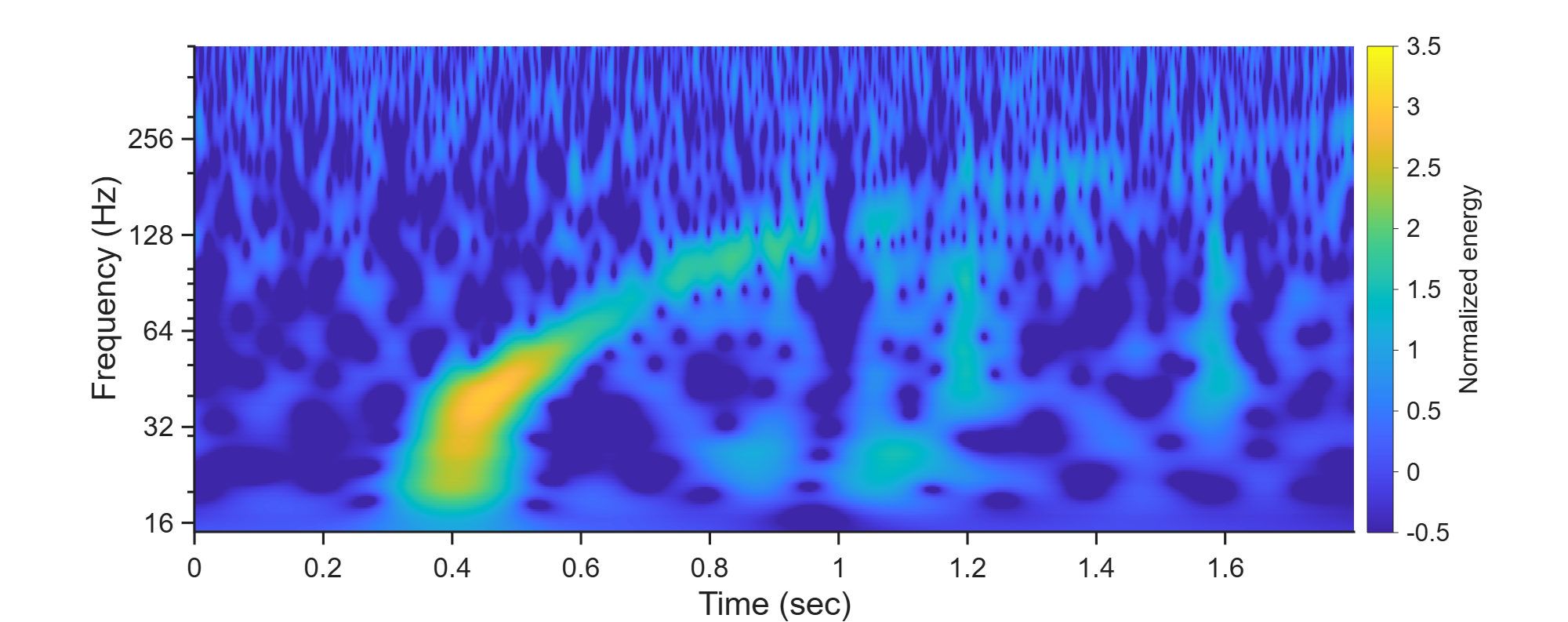}
        \caption{}
        \label{fig:subcomp_lfb_shapes}
    \end{subfigure}
    \hfill
    \begin{subfigure}[b]{0.23\textwidth}
        \includegraphics[width=\linewidth]{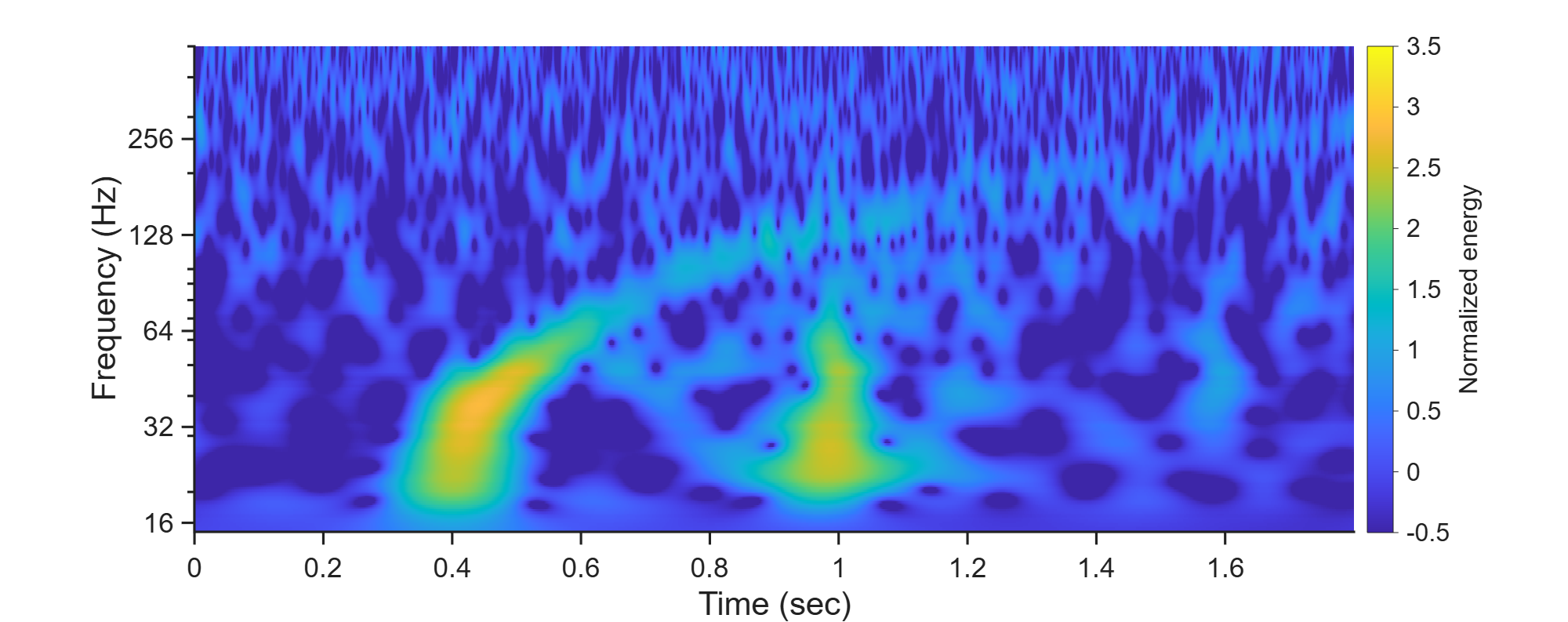}
        \caption{}
        \label{fig:subcomp_lfb_ws}
    \end{subfigure}
    \hfill
    \begin{subfigure}[b]{0.23\textwidth}
        \includegraphics[width=\linewidth]{Figures/FLARE/lfb/residual_qfig.png}
        \caption{}
        \label{fig:subcomp_lfb_shapesws}
    \end{subfigure}

    \caption{Q-transform comparison of subtraction technique across five glitch instances: GW170817, Blip, Koi Fish, Tomte, and low frequency blip, from top to bottom. Each row shows, from left to right, the raw whitened data, the SHAPES residual, the Waveshrink (WS) residual, and the SHAPES+WS residual, all on the same time window. The segment boundary used to build every residual in a row comes from FLARE, held fixed across all three subtraction techniques so that differences between panels reflect the subtraction technique rather than the identification method. Each row uses its own colour scale, chosen to match the power range of that glitch, since a single scale shared across all five rows would suppress the visibility of the lower-power low frequency blip relative to louder cases such as GW170817 and Blip. GW170817 carries no injected chirp and is included for qualitative comparison of residual structure only; the signal-to-noise recovery reported in Table~\ref{tab:snr_comparison} applies to the four real glitch instances beneath it, into which a chirp signal of known parameters was injected to test subtraction performance.}
    \label{fig:subtraction_comparison}
\end{figure*}

\subsection{Comparison of subtraction techniques}
\label{sec:comparison_subtraction}

Figure~\ref{fig:subtraction_comparison} and Table~\ref{tab:snr_comparison} together point to one consistent story. SHAPES removes most of the glitch power on its own, seen in the drop from panel (a) to (b) for GW170817 and from (i) to (j) for Koi Fish, though a faint trace of structure usually remains. WS alone leaves a residual that looks broadly similar to the other two techniques by eye, as in panel (k) for Koi Fish, but the resemblance is only visual. WS alone recovers just 19.48 out of an injected SNR of 30 for Koi Fish, against 29.26–29.81 for SHAPES and 29.48–29.80 for Combined across the three identification methods, a gap the Q-transform alone does not reveal. A residual that looks comparable to the eye can still carry far less of the injected signal, which is why the recovered SNR in Table~\ref{tab:snr_comparison}, not the Q-transform, is the reliable measure of what each technique actually preserves.

Combined lands closest to SHAPES throughout. Table~\ref{tab:snr_comparison} shows Combined matching or beating SHAPES on three of the four injected glitches, Blip, Tomte, and the low frequency blip, and sitting within a narrow margin on Koi Fish, while clearing WS alone by a wide margin everywhere. 

The gap between Combined and SHAPES bears a direct comparison against the noise floor itself. Since $\hat{s}(t)$ in Eq.~\eqref{eq:matched_filter} is unit-norm and the background is unit-variance under the noise model of Eq.~\eqref{eq:noise-model}, the matched-filter statistic carries an intrinsic noise-floor spread of order one SNR unit regardless of subtraction technique. Across the twelve identification-subtraction pairings in Table~\ref{tab:snr_comparison}, the Combined-minus-SHAPES gap ranges from $-0.01$ to $1.52$, with a mean of $0.74$. This spread sits close to the expected noise-floor scale, so part of the apparent advantage of Combined over SHAPES alone reflects the same noise-realization variance already invoked in Appendix~\ref{app:koi_fish_reposition} for the Koi Fish repositioning result, rather than a uniformly large effect of wavelet shrinkage on top of the spline fit. 

Combined appears to apply WS's cleanup only after SHAPES has already isolated most of the glitch, leaving too little residual glitch structure for the shrinkage step to cut into the injected signal the way WS does when applied on its own.

GW170817 shows the same residuals qualitatively. Panel (b) still carries a visible streak after SHAPES; panels (c) and (d) both reduce it further, without the flattening that would signal an overly aggressive fit. No chirp was injected here, so this row is read for residual shape alone, and it does not on its own distinguish WS from the other two, reinforcing that the Q-transform is not where the difference between techniques actually shows up; the SNR values from the four injected glitches are.

Estimation time adds a separate axis to this comparison. WS runs on the full 8~s segment regardless of boundary, so its cost is fixed and small, between 0.0034~s for Tomte and 0.0076~s for Koi Fish, several orders of magnitude below SHAPES in every case. SHAPES itself takes tens of seconds per glitch, and its cost tracks the boundary it is given rather than the identification method's own runtime: averaged over the four injected glitches, SHAPES needs about 52.8~s under AMPS boundaries, 49.2~s under CRISP, and only 26.5~s under FLARE, even though FLARE's own identification stage is the slowest of the three by a wide margin. GW170817 costs SHAPES the most fitting time under every method, 113.95~s under AMPS, 70.34~s under FLARE, and 51.37~s under CRISP, consistent with its wider, more complex boundary. Combined's total cost is SHAPES's time plus a WS pass adding a few thousandths of a second, so the small SNR gain Combined holds over SHAPES alone comes at negligible extra cost.

\section{Conclusion}
\label{sec:conclusion}

We presented three glitch boundary identification methods, AMPS, FLARE, and CRISP, and tested each against three subtraction techniques, SHAPES, wavelet shrinkage, and their combination, across five glitch instances drawn from three observing runs. AMPS relies only on a robust amplitude threshold applied to the whitened strain. FLARE builds its boundary from the best fitness of a spline model fit segment by segment. CRISP localizes a glitch from spectrogram power alone and uses AMPS only to confirm a candidate inside a boundary it has already set.

All three recover an injected chirp signal with a residual SNR close to the injected value for the four broadband glitch cases, and CRISP gives the most consistent boundary width across glitches of markedly different duration, settling near 0.11~s for four of the five instances tested. AMPS remains the fastest identification method by a wide margin, completing in well under a tenth of a second per glitch against tens of seconds for FLARE, while CRISP sits a small factor above AMPS without approaching FLARE's cost.

The subtraction comparison across the three identification methods points to the same ranking. SHAPES combined with wavelet shrinkage gives the highest recovered SNR for Blip, Tomte, and Koi Fish under all three boundary methods, recovering 95 to 97 percent of the injected value on average, and wavelet shrinkage alone trails both SHAPES and the combined technique by a wide margin in every case tested, recovering only 64 percent. The one departure from this pattern is the low frequency blip, where SHAPES alone edges out the combined technique under AMPS, though the two stay close. This narrow departure does not track the broader recovery pattern, in which the low frequency blip performs among the best of the four glitches rather than the worst, so it is best read as a small, boundary-specific fluctuation rather than a consequence of the glitch's own frequency content.

Injection position within the overlap window affects recovered SNR, and Appendix~\ref{app:koi_fish_reposition} shows that Koi Fish's recovered SNR under SHAPES and Combined can exceed the injected value at some positions, a chance effect of the noise realization rather than a property of subtraction~\cite{Gerosa_2024, wang2014ligosurf}. The present study is limited to five glitch instances. Other glitch morphologies, including scattered light, are not yet exercised. The threshold values used throughout this work, including the amplitude threshold in AMPS, the fitness threshold in FLARE, and the two power-mask fractions in CRISP, were each set from a single glitch instance or a small handful of them. These values should be solidified against a much larger and more diverse sample before they can be treated as defaults rather than as instance-specific settings. This scale of evaluation is consistent with prior model-based glitch-subtraction studies~\cite{Merritt_2021,Mohanty_2023, Ghonge_2024}, though unlike SHAPES's single parameter, whose exact value was found to matter little, the threshold values here were tuned directly against the instances used for evaluation, a stronger dependency worth flagging separately. Building that sample by hand is impractical, since instances of the same morphological class carry substantially different glitch SNR and background conditions even within one class. A machine-learning-based parameter search across a larger, automatically labeled sample is a natural next step for fixing threshold values that currently rest on a small number of hand-checked instances.

Future work should extend the comparison across the full range of GravitySpy glitch classes, with particular attention to scattered light and other narrowband or slowly varying morphologies, since the amplitude and periodicity criteria built into CRISP for this purpose have not yet been tested against labelled scattering data. One planned extension is a second, lower amplitude threshold applied to a filtered version of the bounded segment, aimed at separating scattering glitches from the transient classes studied here, since scattering typically shows a weaker and more spread-out amplitude signature than a genuine transient. This criterion is not yet validated and remains untested in the current results. Recent work on adaptive, template-free subtraction of scattering arches suggests one path forward for this extension \cite{Longo_2026}. The interaction between glitch frequency content and the injected signal, which affected recovery differently across the four glitch classes in this work, also warrants a systematic study across a range of injected signal frequencies rather than the single chirp profile used here. Finally, this work tests only one real overlapping signal, GW170817, and relies on a synthetic linear chirp for the remaining four glitches. Extending the comparison to a larger sample of real compact-binary signals, spanning a range of masses and overlap geometries, would test whether the present conclusions hold beyond the single real case and the simplified injection used here as a stand-in. Whether the position-dependence identified in Appendix~\ref{app:koi_fish_reposition} propagates to biases in estimated source parameters for a real overlapping signal, rather than only to recovered SNR, is a separate question this work does not address.

\begin{acknowledgments}
 M A T C acknowledges support from the Presidential Graduate Research Award at the University of Texas Rio Grande Valley, and the Graduate Teaching Assistantship at the Rensselaer Polytechnic Institute.
\end{acknowledgments}


\appendix

\section{Matched-filter check for the GW170817 instance}
\label{app:gw170817_mf}

Section~\ref{sec:demo_data} treats the GW170817 instance as a qualitative check rather than a quantitative recovery test, since no signal wasinjected for this case. As a supplementary check, a matched-filter test was carried out on the SHAPES, wavelet-shrinkage, and combined residuals obtained from the CRISP-bounded segment. The template is a Newtonian, restricted-post-Newtonian chirp with component masses of 1.46 and 1.30 solar masses, close to the values reported for GW170817. A sliding matched filter was applied across each residual, and the peak of the resulting signal-to-noise ratio time series was recorded, following the same convention used elsewhere in this work for the injected-chirp cases.

The data segment used for this check spans 8~s, shorter than the segment needed to hold a chirp starting at the conventional 40~Hz low-frequency cutoff. The template's starting frequency was therefore raised to approximately 63~Hz so that the full chirp fits within the available segment. This choice trades template duration for consistency with the segment length already used for boundary identification in this work, and the resulting statistic should be read as an internal, self-consistent check across the three residuals rather than a value comparable to matched-filter statistics computed over the full inspiral band.

\begin{figure}[htbp!]
    \centering
    \includegraphics[width=\columnwidth]{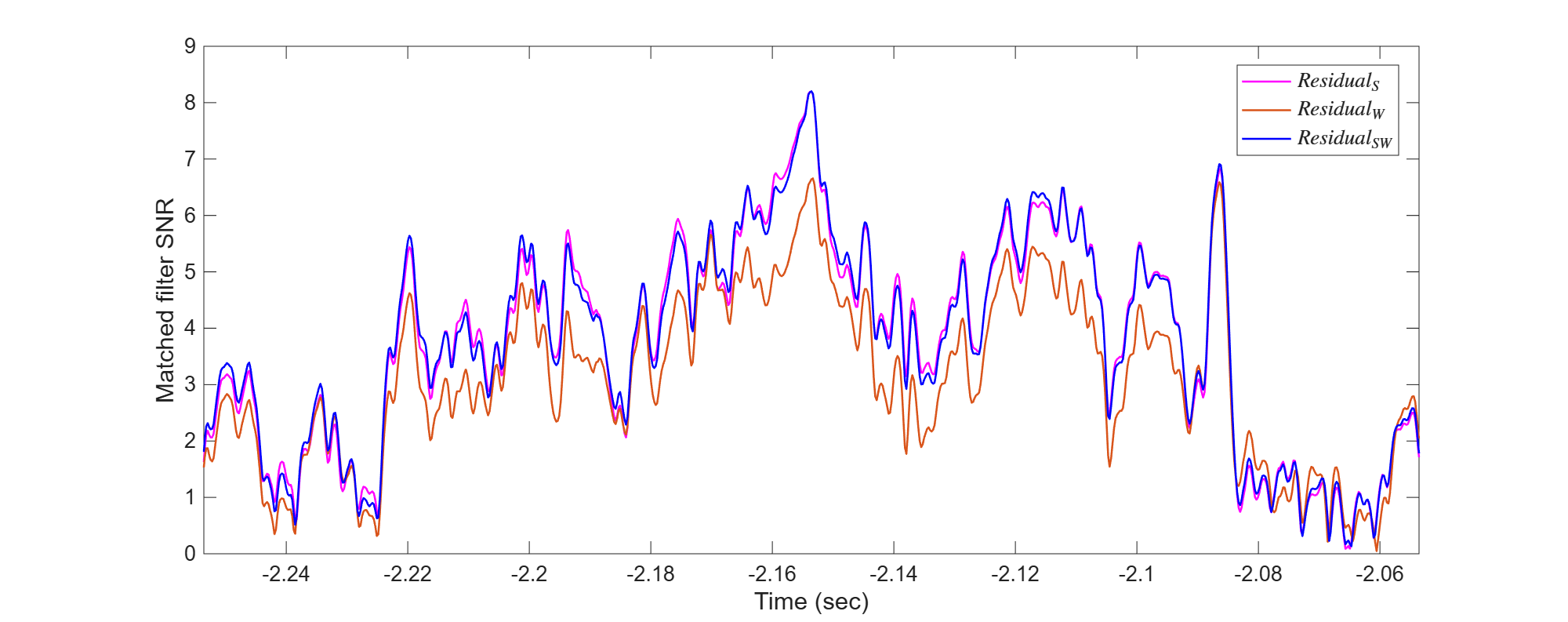}
    \caption{Envelope of the matched-filter signal-to-noise ratio for the CRISP-bounded GW170817 residual, shown for SHAPES alone (magenta), wavelet shrinkage alone (orange), and the combined technique (blue). The horizontal axis is time in seconds relative to the matched-filter lag, and the vertical axis is the matched-filter signal-to-noise ratio. Wavelet shrinkage alone recovers visibly less signal than the other two techniques across the shown window, while SHAPES alone and the combined technique remain close to one another throughout.}
    \label{fig:gw170817_mf}
\end{figure}

Figure~\ref{fig:gw170817_mf} shows the envelope of the matched-filter SNR time series for the three residuals, restricted to a narrow window around the shared peak. All three residuals show a clear, sharply peaked maximum at the same arrival time, confirming that the real signal survives subtraction under all three techniques, though wavelet shrinkage alone recovers less of it. The peak SNR is 8.20 for SHAPES alone, 6.66 for wavelet shrinkage alone, and 8.20 for the combined technique. SHAPES alone and the combined technique track one another closely across the entire window, consistent with their near-identical peak values, while wavelet shrinkage alone sits visibly below both throughout, not only at the peak. This pattern matches the identification-and-subtraction comparison in the main text, where wavelet shrinkage alone consistently trails SHAPES and SHAPES with wavelet shrinkage combined.

\section{Sensitivity of the boundary to the threshold multiplier}
\label{app:boundary_sensitivity}

The boundary used by CRISP and FLARE in Section~\ref{sec:results} is set from a power threshold applied to the spectrogram columns of the candidate segment, scaled by a fixed multiplier before it is compared against the envelope described in Section~\ref{sec:glitch_id_methods}. In Section~\ref{sec:results} this multiplier is fixed at 3.0 for every glitch instance, which keeps the comparison across the five cases consistent, but leaves open how much residual signal power depends on that particular choice. We examine this question for GW170817, the one instance in our sample where the underlying glitch carries a low GravitySpy machine-learning confidence despite its high reported loudness, and where the default multiplier leaves a visible patch of unsubtracted power at low frequency.

\begin{figure}[ht]
\centering

\begin{subfigure}{0.55\linewidth}
    \centering
    \includegraphics[width=\linewidth]{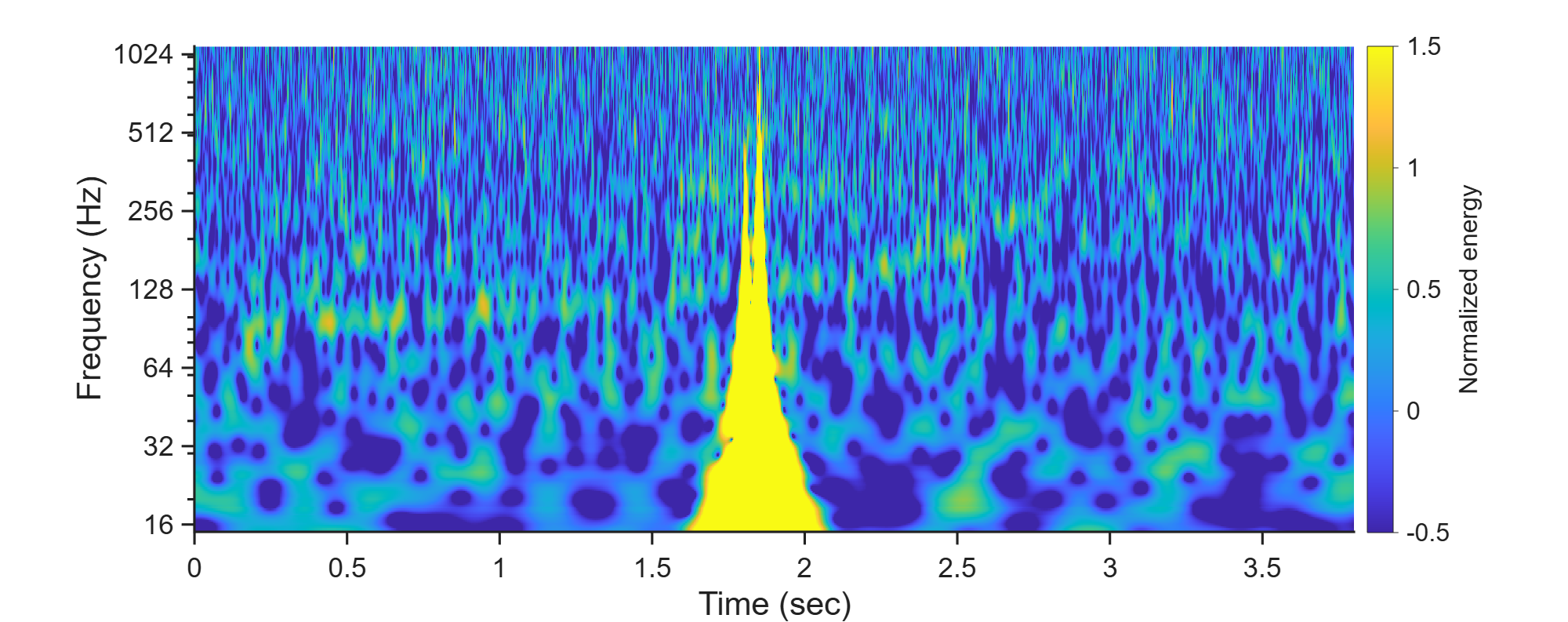}
    \caption{}
    \label{fig:boundary_sensitivity_glitch}
\end{subfigure}

\vspace{0.3cm}

\begin{subfigure}{0.45\linewidth}
    \centering
    \includegraphics[width=\linewidth]{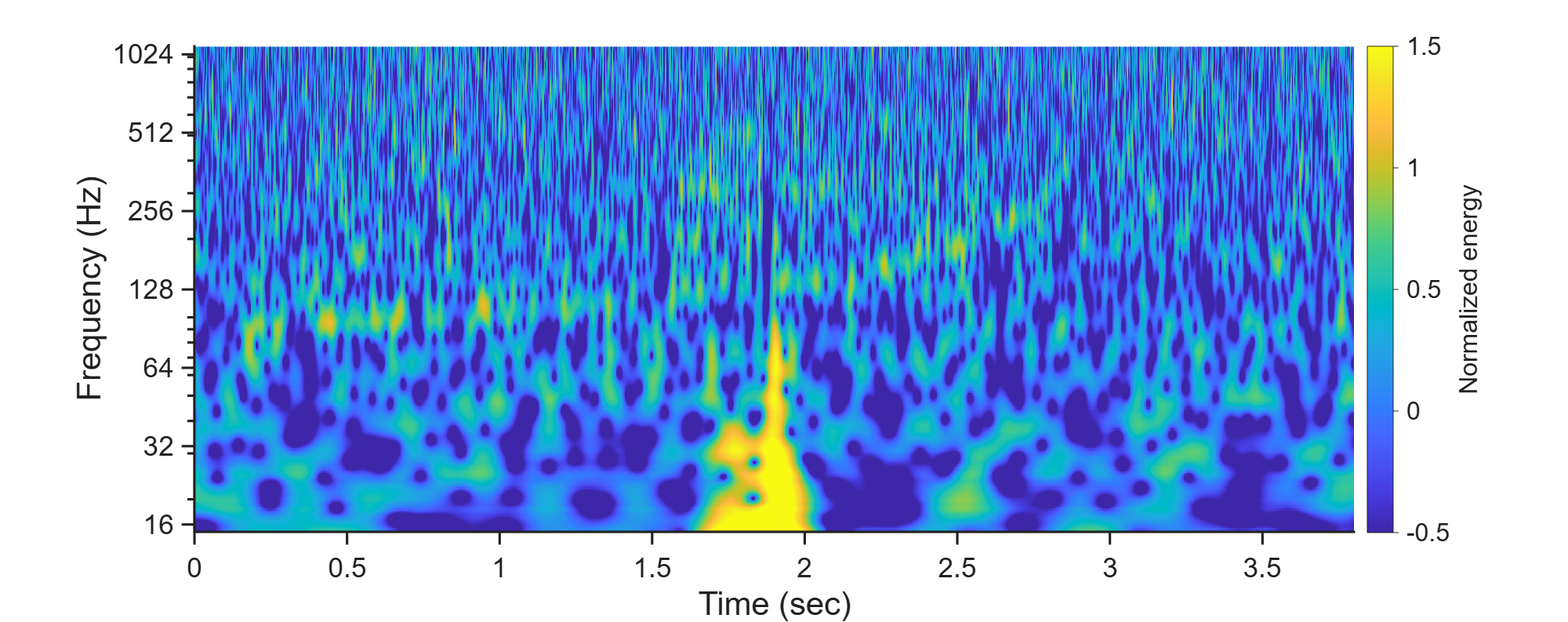}
    \caption{}
\end{subfigure}
\hfill
\begin{subfigure}{0.45\linewidth}
    \centering
    \includegraphics[width=\linewidth]{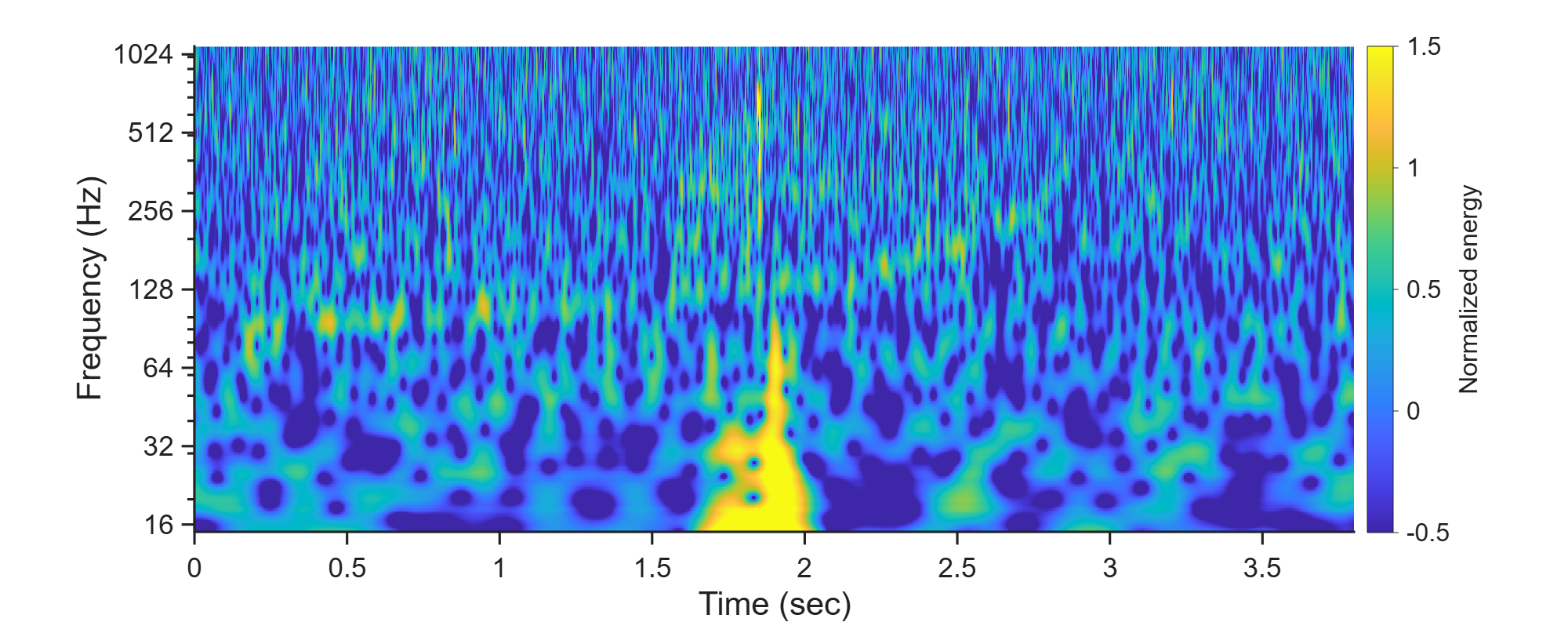}
    \caption{}
\end{subfigure}

\begin{subfigure}{0.45\linewidth}
    \centering
    \includegraphics[width=\linewidth]{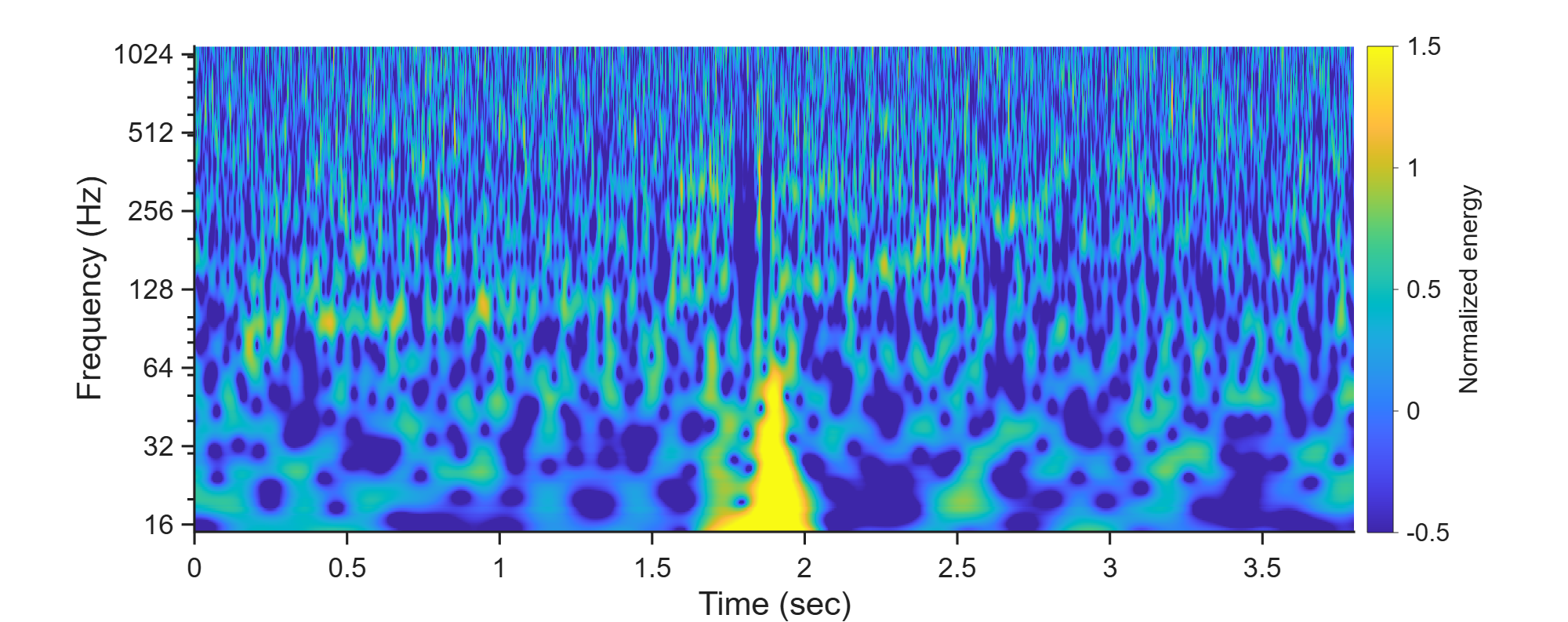}
    \caption{}
\end{subfigure}
\hfill
\begin{subfigure}{0.45\linewidth}
    \centering
    \includegraphics[width=\linewidth]{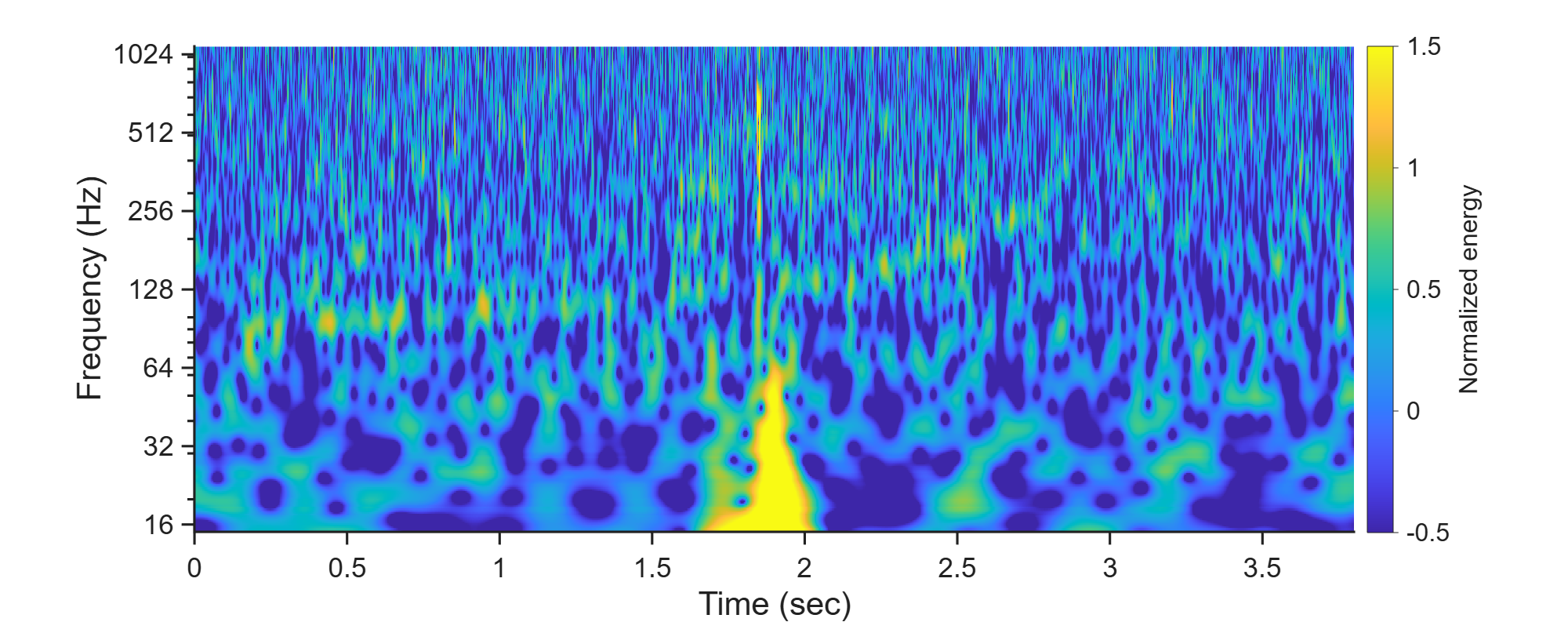}
    \caption{}
\end{subfigure}

\begin{subfigure}{0.45\linewidth}
    \centering
    \includegraphics[width=\linewidth]{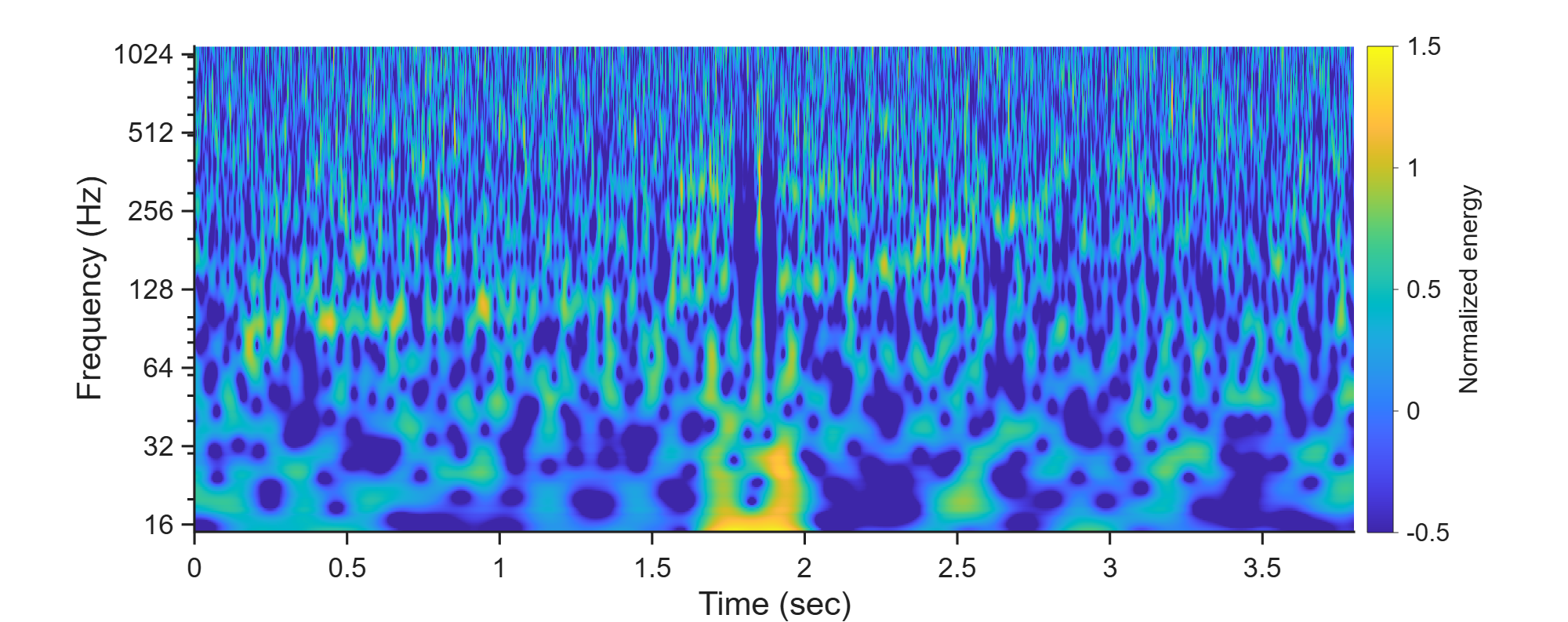}
    \caption{}
\end{subfigure}
\hfill
\begin{subfigure}{0.45\linewidth}
    \centering
    \includegraphics[width=\linewidth]{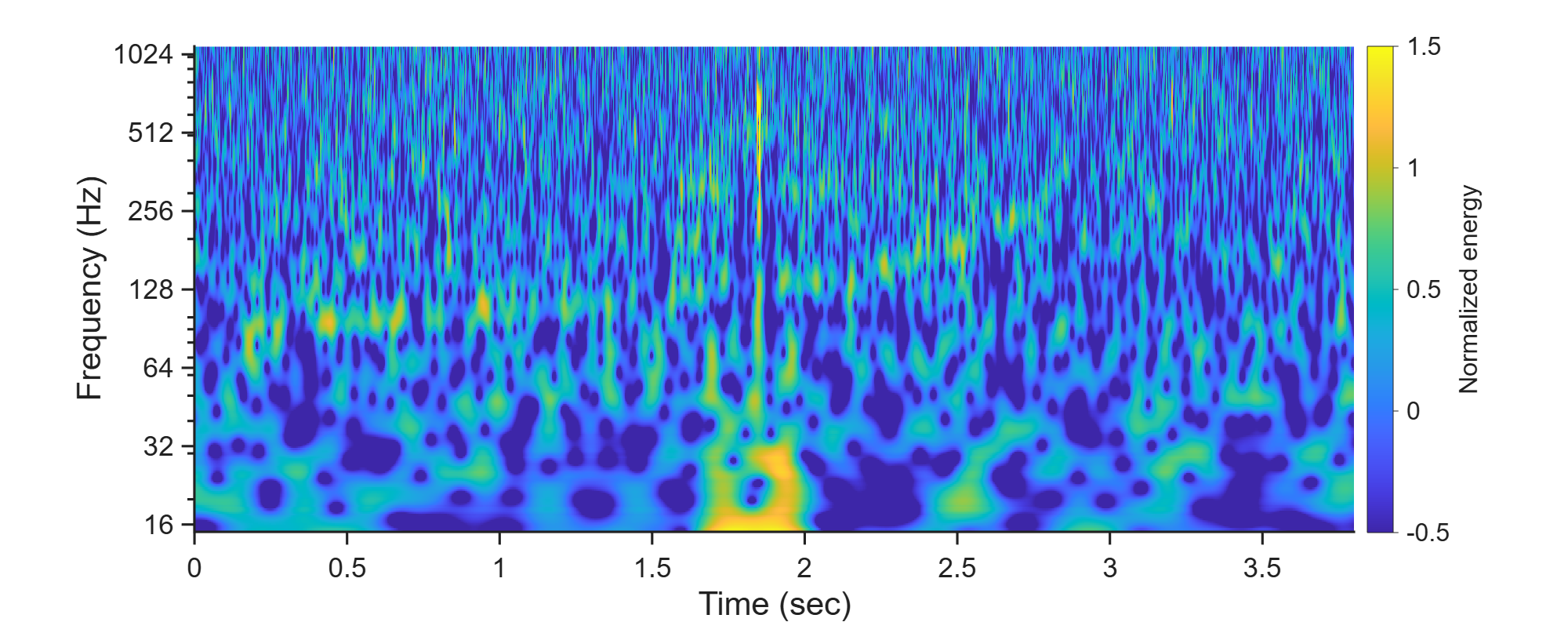}
    \caption{}
\end{subfigure}

\begin{subfigure}{0.45\linewidth}
    \centering
    \includegraphics[width=\linewidth]{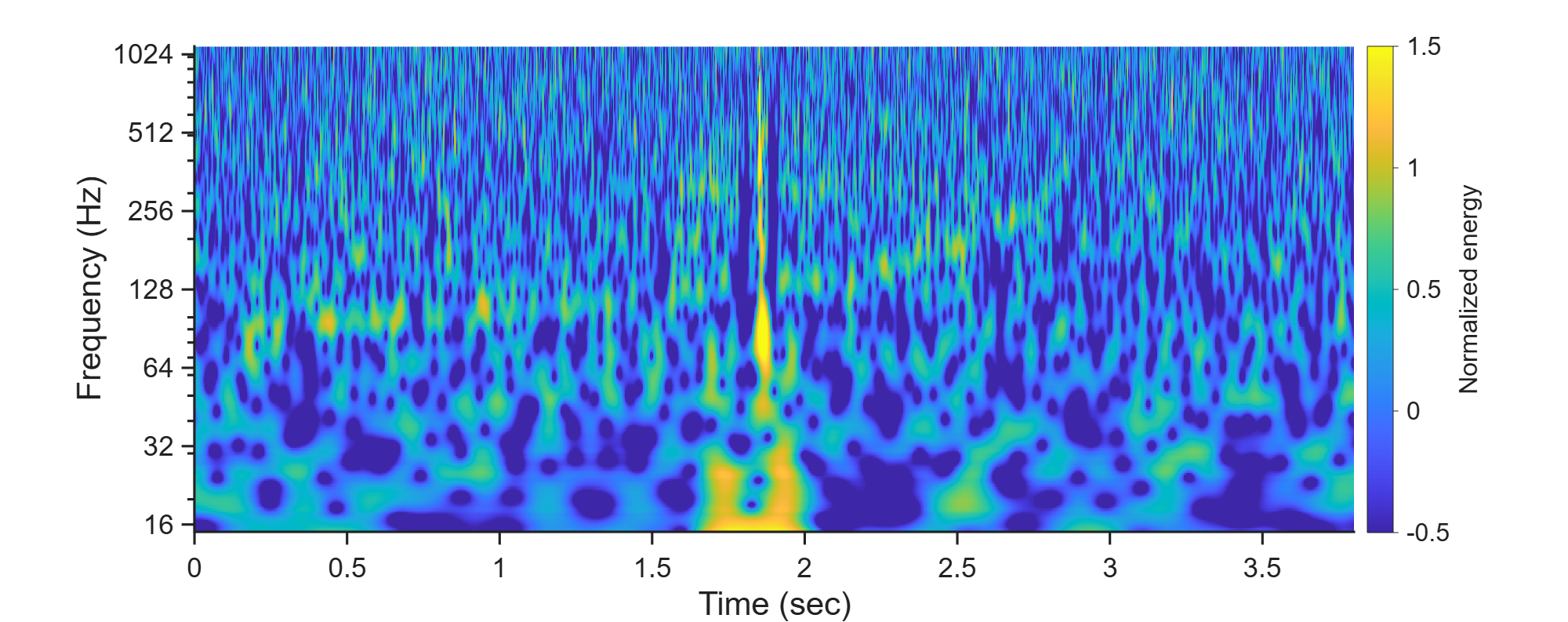}
    \caption{}
\end{subfigure}
\hfill
\begin{subfigure}{0.45\linewidth}
    \centering
    \includegraphics[width=\linewidth]{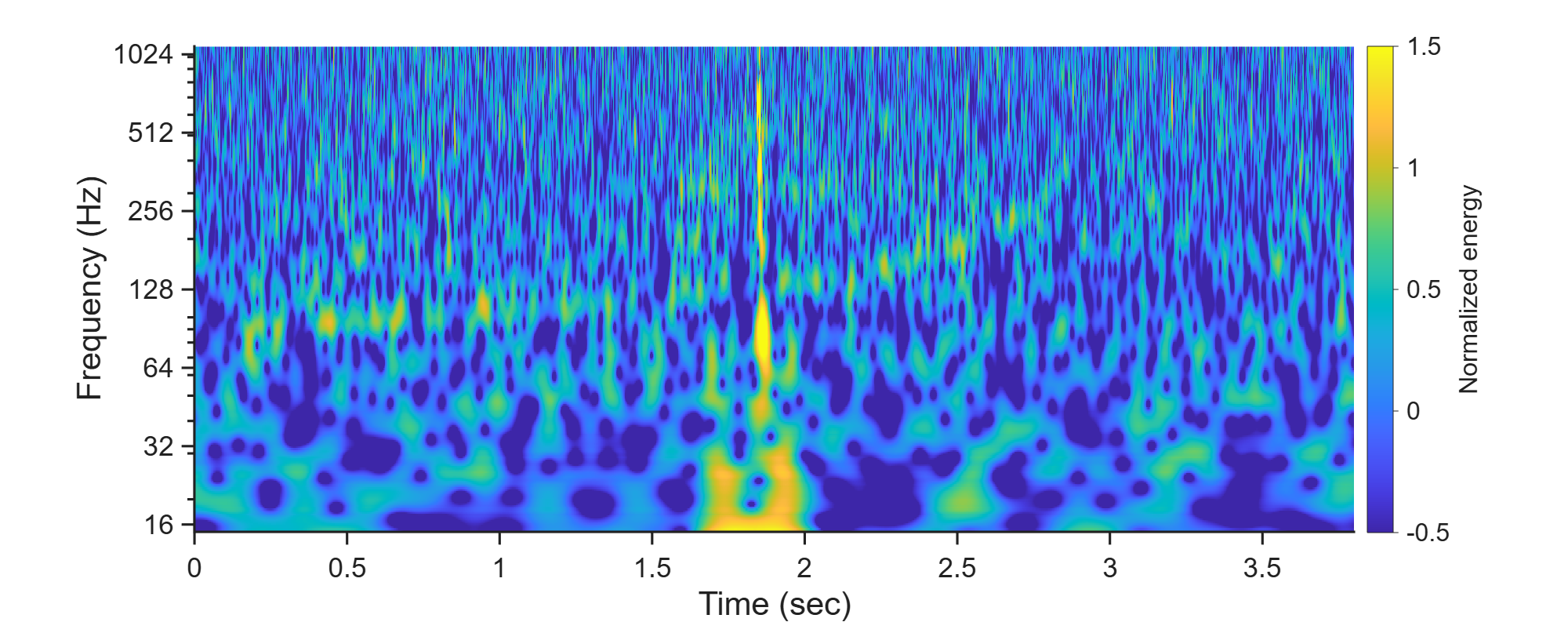}
    \caption{}
\end{subfigure}

\caption{Effect of the boundary threshold multiplier on subtraction of the glitch overlapping GW170817. Top: Q-transform of the glitch before subtraction. Below, each row corresponds to one multiplier, from top to bottom: 3.0 (default), 2.5, 2.0, and 1.5. Left column: Q-transform of the SHAPES residual at the corresponding multiplier. Right column: Q-transform of the SHAPES+WS residual at the corresponding multiplier. In every panel, the horizontal axis is time in seconds and the vertical axis is frequency in Hz, plotted on a logarithmic scale. Color indicates normalized energy, on a logarithmic scale, and the colorbar range is shared across all panels, including the reference panel at top. The low-frequency patch below 40~Hz shrinks as the multiplier decreases from 3.0 to 1.5, while the broadband high-frequency content of the glitch is removed at every multiplier shown.}
\label{fig:boundary_sensitivity}
\end{figure}

Lowering the multiplier from 3.0 to 1.5 widens the boundary from 448 to 640 samples, or from 0.1094~s to 0.1562~s (Table~\ref{tab:boundary_sensitivity}), because a smaller multiplier lowers the power level a spectrogram column must clear to be kept inside the boundary, so more of the glitch's lower-power edges are retained. This widening tracks directly with subtraction quality: Figure~\ref{fig:boundary_sensitivity} shows that the low-frequency patch left below 40~Hz after subtraction at the default multiplier shrinks steadily as the multiplier is lowered toward 1.5, for both SHAPES and SHAPES+WS, while the broadband content of the glitch above roughly 60~Hz is already removed cleanly at every multiplier tested. We did not find a multiplier within this range that removes the low-frequency patch entirely.

\begin{table}[ht]
\centering
\small
\caption{Boundary obtained for the glitch overlapping GW170817 as the threshold multiplier is varied. A smaller multiplier lowers the power level required to keep a spectrogram column inside the boundary, which widens the resulting segment.}
\label{tab:boundary_sensitivity}
\begin{tabular}{|c|c|c|c|}
\hline\hline
Multiplier$(m)$ & Threshold & Samples & Boundary (s) \\
\hline
3.0 & 0.60 & 448 & 0.1094 \\
2.5 & 0.50 & 512 & 0.1250 \\
2.0 & 0.40 & 576 & 0.1406 \\
1.5 & 0.30 & 640 & 0.1562 \\
\hline\hline
\end{tabular}
\end{table}

A wider boundary is not free of cost. It also lets more of the surrounding background into the fitting segment, which was the original reason for keeping the boundary tight, so the trend in Table~\ref{tab:boundary_sensitivity} and Figure~\ref{fig:boundary_sensitivity} should not be read as evidence that a smaller multiplier is a better general-purpose choice, only that this particular glitch's low-frequency residual is sensitive to it.

We attribute the difficulty of this case to the ambiguous morphological status of the GW170817 glitch itself, rather than to a shortcoming specific to CRISP or FLARE. GravitySpy reports high loudness for this glitch but low classifier confidence, meaning the automated pipeline that labels glitch morphology is itself uncertain what class this glitch belongs to. A boundary-identification method tuned and tested against clearly labeled Blip, Tomte, and KoiFish instances has, by construction, less reason to generalize cleanly to a glitch that resists confident classification.

This result reinforces a point already raised in the conclusion. The threshold values used throughout this work, including the multiplier examined here, were set from a small number of instances and should be solidified against a larger and more varied sample before being treated as defaults. We leave a systematic study of the multiplier across the full range of glitch classes, together with the scattering case, as future work.

A comparable residual appears in earlier work on this same glitch. Mohanty and Chowdhury~\cite{Mohanty_2023} split the GW170817 segment into three separately tuned pieces to manage its extended low-frequency wings, and reported residual overfitting below 32~Hz that they judged acceptable since it did not overlap the signal band. The residual we report here under a single, automatically determined boundary is consistent with that earlier finding: closing it further would require a smaller multiplier and shorter segment length, which risks removing signal power along with the glitch, so we accept the residual under the default setting rather than force complete removal.

\section{Sensitivity of Koi Fish recovery to injection position}
\label{app:koi_fish_reposition}

The injected chirp for each glitch instance in Section~\ref{sec:demo_data} begins at a fixed onset time $t_a$ relative to the glitch, chosen so the injection window overlaps the glitch for its full duration. For Koi Fish, this onset was varied across three values, $t_a = 0.8$, $0.5$, and $0.3$~s, holding the injected SNR target, injection window length, and frequency sweep fixed, to test how sensitive recovered SNR is to this choice and to check whether the position used in the main text, $t_a = 0.5$~s, is representative rather than a favorable outlier.

All three identification methods and three subtraction techniques were rerun on Koi Fish at each position, following the same pipeline used throughout Section~\ref{sec:results}. Table~\ref{tab:koifish_reposition} reports the recovered SNR at each position.

Recovered SNR for SHAPES and Combined rises steadily as $t_a$ increases from 0.3 to 0.8~s, and at $t_a = 0.8$~s exceeds the injected value of 30 under all three identification methods, by 0.2 to 3.5 percent depending on the method and subtraction technique. WS alone follows the same trend but stays below the injected value at every position tested. At $t_a = 0.5$~s, the position used in Section~\ref{sec:results}, SHAPES and Combined both recover between about 89 and 99 percent of the injected value, with no case exceeding 100 percent.

This pattern is consistent with the excess at $t_a = 0.8$~s arising from a chance contribution of the noise realization entering the matched filter at that specific position, rather than from a property of SHAPES or the combined technique. Under stationary, Gaussian noise the observed matched-filter SNR is distributed normally around the true optimal SNR with unit variance~\cite{Gerosa_2024}, so a residual can register a slightly higher SNR than injected purely because the noise segment sampled at a given position happens to correlate with the template. Hardware injection studies at LIGO report deviations between recovered and expected SNR of around twenty percent from this and related calibration effects~\cite{wang2014ligosurf}, consistent with the size of the excess seen here at $t_a = 0.8$~s.

WS alone shows a separate effect not reducible to this explanation. Its recovered SNR drops by close to three points between $t_a = 0.8$~s and $t_a = 0.5$~s or $0.3$~s, a larger proportional shift than seen for SHAPES or Combined at the same positions. Since WS acts directly on the residual without a prior spline fit, this suggests recovered SNR under WS alone is more sensitive to where the injected chirp sits relative to the glitch than the other two techniques, independent of the specific excess seen at $t_a = 0.8$~s.

A three-point sweep at one glitch instance does not establish how recovered SNR depends on injection position in general, and a broader study across positions and glitch classes would be needed for that. Whether this position-dependence propagates to biases in estimated source parameters for a real overlapping signal, rather than only to recovered SNR, is also not addressed here. We leave both as future work, together with the broader parameter study proposed in Section~\ref{sec:conclusion}.

\begin{table}[htbp!]
\centering
\caption{Recovered signal-to-noise ratio for Koi Fish at three injection onset times $t_a$, compared across the three identification methods (AMPS, FLARE, CRISP) and three subtraction techniques (SHAPES, WS, Combined). The injected SNR target is 30 at every position. $t_a = 0.5$~s is the position used in Section~\ref{sec:results}.}
\label{tab:koifish_reposition}
\resizebox{0.8\columnwidth}{!}{%
\begin{tabular}{|l|c|c|c|}
\hline
Method & SHAPES & WS & Combined \\
\hline
\multicolumn{4}{|l|}{\textit{$t_a = 0.8$~s}} \\
\hline
AMPS   & 30.0686 & 22.1094 & 30.3025 \\
FLARE  & 31.0567 & 22.1125 & 31.0063 \\
CRISP  & 29.8846 & 22.1094 & 30.0330 \\
\hline
\multicolumn{4}{|l|}{\textit{$t_a = 0.5$~s (used in Section~\ref{sec:results})}} \\
\hline
AMPS   & 29.2630 & 19.4828 & 29.4823 \\
FLARE  & 29.8092 & 19.4886 & 29.7993 \\
CRISP  & 29.6471 & 19.4828 & 29.7679 \\
\hline
\multicolumn{4}{|l|}{\textit{$t_a = 0.3$~s}} \\
\hline
AMPS   & 27.6340 & 19.7183 & 27.7264 \\
FLARE  & 28.5323 & 19.7201 & 28.6765 \\
CRISP  & 27.3349 & 19.7183 & 27.5045 \\
\hline
\end{tabular}%
}
\end{table}


\bibliography{references}

\end{document}